 \documentclass[onecolumn,aps,prd,showpacs,superscriptaddress,floats,floatfix,showkeys,notitlepage,nofootinbib]{revtex4-1}

\usepackage{enumitem}
\usepackage{times}
\usepackage{dcolumn}
\usepackage{mathrsfs}
\usepackage{hyperref}
\usepackage{amsmath}
\usepackage{mathtools}
\usepackage{bbm}
\usepackage{bm}
\usepackage{amsfonts}
\usepackage{amssymb}
\usepackage{mathrsfs}
\usepackage{wasysym}
\usepackage{graphicx}
\usepackage[]{subfigure}
\usepackage{filecontents}
\usepackage[dvipsnames]{xcolor}
\usepackage{bbold}
\usepackage{lipsum}
\usepackage{graphicx}
\usepackage{subfigure}
\usepackage{palatino}
\usepackage{changes}
\usepackage{hyperref}
\hypersetup{colorlinks=true,linkcolor=blue,urlcolor=blue,citecolor=blue}
\usepackage[toc,page]{appendix}
\usepackage[normalem]{ulem}
\usepackage{adjustbox}
\usepackage{latexsym}
\usepackage{amsmath}
\usepackage{amssymb}
\usepackage{amsfonts}
\usepackage{times}
\usepackage{dcolumn}
\usepackage{bm}
\usepackage{tikz}
\usepackage{bigints}
\usepackage{array,tabularx,multirow}
\usepackage[dvipsnames]{xcolor}
\usepackage[tracking=true]{microtype}
\SetTracking{}{500}
\SetTracking{encoding={*}, shape=sc}{40}
\UseRawInputEncoding %for inputenc error%
\allowdisplaybreaks

\usepackage{braket}
\usepackage{tensor}
\usepackage{slashed}
\usepackage{booktabs} % to make nice table
\usepackage{float}

\newcommand{\LF}{\mathcal{L}_{\mathcal{F}}}
\newcommand{\CL}{\mathcal{L}}
\newcommand{\F}{\mathcal{F}}
\newcommand{\E}{\mathcal{E}}
\newcommand{\ed}{\mathrm{d}}
\newcommand{\arccosh}{\operatorname{arccosh}}
\begin{document} \sloppy

\title{Observational Signatures: Shadow cast by %an 
the effective metric of photons for black holes with rational non-linear electrodynamics}

%\author{Akhil Uniyal\orcidlink{0000-0001-8213-646X}}
\author{Akhil Uniyal}
\email{akhil\_uniyal@sjtu.edu.cn}
%\affiliation{Department of Physics, Indian Institute of Technology, Guwahati 781039, India}
\affiliation{Tsung-Dao Lee Institute, Shanghai Jiao Tong University, Shengrong Road 520, Shanghai, 201210, People’s Republic of China}

\author{Sayan Chakrabarti}
\email{sayan.chakrabarti@iitg.ac.in}
\affiliation{Department of Physics, Indian Institute of Technology, Guwahati 781039, India}

\author{Mohsen Fathi}
\email{mohsen.fathi@usach.cl}
\affiliation{Departamento de F\'{i}sica, Universidad de Santiago de Chile,
Avenida V\'{i}ctor Jara 3493,  Estaci\'{o}n Central, 9170124, Santiago, Chile}

\author{Ali \"Ovg\"un}
\email{ali.ovgun@emu.edu.tr}
\affiliation{Physics Department, Eastern Mediterranean University, Famagusta, 99628 North Cyprus via Mersin 10, Turkiye}

\date{\today}
\begin{abstract}

 %In this paper, we investigate a spherically symmetric non-linear electrodynamics black hole and its impact on light rays. We construct the effective metric, finding the radial coordinate within the black hole horizon. Photon trajectories are studied, with increasing magnetic charge expanding the horizon and emission range. Using Event Horizon Telescope data, we constrain parameters and analyze emission profiles. Direct emission dominates while lensing rings contribute less. We compare infalling accretion with the Schwarzschild black hole, showing higher intensity but a larger angular region for the non-linear electrodynamics black hole. This work sheds light on modified spacetimes and their effects on black hole properties.

 This study explores spherically symmetric non-linear electrodynamics black holes and their effects on light propagation. We derive the governing metric, revealing radial coordinate dynamics within the event horizon. We analyze photon trajectories, finding that increasing magnetic charge expands the horizon and emission range. \textcolor{black}{Furthermore with the help of the Event Horizon Telescope results}, we constrain parameters and emission profiles. Direct emission dominates, while lensing rings play a lesser role. Comparing with Schwarzschild black holes, we observe higher intensity but a wider emission region in non-linear electrodynamics black holes. This work enhances our understanding of modified spacetimes and their impact on black hole properties.

%In this paper, we investigate a spherically symmetric non-linear electrodynamic black hole %(NED BH)  solution and its impact on the behavior of light rays. We construct the effective metric and find that the radial coordinate lies within the black hole horizon. Photon trajectories around the black hole are studied, revealing that increasing magnetic charge expands the horizon and emission range. Using data from the Event Horizon Telescope, %(EHT) we constrain the parameters and analyze emission profiles. Direct emission dominates while lensing rings have a smaller contribution. We also compare infalling accretion with the Schwarzschild black hole, showing higher intensity but larger angular region for the non-linear electrodynamic black hole. Overall, this work provides insights into modified spacetimes and their effects on black hole properties.
\end{abstract}

\maketitle

\date{\today}

%\tableofcontents

%%%%%%%%%%%%%%
\section{Introduction}\label{sec:intro}

Black holes (BHs) continue to captivate scientists due to their enigmatic nature and their ongoing challenge to our understanding. The theoretical framework for BHs originates from the pioneering works of Schwarzschild \cite{Schwarzschild:1916} and subsequent contributions by Finkelstein \cite{Finkelstein:1958}. The exploration of BHs gained significant momentum after the discovery of Cygnus X-1 and its subsequent identification \cite{webster_cygnus_1972, bolton_identification_1972}. However, the groundbreaking observations of M87* \cite{EventHorizonTelescope:2019dse} and Sgr A* \cite{EventHorizonTelescope:2022wkp} by the Event Horizon Telescope (EHT) propelled the BH research forward. These observations provided profound insights into the behavior of light in the strong gravitational fields of BHs. \textcolor{black}{It is noteworthy that the concept of searching for a black hole at the galactic center existed previously \cite{Falcke:1999pj}. The idea of reconstructing the black hole's shadow using global interferometers operating in the mm wavelength at the galactic center was initially proposed in Ref. \cite{ZAKHAROV2005479}. These concepts laid the foundation for testing general relativity (GR), as envisioned in the thought experiment by J. Bardeen \cite{1973ApJ...183..237C, Dewitt1973BlackH}, which aims to observe the shadow of the black hole at the galactic center \cite{Zakharov:2023yjl}.} Furthermore, the EHT observations of M87* unveiled the presence of a mysterious magnetic field that may hold clues to the origin of its powerful jets \cite{2021ApJ...910L..12E,2021ApJ...910L..13E,2021PhRvD.103j4047K}. Similarly, assessments of Sgr A* by the EHT explored alternative theories of gravity beyond GR, offering constraints on modified gravity theories \cite{EventHorizonTelescope:2022xqj}, which have been recently employed in Ref. \cite{Vagnozzi:2022moj} to constrain modified theories of gravity. These assessments also shed light on the possibility of compact objects at the cores of galaxies, potentially constituting active galactic nuclei (AGNs). Understanding BHs can provide invaluable insights into the fundamental nature of the Universe, as their extreme gravitational fields serve as unique testing grounds for theories that extend beyond terrestrial laboratories.

Indeed, despite the remarkable successes of GR in passing astrophysical tests \cite{Will:2014kxa}, the theory still leaves unanswered questions, such as the origins of the accelerated expansion of the universe, the flat galactic rotations curves, anti-lensing, anisotropies
on the cosmic microwave background radiation, the coincidence problem and etc. \cite{Rubin1980,Massey2010,Bolejko2013,Riess:1998,Perlmutter:1999,Astier:2012}. Many scientists believe that the aforementioned phenomena arise from the mysterious aspects of the universe, which have not been adequately explained thus far. In order to account for these phenomena, it is believed that modifications to GR are necessary \cite{easson_modified_2004,nojiri_introduction_2007,trodden_cosmic_2007}. 

%\textcolor{black}
{Our study is motivated by the need to understand deviations from the linear superposition of electromagnetic fields, which are well-established at macroscopic and atomic levels but become significant at the subatomic level due to the intense fields near charged particles. This departure from linearity challenges the classical Maxwell electromagnetic theory, leading to singularities. When subjected to strong electromagnetic fields (EM), the behavior of light can be effectively likened to its passage through a dispersive medium. As the electromagnetic field strength nears critical thresholds, such as the critical electric field ($E_\mathrm{cr} \approx 10^{18} \,\mathrm{V}/\mathrm{m})$ or the critical magnetic field ($B_\mathrm{cr} \approx 10^{9}$ T), the influence of external fields on the quantum properties of the vacuum becomes notably pronounced \cite{Novello:1999pg}. These effects can be phenomenally described by classical theories characterized by Lagrangian that exhibit non-linear dependence on the two fundamental electromagnetic invariants.  In the presence of extremely strong electromagnetic fields, such as near critical values, the effects on vacuum quantum properties become notable. 
Additionally, there have been discussions regarding the possibility of removing BH singularities in the framework of GR by employing the Born-Infeld non-linear electrodynamics (NED) model \cite{Born:1934gh}. This approach enables the generation of regular BH spacetimes. The idea was originally sparked by Bardeen, who introduced a regular spherically symmetric BH with a purely magnetic charge using the linear Maxwell theory \cite{bardeen1968non}. \textcolor{black}{Similarly, an analytical expression for the shadow size as a function of charge was recently derived using the Reissner-Nordström (RN) metric \cite{Zakharov:2014lqa}. This expression has been utilized to evaluate black hole models in spaces with extra dimensions \cite{Zakharov:2011zz}. Additionally, the consideration of tidal charges in supermassive black hole candidates, such as M87* and Sgr A*, has led to constraints on these tidal charges. The analytical expression for the shadow size in the RN metric, in conjunction with EHT data, has played a crucial role in obtaining these constraints \cite{Zakharov:2021gbg}.} We are particularly interested in the rational NED.  This concept has given rise to the creation of numerous regular BH spacetimes and continues to be a significant area of research in the field of BH studies (see for examples Refs. \cite{ayon-beato_regular_1998,hendi_slowly_2014,he_2_2017,bronnikov_nonlinear_2018,kuang_nonlinear_2018,wang_thermodynamics_2019,kruglov_shadow_2020,javed_effect_2020,paula_electrically_2020,cataldo_thermodynamics_2021,kruglov_remarks_2021,okyay_nonlinear_2022,bronnikov_regular_2022,assrary_effect_2022,javed_weak_2023,sun_black_2023,canate_transforming_2023,Guzman-Herrera:2023zsv}). In the present study, the investigation includes the consideration of a particular model presented in Ref. \cite{Kruglov:2023cyb}, adding to the existing body of research in this field. Therefore, through the removal of singularities and incorporation of quantum corrections, these models aim to shed light on the behavior of the electromagnetic field near extremely gravitating systems. As a result, they have become a subject of great interest within the scientific community, as they have the potential to explore unresolved phenomena in modern cosmology, such as the Big Bang singularity, cosmic inflation, and the universe's accelerated expansion (see, for example, Refs. \cite{Garcia-Salcedo:2000ujn, Camara:2004, Novello:2003kh, novello_cosmological_2007, Vollick:2008}). 

In this study, driven by the same research interest, our focus lies on investigating the observational signatures of a NED BH. Specifically, we aim to provide precise constraints on the shadow size and the angle of light deflection associated with such a BH.  The investigation of the gravitational lensing effect caused by BHs is an active research area in the fields of astrophysics and cosmology. Weak gravitational lensing refers to the phenomenon where the trajectory of light is slightly deflected when it traverses a region affected by a gravitational field. As a consequence, distant objects such as galaxies and quasars can appear distorted in their images, and in some cases, multiple images of the same object can be formed. The verification of Einstein's theory of relativity through the Eddington experiment, which involved observing gravitational lensing, established this method as a crucial tool in astrophysics. As a result, numerous studies and papers have since focused on utilizing weak gravitational lensing for various astrophysical investigations \cite{Bozza:2001,Bozza:2002,Virbhadra:2007kw,Virbhadra:2008ws,Adler:2022qtb,Virbhadra:2022ybp,Virbhadra:2022iiy,Jusufi:2017mav,Ovgun:2018fnk,Mangut:2021suk}. In fact, the theoretical study of BH shadows and their constraints based on observational data has garnered significant interest among scientists, leading to numerous dedicated publications (see, for instance, References \cite{Kuang:2022ojj,Kuang:2022xjp,Meng:2022kjs,Tang:2022hsu,Abdujabbarov:2012bn, Yumoto:2012kz, Atamurotov:2013sca, Zakharov:2014lqa,Zakharov:2021gbg,Zakharov:2018awx,Konoplya2019,Tsukamoto:2017fxq,Kumar2019,Khodadi:2022pqh,Belhaj:2020rdb,Belhaj:2020okh,Konoplya:2021slg,Konoplya:2020bxa,Konoplya:2019xmn, Chakhchi:2022fl, Devi:2021ctm,Kumaran:2022soh,Symmergent-bh,Symmergent-bh2,Pantig:2022qak,Atamurotov:2022knb,Dymnikova2019,Papnoi:2014aaa, Johannsen:2015hib, Johannsen:2015mdd, Moffat:2015kva, Giddings:2016btb, Cunha:2016bjh, Tsukamoto:2017fxq, Hennigar:2018hza, Cunha:2018acu, Allahyari:2019jqz, Abdikamalov:2019ztb, Ovgun:2019jdo, Shaikh:2019fpu, bambi_testing_2019, vagnozzi_hunting_2019, Kumar:2020hgm, Li2020, Ovgun:2020gjz, khodadi_black_2020, Zhong:2021mty, Zuluaga:2021vjc, Stashko:2021lad, Rahaman:2021kge, Ovgun:2021ttv, Pantig:2022whj, Bisnovatyi-Kogan:2022ujt, Kazempour:2022asl, Pantig:2022ely, roy_superradiance_2022, chen_superradiant_2022,Vagnozzi:2022moj}). Recently, the advent of silhouette imaging by the EHT has amplified the importance of reliable methods for visualizing BHs with accretion disks as their sources of illumination. This significance was initially recognized by Luminet in 1979 \cite{Luminet:1979nyg}, who computed the radiation emitted from a thin accretion disk surrounding a Schwarzschild BH and proposed a ray-traced image of the disk. Generally, this type of accretion is based on models such as the Shakura-Sunyaev \cite{Shakura:1973}, Novikov-Thorne \cite{Novikov:1973}, and Page-Thorne \cite{page_disk-accretion_1974} models, where the disk is assumed to be thin both geometrically and optically. In light of these assumptions and the growing interest in BH imaging, a new method for simulating higher-order light rings in BHs with thin accretion disks was proposed in Ref.\cite{Gralla:2019}. Since then, this method has been employed in several publications (see, for example, Refs.\cite{Chakhchi:2022fls,Guerrero_2021, li_observational_2021, okyay_nonlinear_2022, PhysRevD.105.084057, hu_observational_2022, PhysRevD.106.044070, Guo_2022, WANG2022116026, PhysRevD.105.064031, Uniyal:2022vdu, Uniyal:2023inx,Bapat:2020xfa,Kumaran:2022soh,Kumaran:2023hlu,Pantig:2022gih}), and is also of importance in our paper for the analysis of the shadow of the aforementioned NED BH. This BH possesses a distinctive and intriguing characteristic: it alters the geometric background through which photons propagate. In the context of linear electrodynamics and vacuum, it is well-established that electromagnetic waves travel along the null geodesics of spacetime. However, in the case of self-interacting or NED theories, this is no longer true. Instead, light rays deviate from the null geodesics and traverse the effective metric background, which is modified from the original metric \cite{Novello:1999pg, Novello:2001fv}. This phenomenon is also observed in perturbative theories, where in the high-energy limit, the perturbative effective potential of NED coincides with a function governing the motion of photons in the gravitational field of a central object \cite{Moreno:2002gg, Li:2014fka, Toshmatov:2019gxg}. Thus, the NED BH exhibits a remarkable interplay between its gravitational and electromagnetic properties, leading to deviations from the expected behavior based on linear electrodynamics.

This paper focuses on investigating the detailed impact of the effective metric background on the trajectories of photons, as well as its potential influence on observational %features
signatures in the vicinity of a magnetically charged spherically symmetric NED BH. This effective metric is obtained by modifying the original spacetime metric. To achieve our objective, we structure this paper as follows: Sect. \ref{sec:NEDBH} introduces the physical origin and spacetime structure of the NED BH, providing essential background information. The derivation of the effective metric is presented in Section \ref{sec:effMetric}, which also includes an analysis of the behavior of light rays in the BH's exterior. In Sect. \ref{sec:shadowCons}, we calculate the diameter of the BH's shadow and compare it with observations of M87* and Sgr A*, enabling us to constrain the NED parameters of the spacetime. Next, in Sect. \ref{sec:defAngle}, we employ fully algebraic methods to compute the weak deflection angle of light near the BH. Section \ref{sec:signatures} employs ray-tracing techniques to visualize the BH's shadow when an optically thin accretion disk with different emission profiles is present. Furthermore, in Sect. \ref{sec:signaturesInfall}, we extend the same procedure to visualize the BH's shadow under the condition of infalling accretion. Finally, we conclude our study in Sect. \ref{sec:conclusions}, summarizing our findings and discussing potential future research directions. \textcolor{black}{Throughout the paper, we have considered the natural unit system, in which, $G=c=M=1$.}

\section{BHs with rational %non-linear electrodynamics
NED}\label{sec:NEDBH}

%\textcolor{Orange}{Mohsen esta trabajando en esta parte!}\\

In this study, we consider the rational NED theory proposed by Kruglov in Ref. \cite{Kruglov:2023cyb1}. The Lagrangian density employed in this framework is %\textcolor{black}
{considered such that it obeys the correspondence principle. This principle states that in the weak field limit, the non-linearity should be absent, ensuring that the field equations align with the classical Maxwell equations of classical electrodynamics, as stated in Ref.~\cite{2015AnPhy.353..299K}. Hence, we opt the form}
\begin{equation}
{\cal L} = -\frac{{\cal F}}{2\beta{\cal F}+1}.
 \label{eq:1}
\end{equation}
In the given expression, the parameter $\beta$ is a non-negative quantity with dimensions of (length)$^4$. The quantity ${\cal F}$ is defined as ${\cal F}=(1/4)F_{\mu\nu}F^{\mu\nu}=(B^2-E^2)/2$, where $F_{\mu\nu}=\partial_\mu A_\nu-\partial_\nu A_\mu$ represents the field tensor. The symmetrical energy-momentum tensor is given by \cite{Kruglov:2016ymq}
\begin{equation}
T_{\mu\nu}=-\frac{F_\mu^{~\alpha}F_{\nu\alpha}}{(1+2\beta{\cal F})^{2}}
-g_{\mu\nu}{\cal L},
\label{eq:2}
\end{equation}
by means of which, we derive the energy density
\begin{equation}
\rho=T_0^{~0}=\frac{{\cal F}}{1+2\beta{\cal F}}
+\frac{E^2}{(1+2\beta{\cal F})^2}.
\label{eq:3}
\end{equation}
For a healthy theory, the general principles of causality and unitarity must be upheld. According to the principle of causality, the group velocity of excitations over the background must be less than the speed of light, ensuring the absence of tachyons in the theory. The unitarity principle guarantees the absence of ghosts. These principles are satisfied in the case of $\bm{E}\cdot\bm{B}=0$, if the following inequalities are upheld \cite{Shabad:2011hf}:
\begin{subequations}
    \begin{align}
        & {\cal L}_{\cal F}\leq 0,~~~~{\cal L}_{{\cal F}{\cal F}}\geq 0,\label{eq:4a}\\
        & {\cal L}_{\cal F}+2{\cal F} {\cal L}_{{\cal F}{\cal F}}\leq 0,\label{eq:4b}
    \end{align}
    \label{eq:4}
\end{subequations}
%\[
% {\cal L}_{\cal F}\leq 0,~~~~{\cal L}_{{\cal F}{\cal F}}\geq 0,
%\] \begin{equation}
%{\cal L}_{\cal F}+2{\cal F} {\cal L}_{{\cal F}{\cal F}}\leq 0,
%\end{equation}
where ${\cal L}_{\cal F}\equiv\partial{\cal L}/\partial{\cal F}$.
Therefore, utilizing Eq. \eqref{eq:1}, we can derive
\begin{subequations}
    \begin{align}
        & {\cal L}_{\cal F}= -\frac{1}{(1+2\beta{\cal F})^2},\label{eq:5a}\\
        & {\cal L}_{{\cal F}{\cal F}}=\frac{4\beta}{(1+2\beta{\cal F})^3},\label{eq:5b}\\
        & {\cal L}_{\cal F}+2{\cal F} {\cal L}_{{\cal F}{\cal F}}=\frac{6\beta{\cal F}-1}{(1+2\beta{\cal F})^3}.\label{eq:5c}
    \end{align}
    \label{eq:5}
\end{subequations}
%\[
%{\cal L}_{\cal F}= -\frac{1}{(1+2\beta{\cal F})^2},
%\]
%\begin{equation}
%{\cal L}_{\cal F}+2{\cal F} {\cal L}_{{\cal F}{\cal F}}=\frac{6\beta{\cal F}-1}{(1+2\beta{\cal F})^3},~~~~
%{\cal L}_{{\cal F}{\cal F}}=\frac{4\beta}{(1+2\beta{\cal F})^3}.
%\label{eq:5}
%\end{equation}
Based on Eqs. \eqref{eq:4} and \eqref{eq:5}, we can deduce that the principles of causality and unitarity are satisfied when $6\beta{\cal F}\leq 1$ ($\beta\geq 0$). Consequently, when $\bm{E}=0$, we have $\beta{B}^2\leq 1/3$.
Considering a static magnetic BH %\footnote{In the paper M.-S. Ma, Ann. Phys. \textbf{362}, 529 (2015) the author also considered the static magnetic BH based on NED proposed in \cite{Krug2}. However, here we use unitless variables that are more convenient for the analyses of the BH thermodynamics. In addition, we analyse more general case when the BH besides the electromagnetic mass possesses the Schwarzschild mass (having non-electromagnetic nature).}.
and taking into account the absence of electric charge ($q_e=0$) and assuming ${\cal F}=q_m^2/(2r^4)$ (where $q_m$ represents the magnetic charge), we can derive the expression for the magnetic energy density from Eq. \eqref{eq:3} as follows:
\begin{equation}
\rho_M=\frac{B^2}{2(\beta B^2+1)}=\frac{q_m^2}{2(r^4+\beta q_m^2)}.
\label{eq:20}
\end{equation}
Now, let us consider the line element 
\begin{equation}
\ed s^2 = -A(r) \ed t^2 + \frac{1}{A(r)} \ed r^2 + r^2 \left(\ed\theta^2 + \sin^2\theta \ed\phi^2\right),
\label{eq:21s}
\end{equation}
with the %metric 
lapse function being defined by
\begin{equation}
A(r) = 1 - \frac{2M(r)}{r},
\label{eq:21m}
\end{equation}
in which the mass function is
\begin{equation}
M(r)=m_0+\int_0^r \rho(r) r^2 \ed r=m_0+m_M-\int_r^\infty \rho(r) r^2 \ed r.
\label{eq:21ma}
\end{equation}
In this context, the BH's total mass is given by the sum of the Schwarzschild mass $m_0$ and the magnetic mass $m_M=\int_0^\infty \rho(r) r^2 \ed r$. Thus, utilizing Eqs. \eqref{eq:20} and \eqref{eq:21ma}, we can express the mass function as 
\begin{equation}
M(x)=m_0+\frac{q_m^{3/2}}{8\sqrt{2}\beta^{1/4}}\left[\ln\frac{x^2-\sqrt{2}x+1}{x^2+\sqrt{2}x+1}
+2\arctan\left(\sqrt{2}x+1\right)-2\arctan\left(1-\sqrt{2}x\right)\right],
\label{eq:21}
\end{equation}
where $x=r/\sqrt[4]{\beta q_m^2}$, \textcolor{black}{and for $x>0$, the expression inside the logarithm is positive.} %\textcolor{red}{...$x$ diverges at $q_m=0$!}. 
On the other hand, the magnetic mass of the BH is determined as follows:
\begin{equation}
m_M=\int_0^\infty\rho_M(r)r^2\ed r=\frac{\pi q_m^{3/2}}{4\sqrt{2}\beta^{1/4}}\approx 0.56\frac{q_m^{3/2}}{\beta^{1/4}}.
\label{eq:22}
\end{equation}
As expected, when $q_m=0$, the magnetic mass $m_M$ becomes zero, resulting in the Schwarzschild BH. Then, the %metric 
lapse function can be obtained by employing Eqs. \eqref{eq:21m} and \eqref{eq:21s}, resulting in the following expression:
\begin{equation}
A(x)=1-\frac{2m_0G}{\sqrt[4]{\beta q_m^2}x}-\frac{q_mG}{4\sqrt{2\beta}x}\left[\ln\frac{x^2-\sqrt{2}x+1}{x^2+\sqrt{2}x+1}+2\arctan\left(\sqrt{2}x+1\right)-2\arctan\left(1-\sqrt{2}x\right)\right].
\label{eq:A(r)}
\end{equation}
In the limit of $r\rightarrow\infty$, the %metric 
lapse function \eqref{eq:A(r)} can be approximated as
\begin{equation}
A(r)=1-\frac{2mG}{r}+\frac{q_m^2G}{r^2}+{\cal O}(r^{-5}),
%~~~~r\rightarrow \infty,
\label{eq:24}
\end{equation}
where $m=m_0+m_M$.
Consequently, one can deduce from Eq. \eqref{eq:24} that the correction to the Reissner-Nordstr\"{o}m (RN) solution, is of the order ${\cal O}(r^{-5})$. Additionally, when $m_0=0$ and $r\rightarrow 0$, Eq. \eqref{eq:A(r)} indicates the presence of the asymptotic 
\begin{equation}
A(r)=1-\frac{Gr^2}{\beta}+\frac{Gr^6}{7\beta^2q_m^2}-\frac{Gr^{10}}{11\beta^3q_m^4}+{\cal O}(r^{12}),
%~~~~r\rightarrow 0.
\label{eq:25}
\end{equation}
possessing a de Sitter core. Note that, the solution given by Eq. \eqref{eq:25} is regular, as it approaches unity as $r$ tends to zero. However, when $m_0\neq 0$, the solution becomes singular, leading to $A(r)$ diverging to infinity.

With the introduction of the geometrical structure of the BH spacetime under consideration, we can now move forward to the main objectives of this study. We begin by examining the dynamics of light rays within the effective geometry of the BH.

%%%%%%%%%%%%%%%%%%%%%%%%%%%%%%%
%\section{Physical Properties of the magnetic BH}
%\label{SecIVB}
\section{%Shadow of NED BH 
Light propagation around the NED BH exterior in the effective metric}\label{sec:effMetric}
\label{SecIIB}

In the context of NED, electromagnetic fluctuations travel along an \textit{effective} light cone \cite{PhysRevD.91.083008,Schee_2015,Stuchlik:2015IJMPD,2017PhLB..771..597K,2018PhRvD..97h4058T,Schee_2019,schee_profiled_2019,PhysRevD.91.083008,stuchlik_shadow_2019,stuchlik_generic_2019,dePaula:2023ozi}, which generally differs from the standard geometrically-defined light cones \cite{Plebanski:1970zz, Boillat:1970gw}. Notably, for a general theory of NED, characterized by two independent four-dimensional relativistic invariants, $\bm{F}$ (as defined above) and $\bm{F} \star \bm{F}$, there exist (in general) two effective light cones, each associated with a specific polarization. This phenomenon is referred to as \textit{birefringence}, and it supports the interpretation of electromagnetic fluctuations propagating on a NED background as a medium (independent of their coupling to gravity). In the case of NED models solely dependent on $\bm{F}$ (with no dependence on $\bm{F} \star \bm{F}$), birefringence does not occur in general\footnote{However, birefringence phenomena can arise in NED models that solely depend on $\bm{F}$ when external magnetic fields are present \cite{Gaete:2017cpc, Gaete:2021ytm}.}. In such scenarios, the single effective light cone can be geometrically described by considering photons propagating along null geodesics of an effective metric tensor $g^{\mu\nu}_{\rm{eff}}$, which relies on the contributions of the NED source to the energy-momentum tensor \cite{Plebanski:1970zz, Boillat:1970gw, Gutierrez:1981ed, Novello:1999pg}. The expression for the effective metric tensor is given as \cite{Novello:1999pg}
\begin{equation}
\label{EFF_GEO}
g^{\mu\nu}_{\rm{eff}} = g^{\mu\nu} \mathcal{L}_\F - 4 \mathcal{L}_{\F\F} F^\mu_\alpha F^\alpha_\nu.
\end{equation}
And, therefore, for a magnetically charged spherically symmetric BH, the line element will take the following form:
\begin{equation}
\label{LE_EG1}
\ed s^{2}_{\rm{eff}} = g_{\mu\nu}^{\rm{eff}}\ed x^{\mu}\ed x^{\nu} = \frac{1}{\mathcal{L}_\F}\left( g_{tt}\ed t^{2}+g_{rr} \ed r^{2} \right)+\frac{g_{\theta \theta}}{\Phi} \ed\theta^{2}+\frac{g_{\phi \phi}}{\Phi} \ed\phi^{2},
\end{equation}
where
\begin{equation}
\label{LE_EG2}\Phi=\mathcal{L}_\F + 2 \F \mathcal{L}_{\F\F}.
\end{equation}
In this case, the Lagrangian associated with the geodesic motion in the spacetime described by the line element \eqref{LE_EG1} is defined as
\begin{equation}
    \mathscr{L}= \dfrac{1}{\mathcal{L}_\F}\left(g_{tt}\Dot{t}^{2}+g_{rr} \Dot{r}^{2} \right)+\frac{g_{\theta \theta}}{\Phi} \Dot{\theta}^2+\frac{g_{\phi \phi}}{\Phi} \Dot{\phi}^2,
\end{equation}
where the dot represents the derivative with respect to the affine parameter. Now, considering the equatorial plane (i.e. $\theta=\pi/2$), we can express the equations of motion for null geodesics as 
\begin{eqnarray}
&& \dot{t} = -\dfrac{\E\mathcal{L}_\F}{g_{tt}}, \label{eqm1_EG}\\
%\label{eqm2_EG}\dot{r} &= -A(r)\Phi \bar{p}_{r}  ,\\
&& \dot{\phi} =  \dfrac{L \Phi}{g_{\phi \phi} }, \label{eqm3_EG}\\
&& \dfrac{1}{\mathcal{L}_\F}\left( g_{tt} \Dot{t}^{2}+g_{rr} \Dot{r}^{2} \right)+\frac{g_{\phi \phi}}{\Phi} \Dot{\phi}^2 = 0, \label{eqm2_EG}
\end{eqnarray}
where $\E$ and $L$ represent the energy and angular momentum associated with the null geodesics, respectively. Combining these equations yields
\begin{equation}
    \left(\frac{dr}{d\phi} \right)^2= %V(r) = 
    - \frac{g_{\phi \phi} \mathcal{L}_\F}{g_{rr} \Phi} - \frac{\E^2  g_{\phi \phi}^2 \mathcal{L}_\F^2}{L^2 g_{tt} g_{rr} \Phi^2 }.
    \label{eq:drdphi_0}
\end{equation}
To determine the turning point where circular orbits occur, we initially employ the condition of $\dot{r}=0$. By utilizing Eqs. \eqref{eqm1_EG}--\eqref{eqm2_EG}, we can derive the impact parameter associated with the null geodesics as 
%that at such orbits,  $V(r)=0=V'(r)$, where the prime denotes the derivative with respect to $r$. The first of these equations will yield the impact parameter associated with the null geodesics as
\begin{equation}
    b=\frac{L}{\E}=\sqrt{-\frac{g_{\phi \phi} \mathcal{L}_\F}{g_{tt} \Phi}}.
    \label{eq:impactParameter}
\end{equation}
The impact parameter plays a crucial role in determining the size of the BH shadow. Considering the effective metric \eqref{EFF_GEO}, and the relationship $g_{rr}=-g_{tt}^{-1}=A(r)^{-1}$, one can recast Eq. \eqref{eq:drdphi_0} as
\begin{equation}
\left(\frac{\ed r}{\ed\phi}\right)^2=\frac{r^2\CL_\F A(r)}{\Phi}\left[
\frac{h^2(r)}{b^2}-1
\right],
    \label{eq:drdphi_1}
\end{equation}
in which 
\begin{equation}
    h^2(r) = -\frac{\CL_\F}{\Phi}\frac{r^2}{A(r)}.
    \label{eq:h2(r)}
\end{equation}
Note that for marginally stable circular orbits, an additional condition must be imposed, namely $\ddot{r}=0$, which leads to the following result:
\begin{equation}
\frac{2b^2\Phi}{r^3 \CL_\F A(r)}-\frac{2 A'(r)}{A^3(r)}+\frac{b^2\Phi A'(r)}{r^2 \CL_\F A^2(r)}+\frac{b^2\Phi\CL'_\F}{r^2\CL_\F^2 A(r)}-\frac{b^2\Phi'}{r^2\CL_\F A(r)} = 0.
    \label{eq:rddot0}
\end{equation}
in which prime denotes differentiation with respect to $r$. Substituting the expression \eqref{eq:impactParameter} into the above relation yields
\begin{equation}
\Big[
r\Phi \CL'_\F+\CL_\F\Big(
2\Phi-r\Phi'
\Big)
\Big]A(r)-r\CL_\F\Phi A'(r)=0,
    \label{eq:rddot1}
\end{equation}
which is equivalent to the condition
\begin{equation}
\frac{\ed}{\ed r}h^2(r) = 0,
    \label{eq:dh2}
\end{equation}
that governs the radius of the photon sphere, denoted as $r_\mathrm{ph}$, where unstable circular orbits occur. 
Adopting the lapse function \eqref{eq:A(r)}, and expanding up to the fourth order of $r$, the condition \eqref{eq:dh2} yields the depressed quartic
\begin{equation}
r^4 +\mathrm{a} r + \mathrm{b} = 0,
    \label{eq:quartic_rph_0}
\end{equation}
%\begin{equation}
%r_{\mathrm{ph}} = \frac{2}{5}  \left(\sqrt{11}+2 \right) G m_0,
%    \label{eq:rph}
%\end{equation}
%which represents the radius of the photon sphere.
where 
\begin{subequations}
    \begin{align}
        & \mathrm{a} = -\frac{3\beta q_m^2}{52 G m_0},\\
        & \mathrm{b} = \frac{9 G m_0 \beta q_m^2}{52 G m_0}.
    \end{align}
    \label{eq:a,b}
\end{subequations}
The above equation has the four solutions %{\mathrm{ph}_1}
\begin{eqnarray}
    && r_{p_1} = \frac{1}{2}\left[-\bar{\mathrm{c}}+\sqrt{\bar{\mathrm{c}}^2-4\bar{\mathrm{d}}}\,\right],\label{eq:rp_1}\\
    && r_{p_2} = \frac{1}{2}\left[-\bar{\mathrm{c}}-\sqrt{\bar{\mathrm{c}}^2-4\bar{\mathrm{d}}}\,\right],\label{eq:rp_2}\\
    && r_{p_3} = \frac{1}{2}\left[-\bar{\mathrm{e}}+\sqrt{\bar{\mathrm{e}}^2+
    4\bar{\mathrm{f}}}\,\right],\label{eq:rp_3}\\
    && r_{p_4} = \frac{1}{2}\left[-\bar{\mathrm{e}}-\sqrt{\bar{\mathrm{e}}^2-4\bar{\mathrm{f}}}\,\right],\label{eq:rp_4} 
\end{eqnarray}
where
\begin{subequations}
    \begin{align}
    &   \bar{\mathrm{c}} = \sqrt{\bar{\mathrm{u}}_1},\\
    &  \bar{\mathrm{e}} = - \sqrt{\bar{\mathrm{u}}_1},\\
    & \bar{\mathrm{d}} = \frac{\bar{\mathrm{u}}_1}{2}-\frac{\mathrm{a}}  {2\sqrt{\bar{\mathrm{u}}_1}},\\
    & \bar{\mathrm{f}} = \frac{\bar{\mathrm{u}}_1}{2}+\frac{\mathrm{a}}{2\sqrt{\bar{\mathrm{u}}_1}},
    \end{align}
    \label{eq:cdef}
\end{subequations}
in which
\begin{equation}
\bar{\mathrm{u}}_1 = \sqrt{\frac{\bar\xi_2}{3}}\cosh\left(\frac{1}{3}\arccosh\left(3\bar\xi_3\sqrt{\frac{3}{\bar\xi_2^3}}\,\right)\right),
    \label{eq:baru}
\end{equation}
with $\bar\xi_2=16\mathrm{b}^2$ and $\bar\xi_3 = 4\mathrm{a}^2$. Note that, when we expand the differential equation \eqref{eq:dh2} up to the third order of $r$, it simplifies to the first order equation $r=3G m_0$. This equation provides the radius of the photon sphere around a Schwarzschild BH. Upon checking the solutions in Eqs. \eqref{eq:rp_1}--\eqref{eq:rp_4}, it becomes evident that $r_{p_3}$ and $r_{p_4}$ are of complex values (i.e. $r_{p_3}, r_{p_4} \in \mathbb{C}$), and therefore, we disregard them. Plotting the two remaining solutions in Fig.~\ref{fig:rph1234}, we observe that $r_{p_1}<0$, while $r_{p_2}>0$. Thus, we can define the radius of the photon sphere for the BH as $r_{\mathrm{ph}}=r_{p_2}$. 
\begin{figure}[t]
    \centering
    \includegraphics[width=9cm]{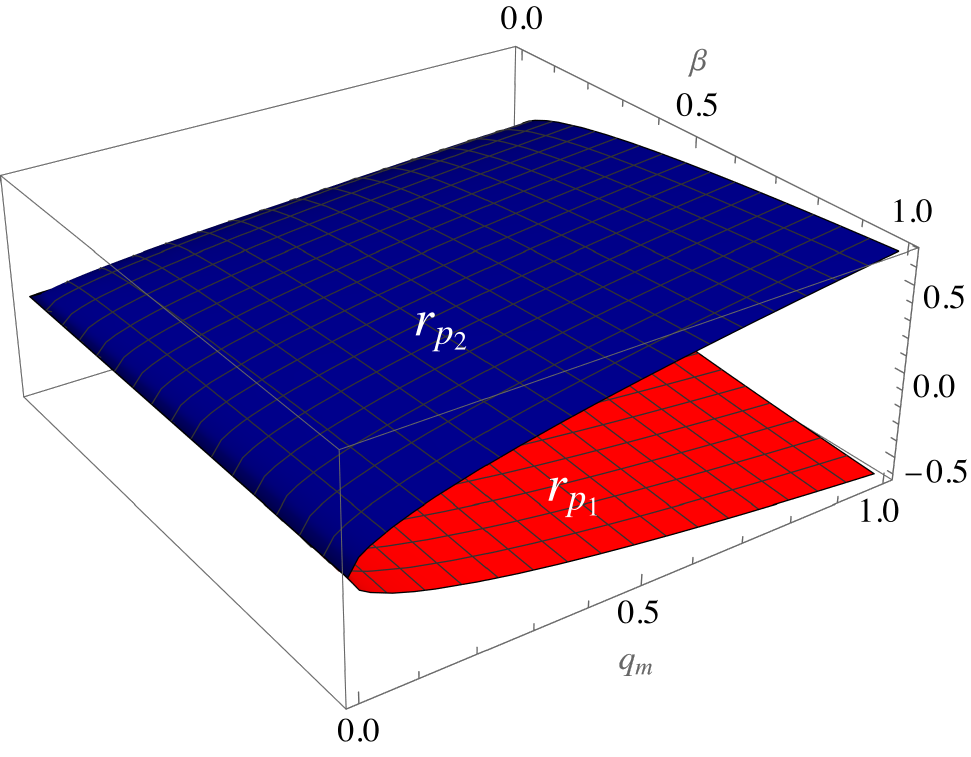}
    \caption{The behavior of $r_{p_i}, i=1,2$, concerning changes in the magnetic charge $q_m$ and the $\beta$-parameter.}
    \label{fig:rph1234}
\end{figure}

It is also essential to highlight that preserving the sign of the effective metric background in Eq. \eqref{LE_EG1}, requires the imposition of a specific condition on the %metric 
lapse function. To achieve this, we rigorously solve the equations and derive the precise condition governing the radial distance. This condition leads to the establishment of a minimum permissible value for the radial coordinate, denoted by $r_{\mathrm{eff}}= (3\beta q^2)^{1/4}$. We created a three-dimensional plot to explore the relationship between the parameters $\beta$
and $q$ concerning $r_{\mathrm{eff}}$ (on the left) and the BH horizon $r_\mathrm{h}$ (on the right) in Fig. \ref{fig:1aa}. Our observations reveal that, in our particular scenario, $r_\mathrm{h}$ consistently exceeds $r_{\mathrm{eff}}$, thereby allowing us to study photon motion directly from the BH horizon. However, in cases where $r_{\mathrm{eff}}>r_\mathrm{h}$, the photon motion should be investigated away from $r_{\mathrm{eff}}$ rather than the horizon. Consequently, the effective metric background imposes this additional constraint on the photon trajectories around the BH.
\begin{figure*}[htbp]
    \centering
    \includegraphics[width=.49\textwidth]{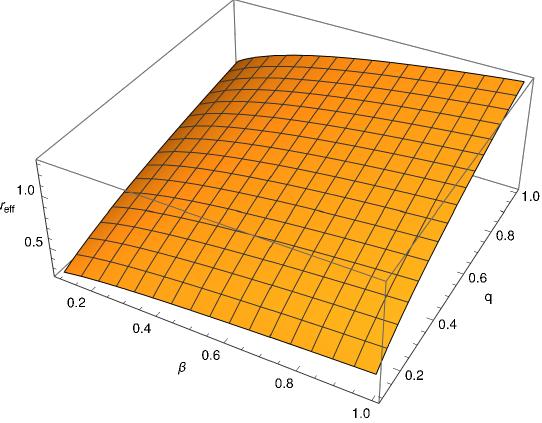} %\hfill
    \includegraphics[width=.49\textwidth]{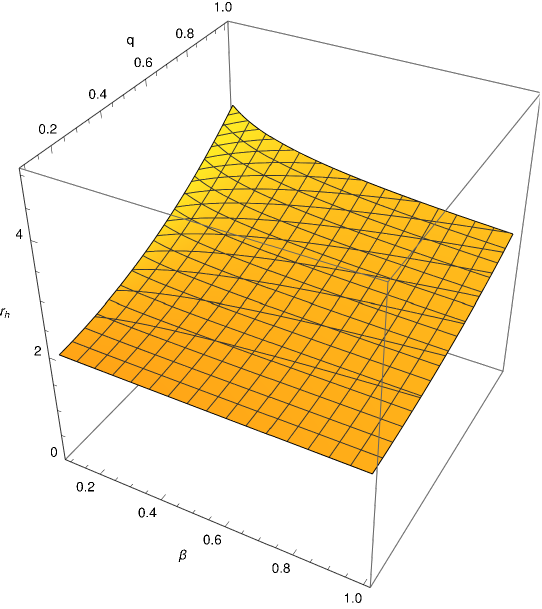} %\hfill
    \caption{Three-dimensional plots to illustrate the dependence of the minimum allowed values of $r_{\mathrm{eff}}$ (left panel) and the BH horizon $r_\mathrm{h}$ (right panel)  on the $\beta$-parameter and the magnetic charge $q_m$.
    %3D plots to understand the minimum allowed value of the radial distance ($r_{eff}$) (left) and BH horizon ($r_h$) (right) depending on the NED parameter ($\beta$) and magnetic charge ($q$).
    }
    \label{fig:1aa}
\end{figure*}

In this context, the most important phenomenon revolves around the propagation of light rays in the effective metric, where the aforementioned parameters play a significant role. By solving the equations of motion \eqref{eqm1_EG}--\eqref{eqm2_EG} in the effective spacetime geometry, we have simulated the orbit of light rays in the equatorial plane, as depicted in the bottom panels of Fig. \ref{fig:15}. The top panels of the figure illustrate the contribution of these rays to the formation of the photon and lensing rings. The horizontal axis represents the impact parameter, $b$, while the vertical axis, $n$, corresponds to the number of times the rays cross the BH's plane. In other words, it shows the number of half-orbits that the light rays undergo during their trajectories.  The diagrams were generated for a fixed $\beta$-parameter and three different values of the magnetic charge. As evident from the figures, with an increase in the magnetic charge, the size of the horizon expands, leading to a higher likelihood of light rays being absorbed by the BH. We have categorized the photon trajectories into three groups, following Ref.~\cite{Gralla:2019xty}. For light rays with $n<3/4$, we observe the direct emission profile, where the distant observer receives the light directly from the light source (such as the BH's accretion disk). In the range $3/4<n<5/4$, the observer receives a lensed image of the backside of the source, as the light rays cross the BH's plane twice. This phenomenon corresponds to the formation of the lensing ring. For $n>5/4$, the light rays cross the BH's plane more than twice, resulting in the formation of photon rings of higher order. The top panels of Fig.~\ref{fig:15} depict the thickness of the three aforementioned categories for the adopted values of the BH parameters. 

We continue our discussion by validating our study with real astrophysical data from the EHT, enhancing the significance of our findings for understanding BH properties and light propagation.

%we have shown a two-dimensional plot (in the lower panel) for the photon trajectories in the case of the effective metric background and their contribution to the emission profiles (in the upper panel). 
%The y-axis shows the number of orbits $n(\gamma)$ that a photon crosses the equatorial plane around the BH before leaving the effective gravitational field of the BH and the x-axis is the impact factor ($b$). 
%We fixed $\beta=0.1$ and varies $q$ and showed that with increasing $q$, the horizon, as well as impact factor increases and photon trajectories, follow the specific pattern depending on it crossing the equatorial plane around the BH. The plot is for the face on view and we divide the photon trajectories into three categories as done by Wald et al. \cite{Gralla:2019xty}. Null rays with $n<3/4$ indicate direct emission while crossing the equatorial plane only once, null rays with $3/4<n<5/4$ indicate lensing ring while crossing the equatorial plane twice and null rays with $n>5/4$ indicate photon ring while crossing the equatorial plane more than twice.
%
\begin{figure*}[htbp]
    \centering
    \includegraphics[width=.3\textwidth]{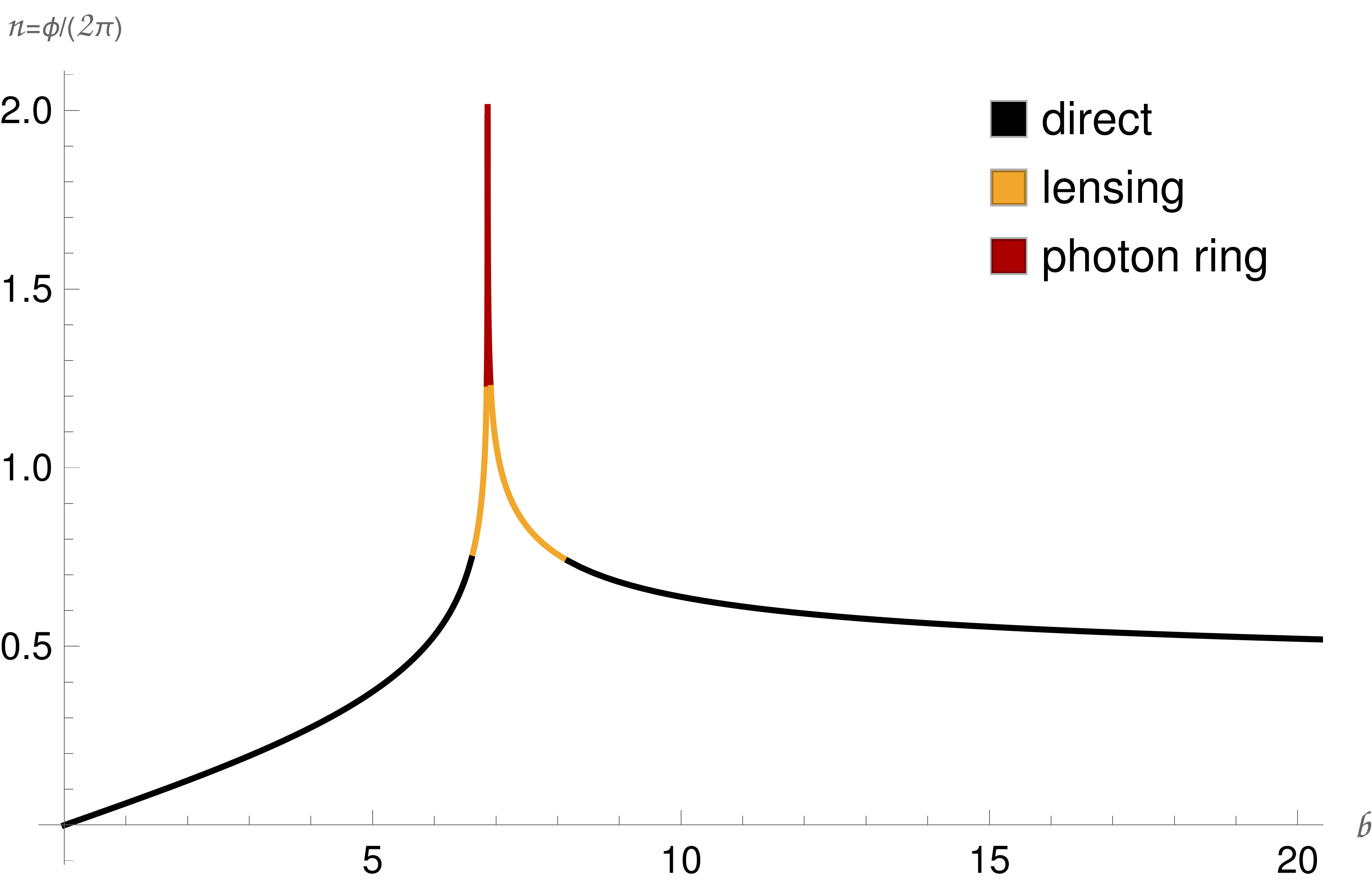} \hfill
    \includegraphics[width=.3\textwidth]{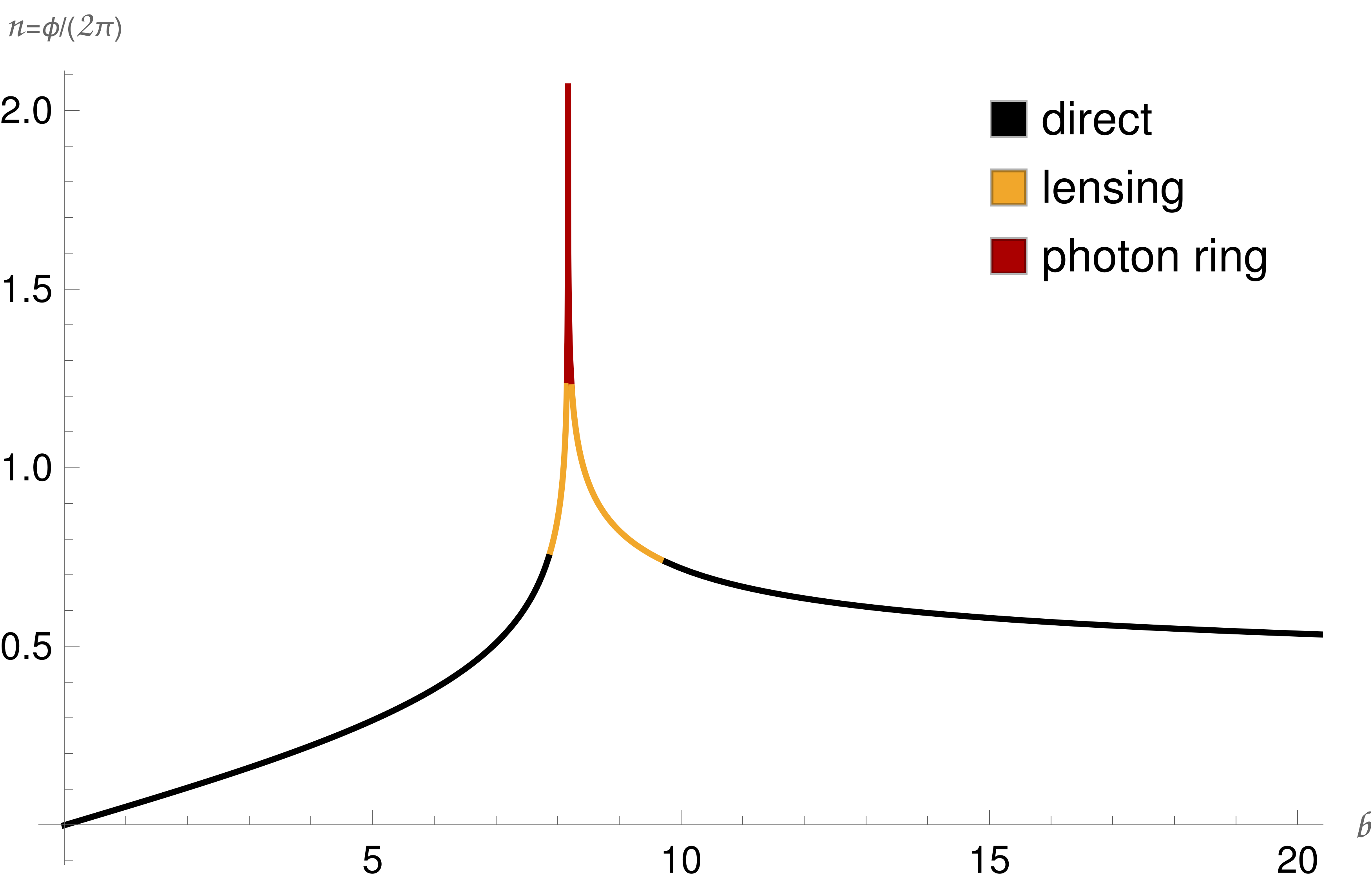} \hfill
    \includegraphics[width=.3\textwidth]{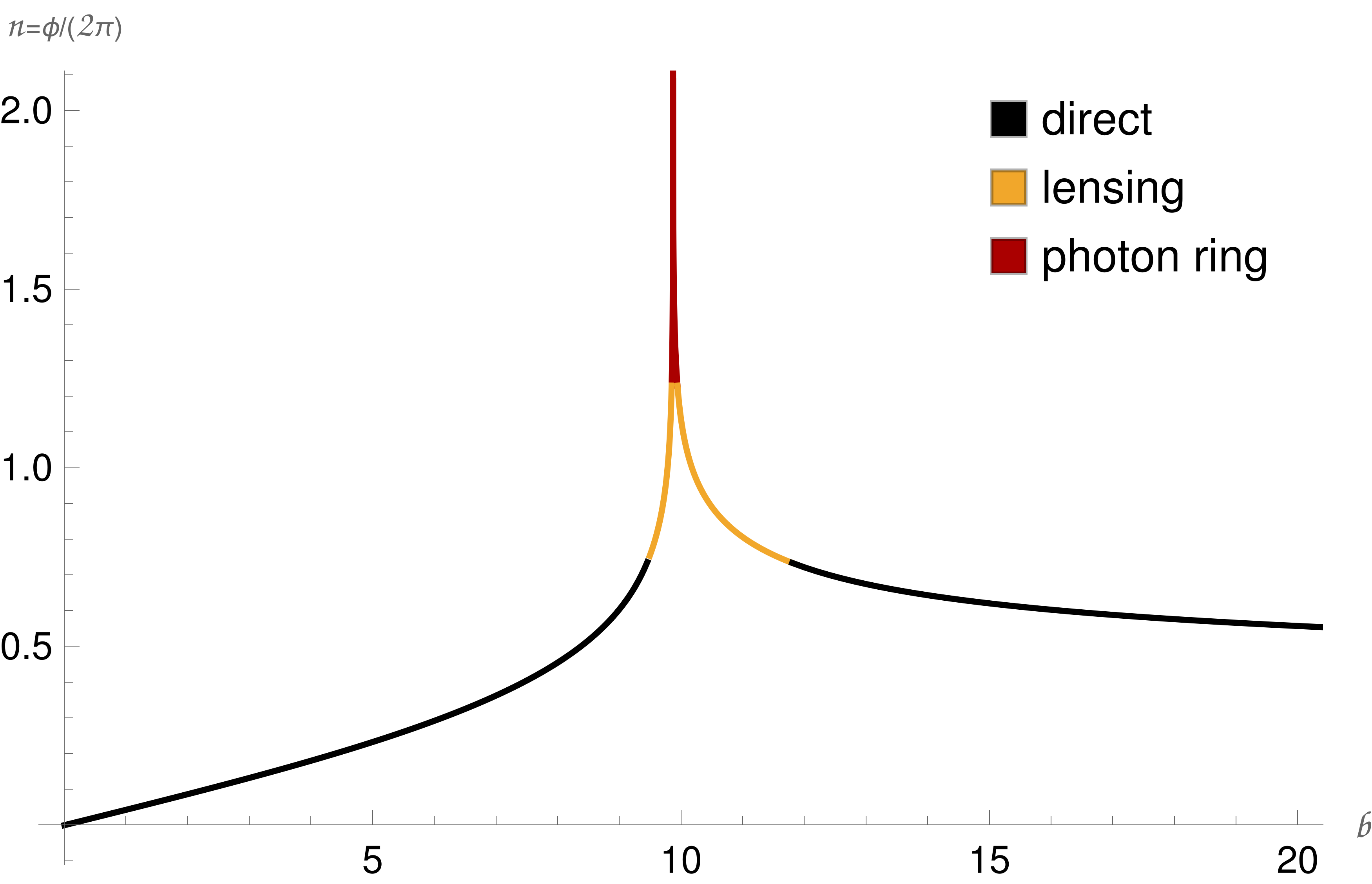} \hfill
    \vfill
    \centering
    \includegraphics[width=.3\textwidth]{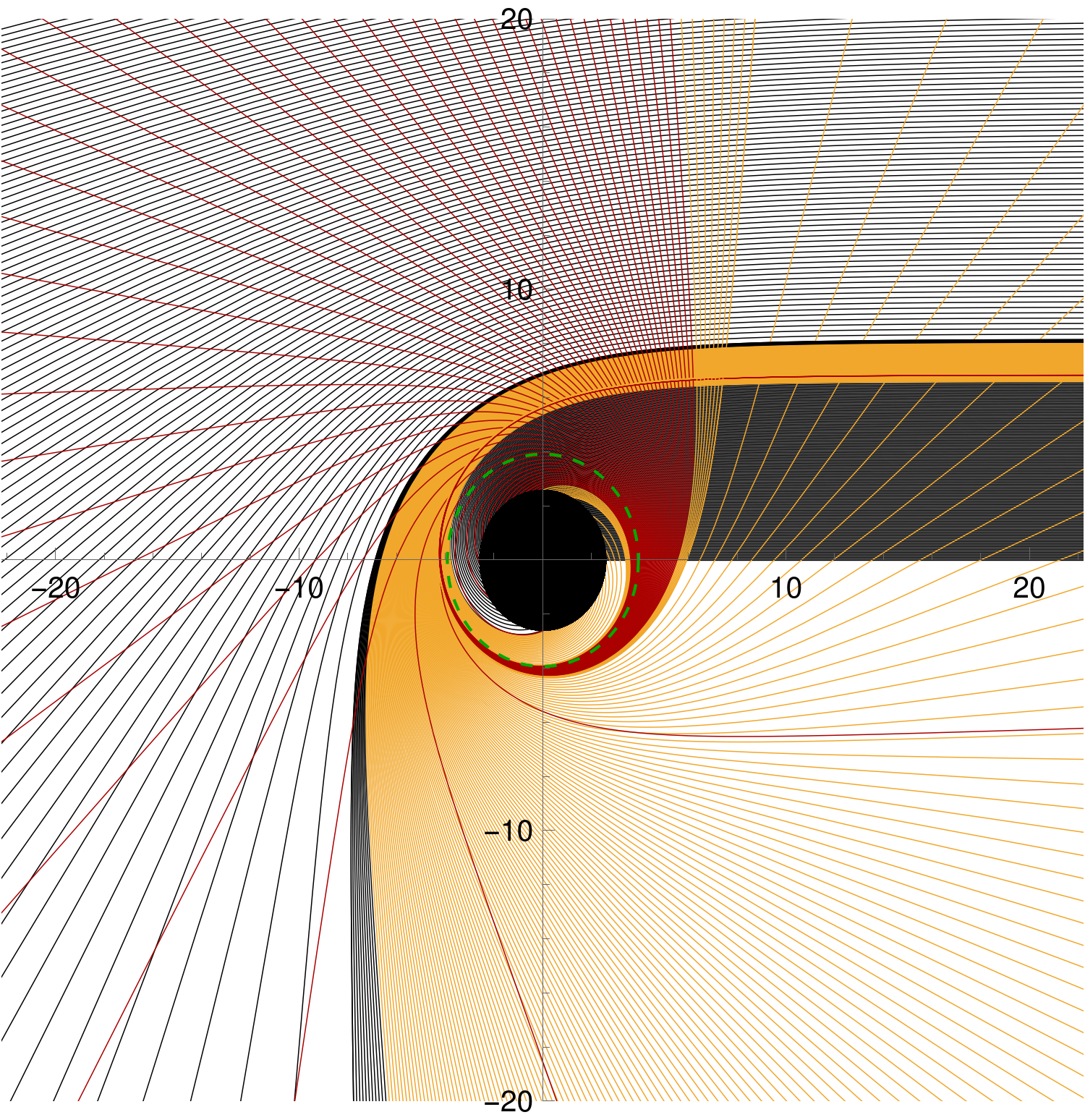} \hfill
    \includegraphics[width=.3\textwidth]{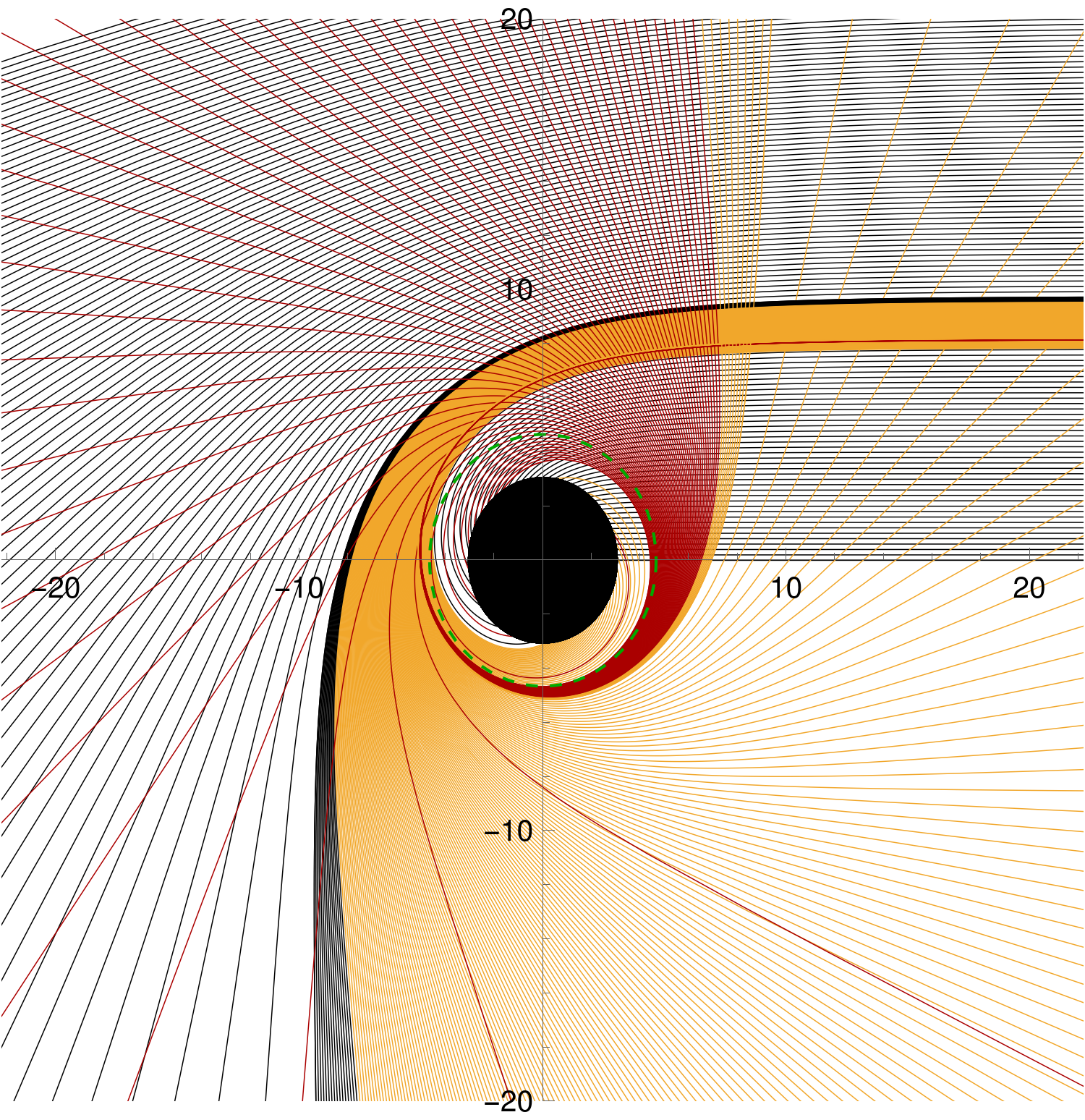} \hfill
    \includegraphics[width=.3\textwidth]{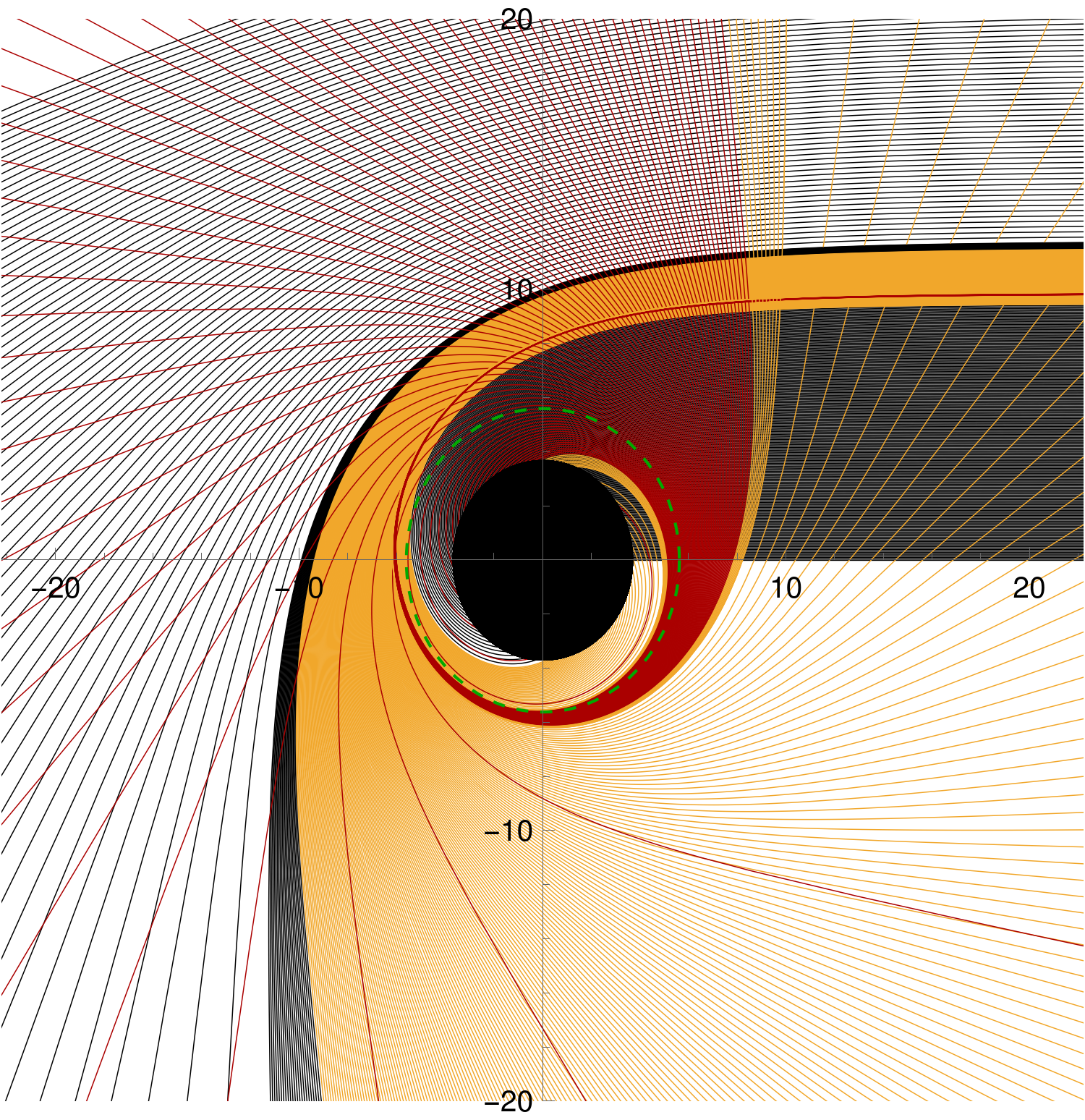} \hfill
    \caption{The behavior of photons in the effective geometry plotted for $\beta=0.1$ and, from left to right, for $q_m=0.6, 0.8$, and $1.0$. The red, orange, and black lines correspond to the photon ring, lensing ring, and direct emission, respectively. The green dotted circle represents the radius of unstable photon orbits, $r_{\mathrm{ph}}$. The black disk indicates the event horizon of the BH. In the top panels, the fractional number of orbits ($n = \phi/(2\pi)$) is displayed, where $\phi$ represents the total change in the azimuth angle outside the horizon. The bottom panel illustrates selected photon trajectories, treating $r$ and $\phi$ as Euclidean polar coordinates.}
    \label{fig:15}
\end{figure*}

\section{Constraints from the M87* and S\lowercase{gr} A*}\label{sec:shadowCons}

In this section, we aim at constraining the magnetic charge and NED parameter by utilizing observed data from the EHT. For M87*, the angular diameter of the BH shadow is $\theta_\text{M87*} = 42 \pm 3 \,\mathrm{\mu as}$, the distance to the BH is $d_{s}^\text{M87*} = 16.8$ Mpc, and the mass is $M_\text{M87*} = (6.5 \pm 0.90) \times 10^9 M_\odot$ \cite{EventHorizonTelescope:2019dse}. For Sgr A*, the angular diameter of the shadow is $\theta_\text{Sgr A*} = 48.7 \pm 7 \mathrm{\mu as}$, the distance to the BH  is $d_{s}^{\text{Sgr A*}} = 8277\pm33$ pc, and the BH mass is $M_\text{Sgr A*} = (4.3 \pm 0.013) \times 10^6 M_\odot$ \cite{EventHorizonTelescope:2022wkp}. By using this data and the formula from Ref. \cite{Bambi:2019tjh}, the BH shadow diameter can be calculated as
%In this section, we tried to constrain the magnetic charge and NED parameter using the observed diameter of the  M87* and Sgr $A^{\ast}$ provided by the EHT. The EHT has reported that the angular diameter of the M87* BH shadow is $\theta_\text{M87*} = 42 \pm 3 \:\mu as$, distance of M87* from the Earth is measured as $d_{s}^{M87^{\ast}} = 16.8$ Mpc, and mass of the M87* is $M_\text{M87*} = 6.5 \pm 0.90$x$10^9 \: M_\odot$ in one of their series of papers on M87* \cite{EventHorizonTelescope:2019dse}. Similarly, as reported in one of the series papers on Sgr $A^{\ast}$ \cite{EventHorizonTelescope:2022wkp} that the angular diameter of the Sgr $A^{\ast}$ shadow is $\theta_\text{Sgr $A^{\ast}$} = 48.7 \pm 7 \:\mu as$, the distance of the Sgr $A^{\ast}$ from the Earth is $d_{s}^{Sgr A^{\ast}} = 8277\pm33$ pc and mass of the Sgr $A^{\ast}$ BH is $M_\text{Sgr $A^{\ast}$} = 4.3 \pm 0.013$x$10^6 \: M_\odot$. Hence, using the above data and the following formula, one can calculate the diameter of the BH shadow \cite{Bambi:2019tjh},
\begin{equation}
    d_\text{sh} = \frac{d_{s}  \theta}{M}.
    \label{eq:dsh}
\end{equation}
Using the above formula, one can calculate the shadow diameters for M87* and Sgr A* as $d^\text{M87*}_\text{sh} = (11 \pm 1.5)M$ and $d^\text{Sgr A*}_\text{sh} = (9.5 \pm 1.4)M$, respectively. Next, utilizing the impact parameter \eqref{eq:impactParameter} and the previously obtained value of $r_{\mathrm{ph}}$, we can calculate the shadow radius as $r_{\mathrm{sh}}=b_{\mathrm{ph}}$, where $b_{\mathrm{ph}} = b(r_{\mathrm{ph}})$. Consequently, the theoretical shadow diameter in the effective metric can be expressed as $d_{\mathrm{sh}}^{\mathrm{theo}} = 2 b_{\mathrm{ph}}$. Figure \ref{fig:shacons} illustrates the profile of this quantity over a potential range of $\beta$ and $q_m$, which could impact the photons due to the effective metric background. This comparison includes the shadow diameters of M87* and Sgr A*, for which we precisely selected appropriate parameter values falling within the range of $1\sigma$ and $2\sigma$ uncertainties. Notably, Sgr A* imposes a more stringent constraint on the parameter than M87*.
\begin{figure*}
  \centering
    \includegraphics[width=0.48\textwidth]{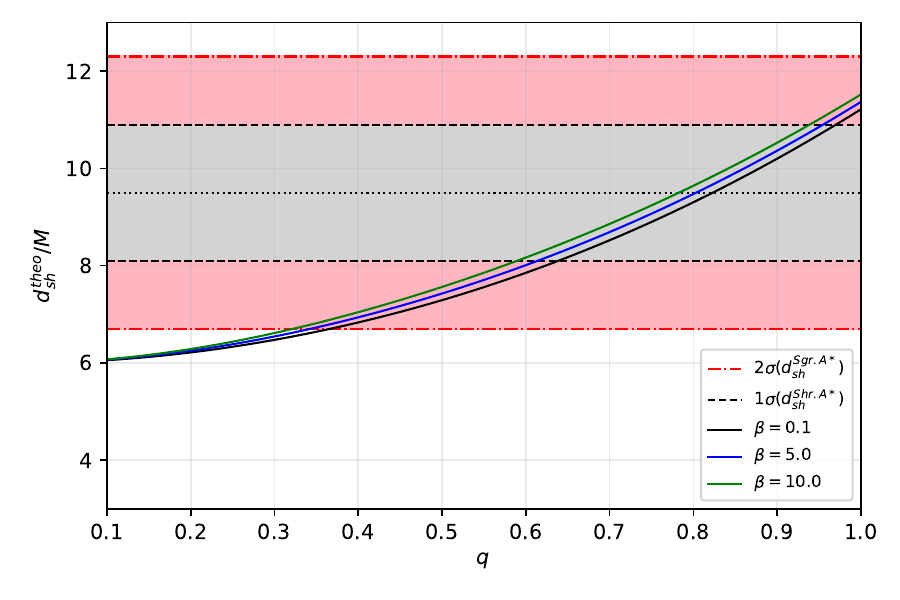}
    \includegraphics[width=0.48\textwidth]{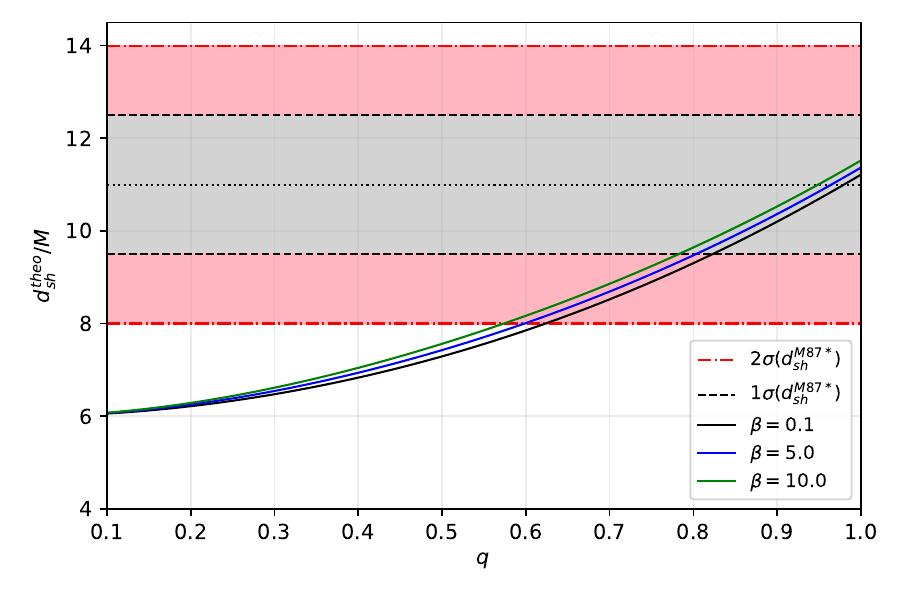}
    \caption{Constraining $q_m$ and $\beta$ for Sgr A* (left) and M87* (right), showing $1\sigma$ and $2\sigma$ levels.}
    \label{fig:shacons}
\end{figure*}

It is essential to note that observing BHs relies on gravitational lensing and the process of light deflection. Therefore, analyzing how the NED BH affects light ray trajectories is crucial. This analysis involves calculating the light deflection angle, which will be addressed in the next section.

%%%%%%%%%%%%%%%%%%%%
\section{Deflection angle of the NED BH in the effective geometry}\label{sec:defAngle}

Considering the line element \eqref{LE_EG1}, the canonical equations \eqref{eqm1_EG} and \eqref{eqm3_EG} together with the null condition \eqref{eqm2_EG}, result in the angular equation of motion 
\begin{equation}
\left(\frac{dr}{d\phi}\right)^2 = -\frac{g_{\phi\phi}^2 \LF^2 }{b^2\Phi^2}-\frac{g_{\phi\phi} \LF}{g_{rr} \Phi}. 
    \label{eq:drdphi-0}
\end{equation} 
Assuming the complete spherical symmetry% and working in the $(+ - - \,-)$ sign convention, for which $g_{tt} = - g^{-1}_{rr} = A(r)$ and $g_{\phi\phi} = - r^2$
, the above relation yields the integral equation
\begin{equation}
\phi-\phi_0 = \int_{r_d}^\infty\frac{dr}{\sqrt{-\frac{r^4\LF^2}{b^2\Phi^2}-\frac{r^2\LF A(r)}{\Phi}}}\equiv\int_{r_d}^\infty \frac{dr}{\sqrt{\mathcal{P}(r)}},
    \label{eq:intP(r)}
\end{equation}
regarding the changes in the azimuth angle for deflecting light ray trajectories, where $\phi_0$ is the initial azimuth angle, $r_d$ is the minimum distance to the BH at which the deflection occurs, and $\mathcal{P}(r)$ is the characteristic polynomial, which in the weak field limit and by taking into account the expressions in Eq.~\eqref{eq:5} and the lapse function \eqref{eq:A(r)}, is given by
\begin{equation}
\mathcal{P}(r)=\delta_0 r\left(4r^3-g_2r-g_3\right)+\mathcal{O}(r^5),
    \label{eq:P(r)}
\end{equation}
expanded up to the fourth order in $r$, where
\begin{subequations}
\begin{align}
    & \delta_0 = -\frac{1}{12}\left(\frac{1}{3b^2}+\frac{G q_m}{3\sqrt{\beta}}\right),\\
    & g_2 = -\frac{1}{3\delta_0},\\
    & g_3 = -\frac{2G m_0}{3\sqrt{q_m}\beta^{1/4}\delta_0}.
\end{align}
\end{subequations}
It is then readily verified for the discriminant of the cubic in the parenthesis of Eq.~\eqref{eq:P(r)}, that $\Delta=g_2^3-27g_3^2 < 0$. Hence, the quartic $\mathcal{P}(r) = 0$ has only one positive real root $r_d>0$, and two complex conjugate roots $r_1=r_2^*$, which are given by
\begin{eqnarray}
&& r_d = \sqrt{\frac{g_2}{3}}\cos\left(\frac{1}{3}\arccos\left(3g_3\sqrt{\frac{3}{g_2^3}}\right)\right),\label{eq:rd}\\
&& r_1 = \sqrt{\frac{g_2}{3}}\cos\left(\frac{1}{3}\arccos\left(3g_3\sqrt{\frac{3}{g_2^3}}\right)+\frac{4\pi}{3}\right),\label{eq:r1}\\
&& r_2 = \sqrt{\frac{g_2}{3}}\cos\left(\frac{1}{3}\arccos\left(3g_3\sqrt{\frac{3}{g_2^3}}\right)+\frac{2\pi}{3}\right).\label{eq:r2}\\
\end{eqnarray}
Now to obtain the deflection angle 
\begin{equation}
\Theta = 2(\phi-\phi_0)-\pi,
    \label{eq:Theta0}
\end{equation}
for the light rays at the distance $r_d$ from the BH, we directly  integrate the Eq.~\eqref{eq:intP(r)}, which yields 
\begin{equation}
\Theta =-\frac{1}{\sqrt{\delta_1}}\wp^{-1}\left(\frac{\chi_1}{12}\right)-\pi,
    \label{eq:intP(r)-sol}
\end{equation}
where the inverse Weierstrassian $\wp$ elliptic function with the invariants $\zeta_2$ and $\zeta_3$, is defined in terms of the integral 
\begin{equation}
\wp^{-1}(y)\equiv\wp^{-1}(y;\zeta_2,\zeta_3)=\int_\infty^y\frac{dy}{\sqrt{4y^3-\zeta_2 y-\zeta_3}}.
    \label{eq:Wp-1}
\end{equation}
In the solution \eqref{eq:intP(r)-sol} we have defined
\begin{equation}
 \delta_1 = 1-\frac{r_1}{r_d}-\frac{r_2}{r_d}+\frac{r_1 r_2}{r_d^2},
 \label{eq:delta1}
\end{equation}
and the Weierstrass invariants are given as
\begin{subequations}
\begin{align}
    & \tilde{g}_2 = -\frac{1}{4}\left(\chi_1-\frac{\chi_2^2}{3}\right),\\
    & \tilde{g}_3 = -\frac{1}{16}\left(\chi_0-\frac{\chi_1\chi_2}{3}+\frac{2\chi_2^3}{27}\right),
\end{align}
\label{eq:tg2tg3}
\end{subequations}
where
\begin{subequations}
\begin{align}
    & \chi_0 = \frac{1}{\delta_1},\\
    & \chi_1 = \frac{1}{\delta_1}\left(3-\frac{r_1}{r_d}-\frac{r_2}{r_d}\right),\\
    & \chi_2 = \frac{1}{\delta_1}\left(3-\frac{2r_1}{r_d}-\frac{2r_2}{r_d}+\frac{r_1r_2}{r_d^2}\right).
\end{align}
\label{eq:chi012}
\end{subequations}
In Fig.~\ref{fig:Theta-beta}, the behavior of the deflection angle $\Theta$ has been demonstrated for some allowed values of the $q_m$-parameter, as constrained in Fig.~\ref{fig:shacons}.
\begin{figure}
    \centering
    \includegraphics[width=0.3\textwidth]{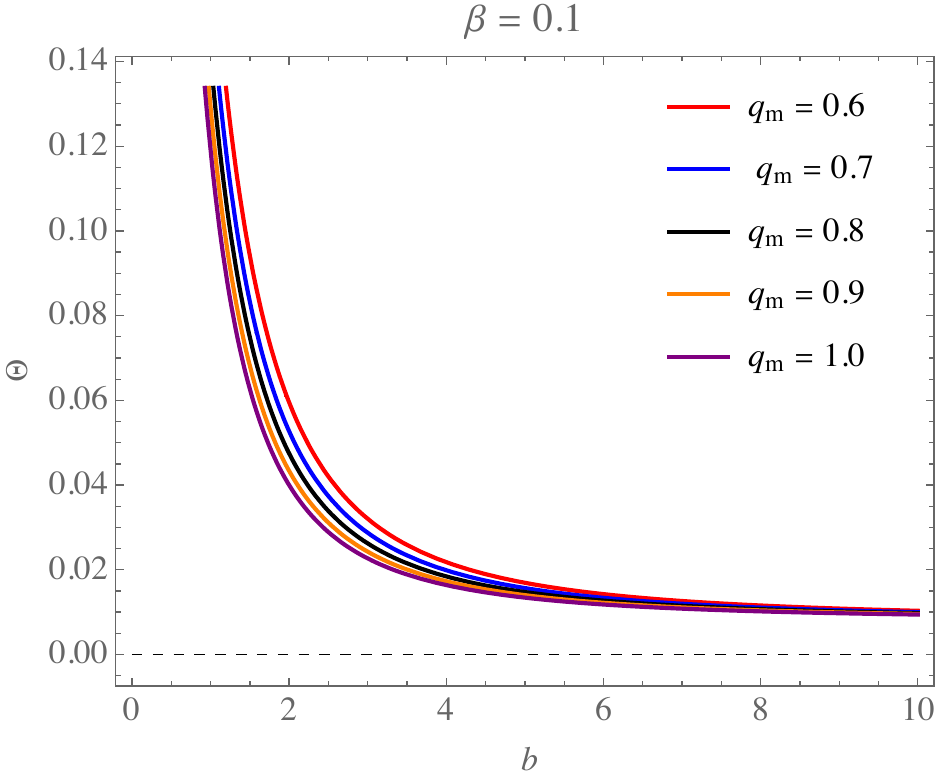}\qquad
      \includegraphics[width=0.3\textwidth]{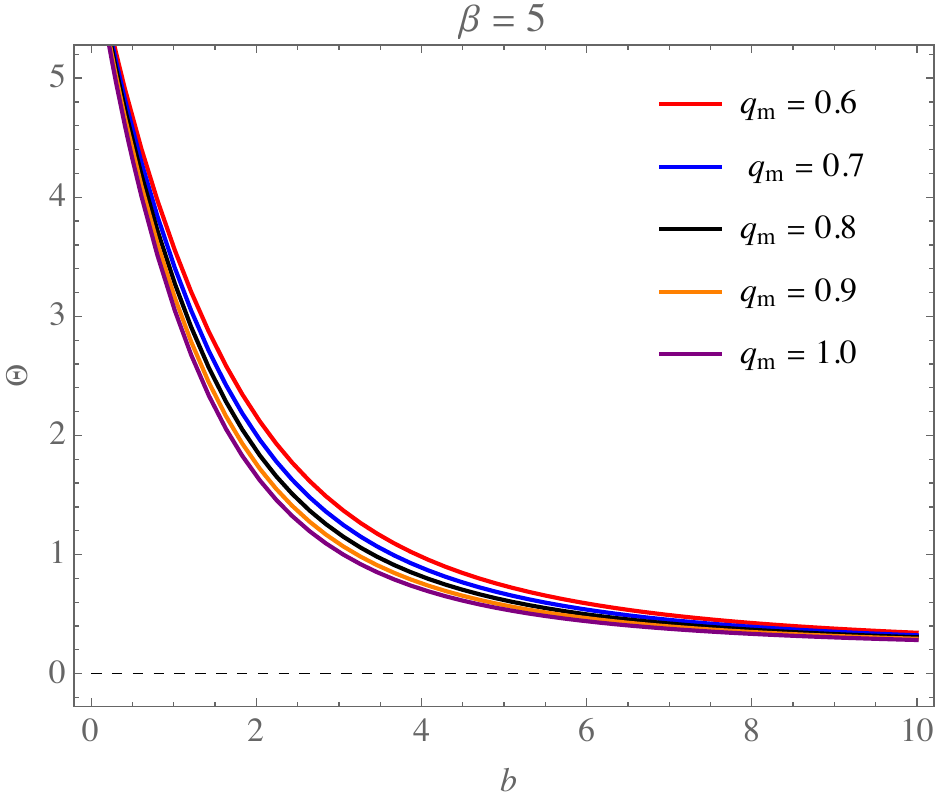}\qquad
        \includegraphics[width=0.3\textwidth]{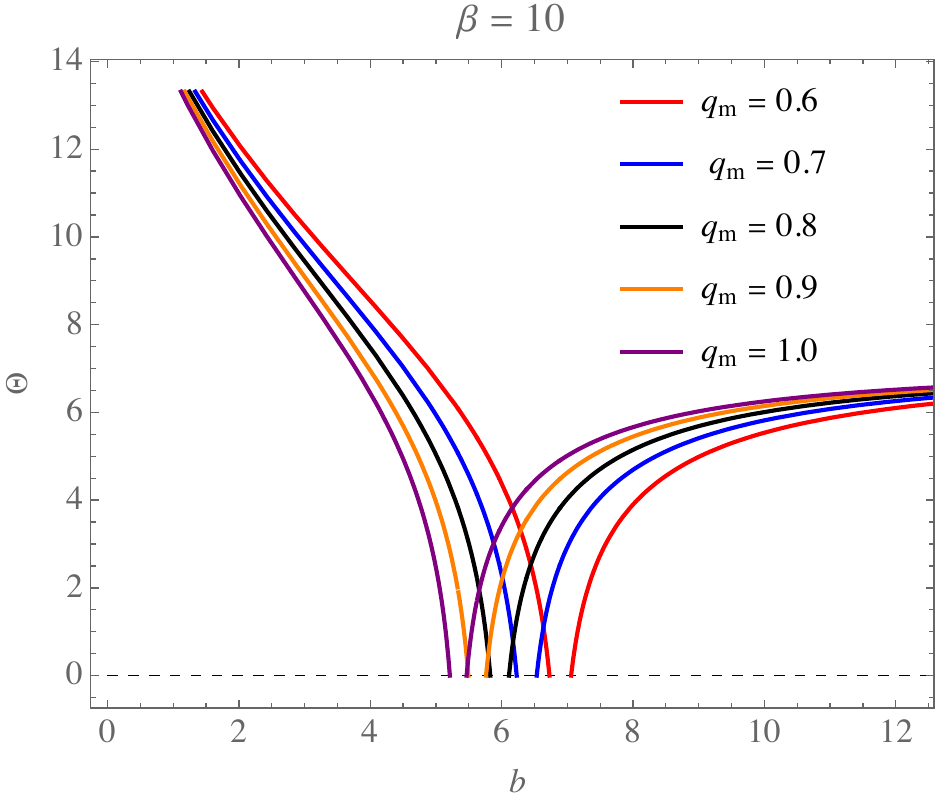}
    \caption{The evolution of the deflection angle $\Theta$ versus the changes in the impact parameter $b$ for the three cases of $\beta = 0.1, 5$ and $10$, plotted for $\phi_0=0$ and five different values of $q_m$, in accordance with the constraints obtained in Fig. \ref{fig:shacons}. }
    \label{fig:Theta-beta}
\end{figure}
As expected, the deflection drops intensely with the increase in the impact parameter and approaches zero. For the special case of $\beta=10$, however, after vanishing completely at a certain impact parameter, the deflection angle raises up to a smooth curve and then reaches a constant value as the impact parameter increases.
%\textcolor{Orange}{Me parece que estamos listos con el angulo de defleccion, pero estoy atento a los comentarios de los colaboradores!}

Having completed a comprehensive analytical study of photon trajectories in the effective geometry of the NED BH, we now turn our focus to the next two sections, where we will employ a thin accretion model around the BH and utilize a ray-tracing method to simulate the shadow and rings. This approach will help us understand the observational signature of the BH.

%%%%%%%%%%%%%%%%%%%%%
\section{Observed emission profile using the direct emission, lensing, and photon ring characteristics}\label{sec:signatures}

In this section, we will examine the overall characteristics of observed emissions by considering specific emission profiles from the accretion disk in the equatorial plane. As the brightness decreases while the accretion disk extends outward from the BH, we define three distinct emitted intensity profiles ($I_\text{EM}$) based on their decay rate concerning the radial coordinate $r$, as well as the inner disk radius.
 
\begin{itemize}
  \item Model 1:\[
    I_\text{EM1}(r)= 
\begin{cases}
    \left( \frac{1}{r-(r_\text{ISCO}-1)} \right)^2   ,&  r\geq r_\text{ISCO}\\
    0,              & r \leq r_\text{ISCO}
\end{cases}
\]
  \item Model 2:\[   I_\text{EM2}(r)= 
\begin{cases}
    \left( \frac{1}{r-(r_\text{ph}-1)} \right)^3   ,&  r\geq r_\text{ph}\\
    0,              & r \leq r_\text{ph}
\end{cases}
\]
   \item Model 3:\[   I_\text{EM3}(r)= 
\begin{cases}
    \frac{1-\arctan(r-(r_\text{ISCO}-1))}{1-\arctan(r_\text{ph})}    ,&  r\geq r_\text{h}\\
    0.              & r \leq r_\text{h}
\end{cases}
\]
\end{itemize}
These models exhibit specific properties: the first model initiates from the ISCO (innermost stable circular orbit) position ($r_\text{ISCO}$), with the inner disk boundary set at the ISCO. The second model begins from the photon radius ($r_\text{ph}$), where the inner disk boundary is positioned at the photon sphere. Lastly, the third model originates from the horizon ($r_\text{h}$). The second model exhibits rapid decay, while the third model decays very slowly compared to the other models. To calculate the observed intensity, we utilize Liouville's theorem and express the observed intensity ($I_{\nu'}^\text{obs}$) in terms of the emitted intensity from the disk ($I^{\mathrm{em}}_{\nu}$) as 
\begin{equation}
    I_{\nu'}^\text{obs}=g^3 I_\nu^\text{em},
\end{equation}
where $g=\nu_o/\nu_e=\sqrt{g_{tt}}$ is the redshift factor. Now considering $d\nu'=g d\nu$ and integrating over all the frequencies of $I_{\nu'}^\text{obs}$ as,
\begin{equation}
    I^\text{obs}=\int I^{obs}_{\nu'} \ed\nu'=g^4 I_\text{em}.
\end{equation}
where we have used $I_\text{em}=\int I_\nu^{\mathrm{em}} \ed\nu$. Hence, the observed intensity for an observer will be
\begin{equation}
    I(r)=\sum_{n} I^\text{obs}(r)|_{r=r_m(b)},
\end{equation}
where transfer function $r_m(b)$ represent that $m^{\mathrm{th}}$ intersections with the equatorial plane. The slope of this function provides the demagnification factor, which reveals the contributions of direct emission, photon, and lensing ring to the observed emission intensity profile. This factor has been extensively studied in Refs. \cite{Gralla:2019xty,Zeng:2020vsj,Uniyal:2022vdu,Uniyal:2023inx}.
\begin{figure*}[htbp]
    \centering
    \includegraphics[width=.3\textwidth]{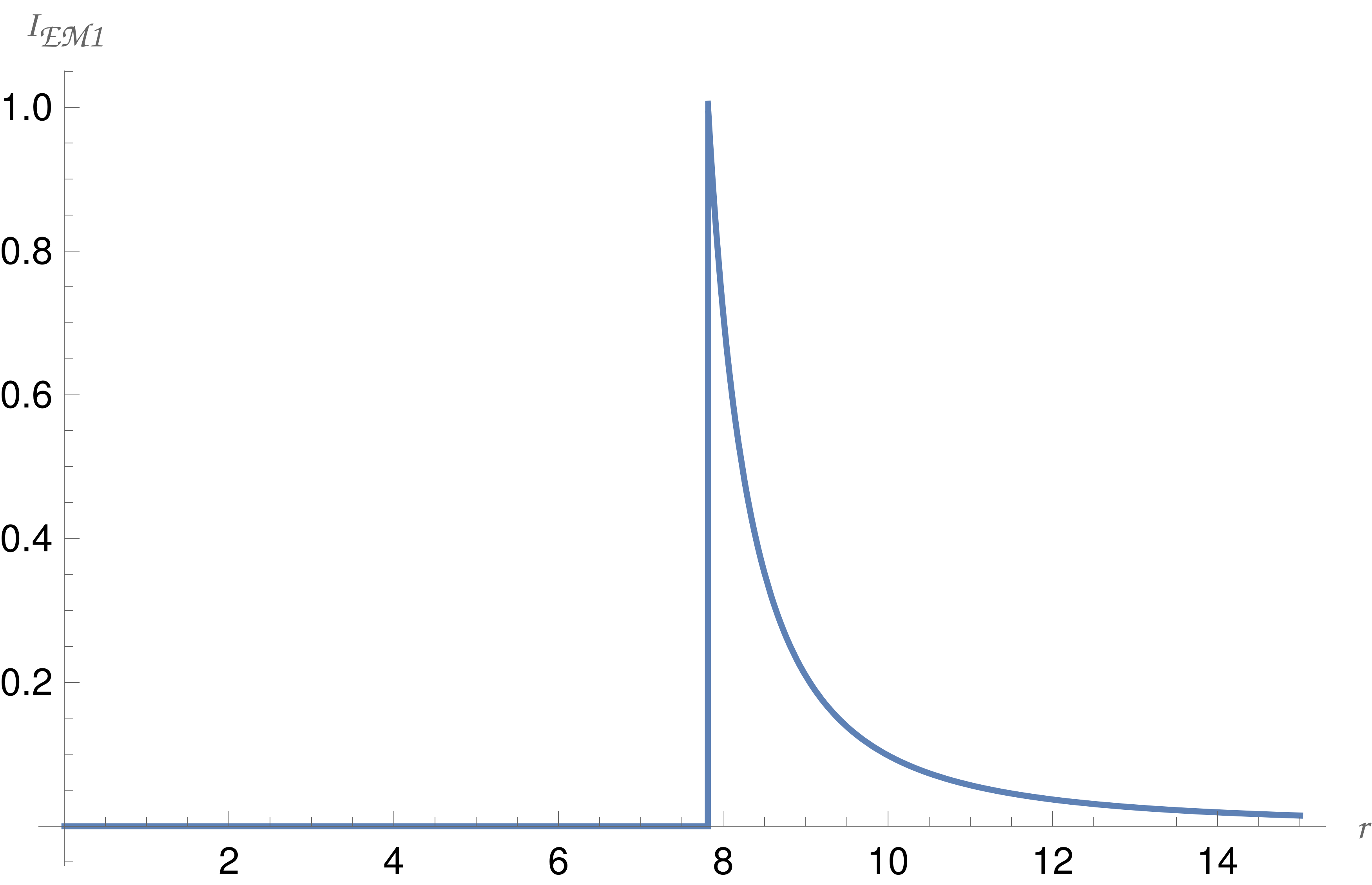}
    \includegraphics[width=.3\textwidth]{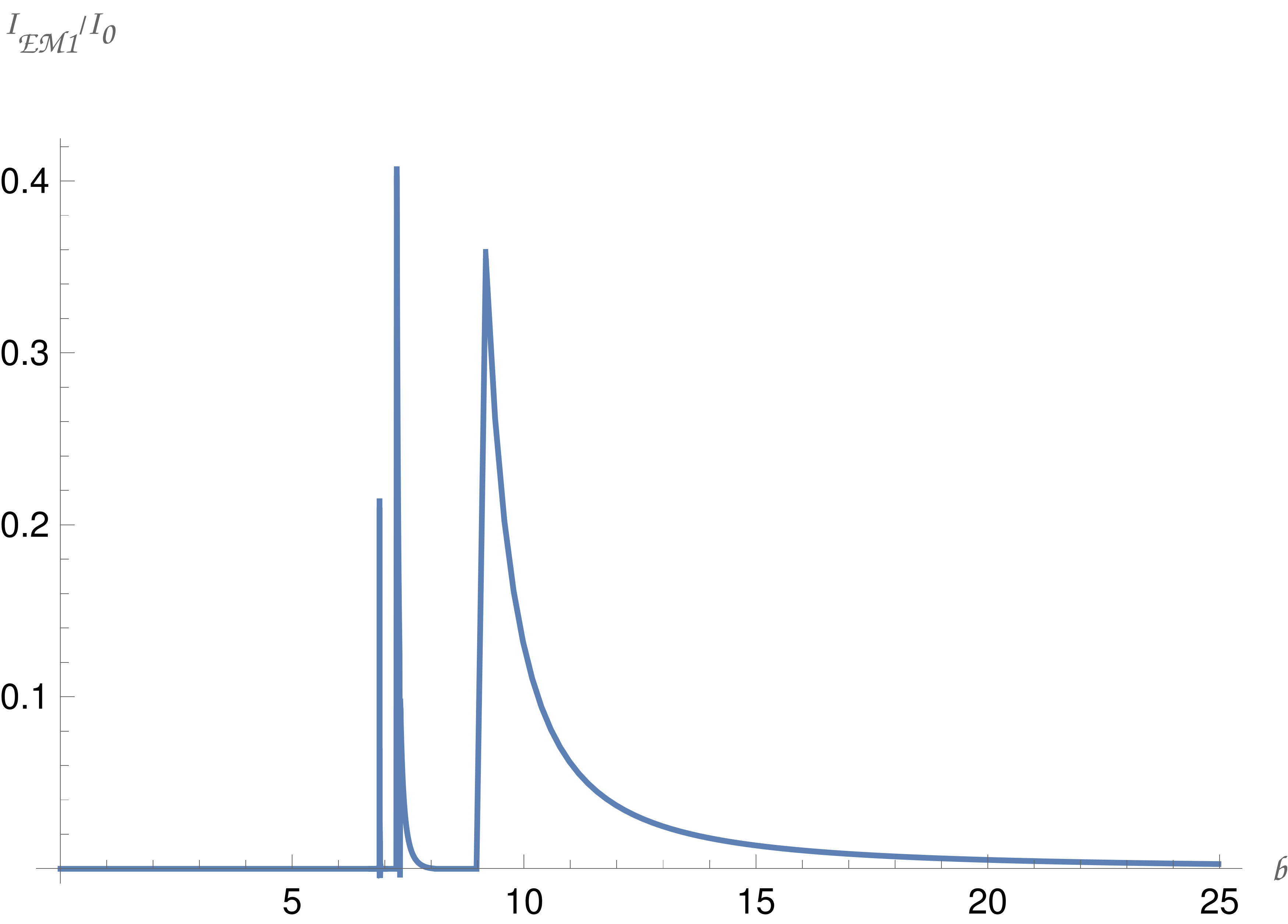}
    \includegraphics[width=.3\textwidth]{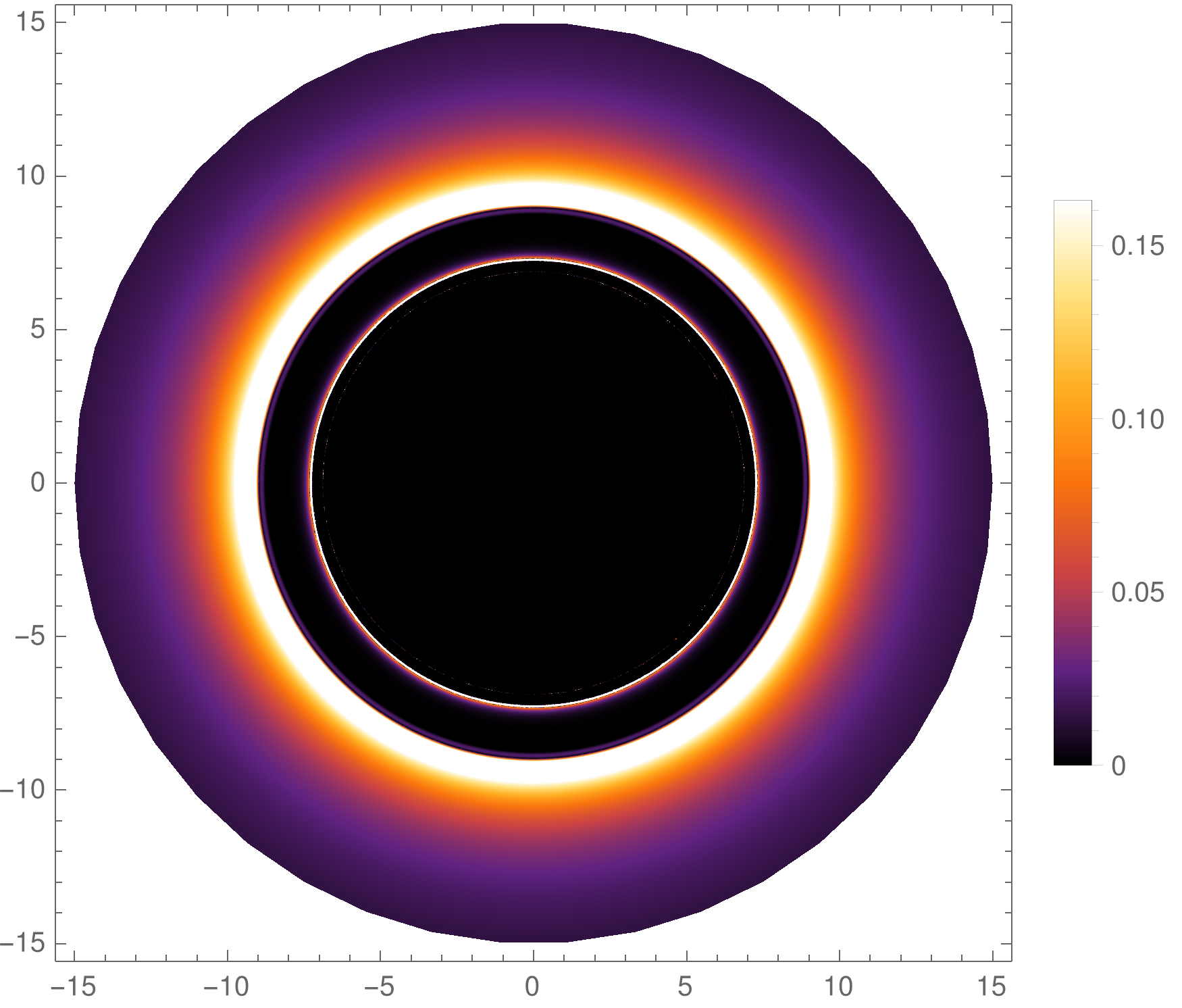}
    \vfill
    \centering
    \includegraphics[width=.3\textwidth]{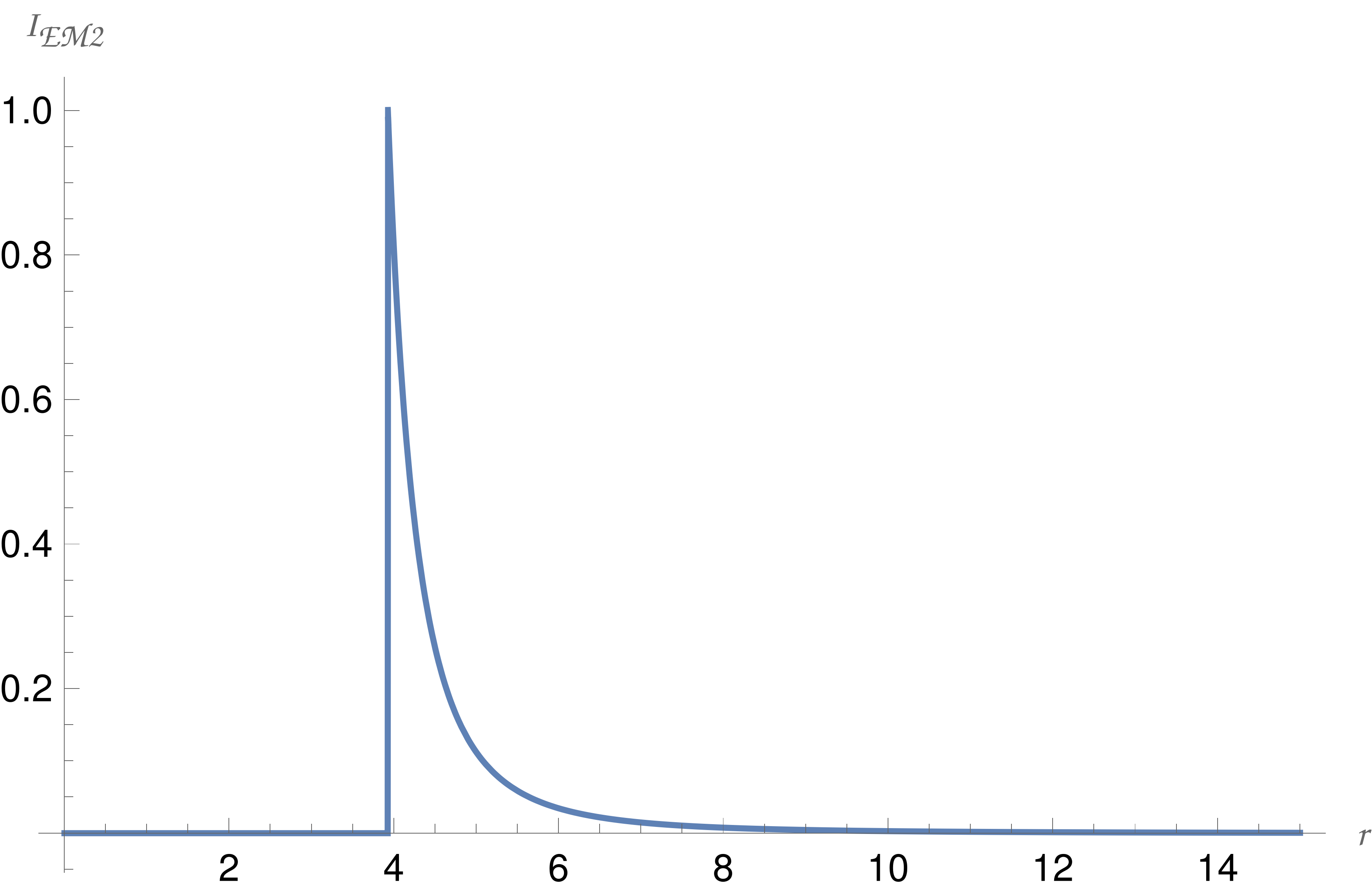}
    \includegraphics[width=.3\textwidth]{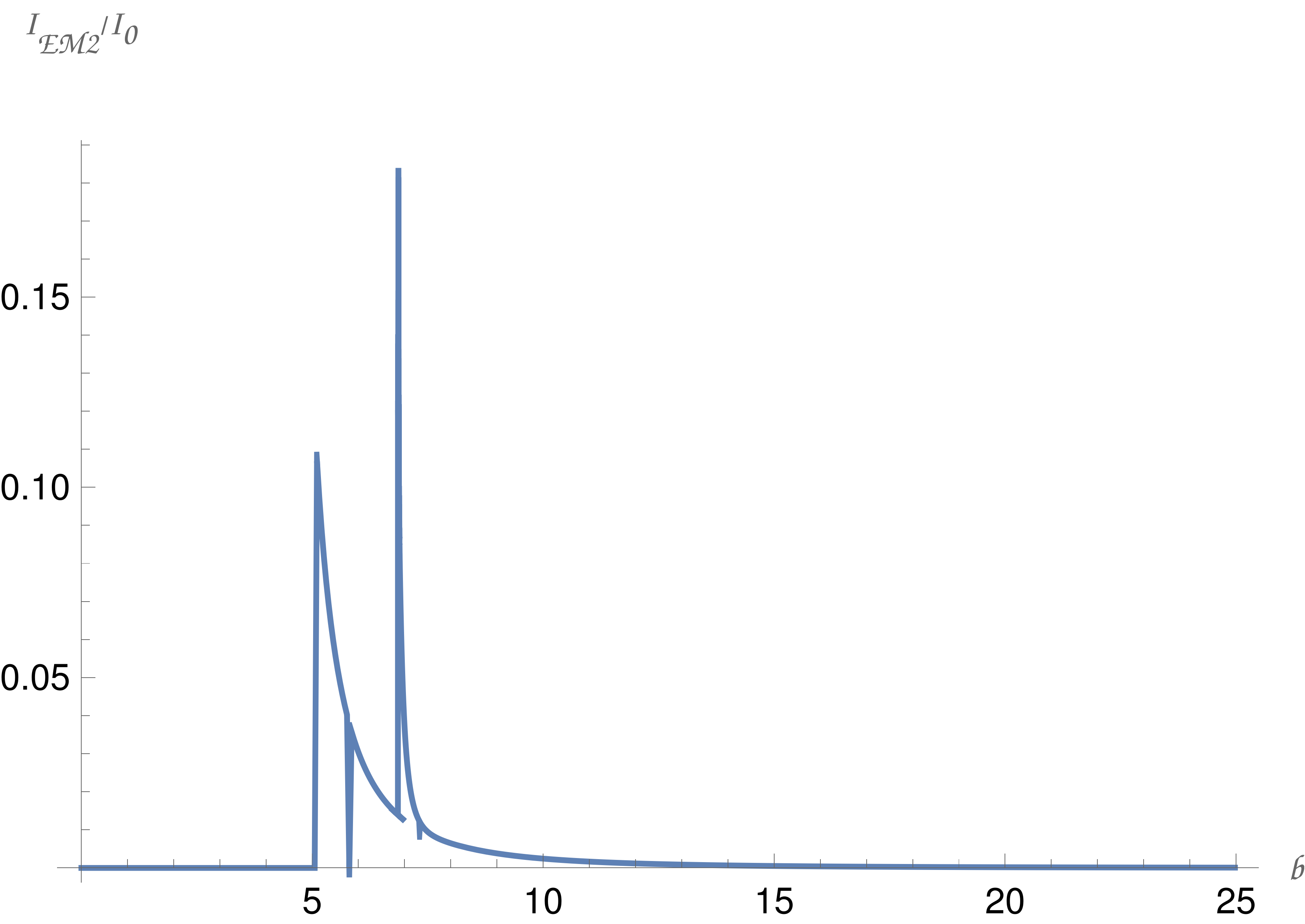}
    \includegraphics[width=.3\textwidth]{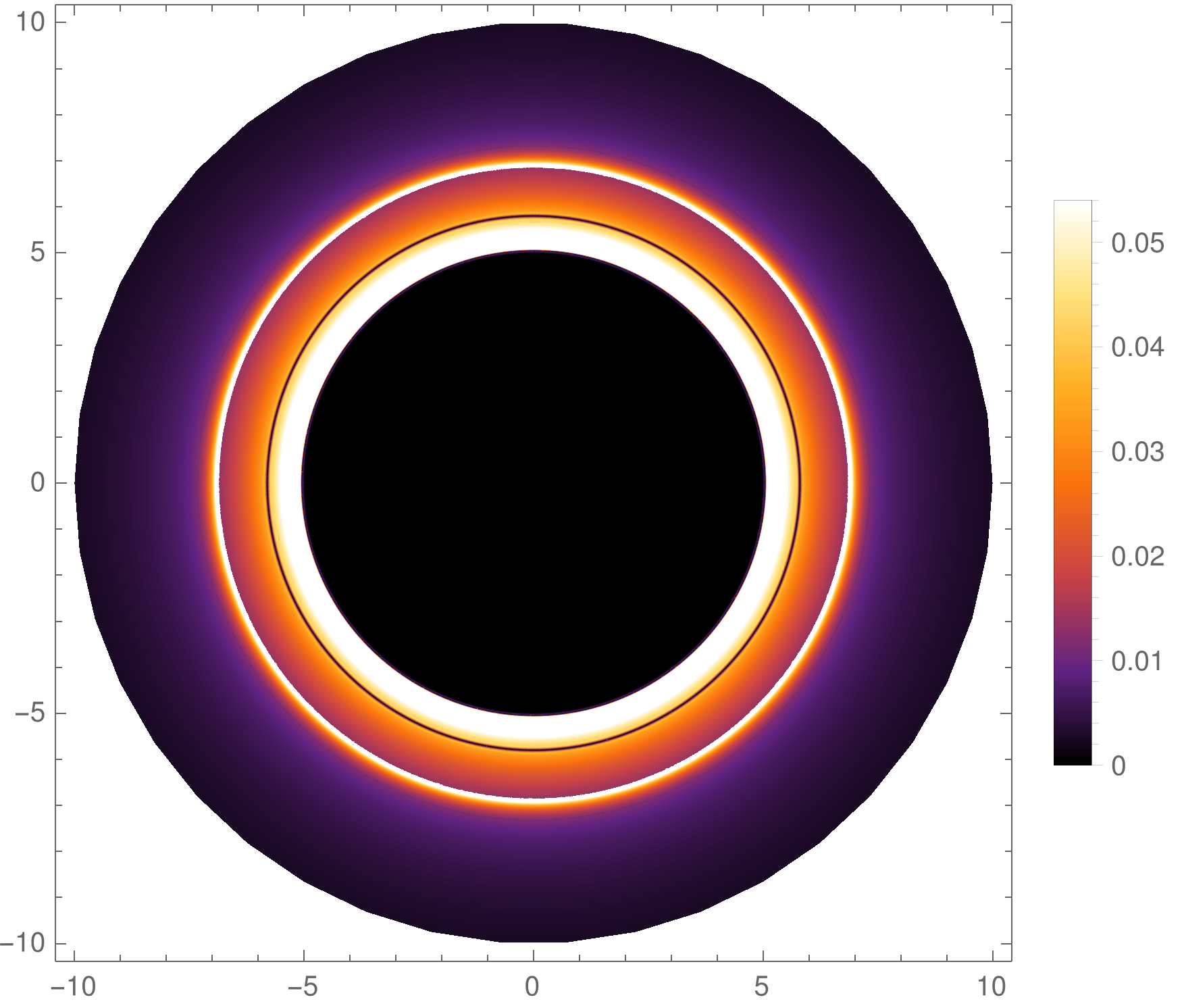}
    \vfill
    \centering
    \includegraphics[width=.3\textwidth]{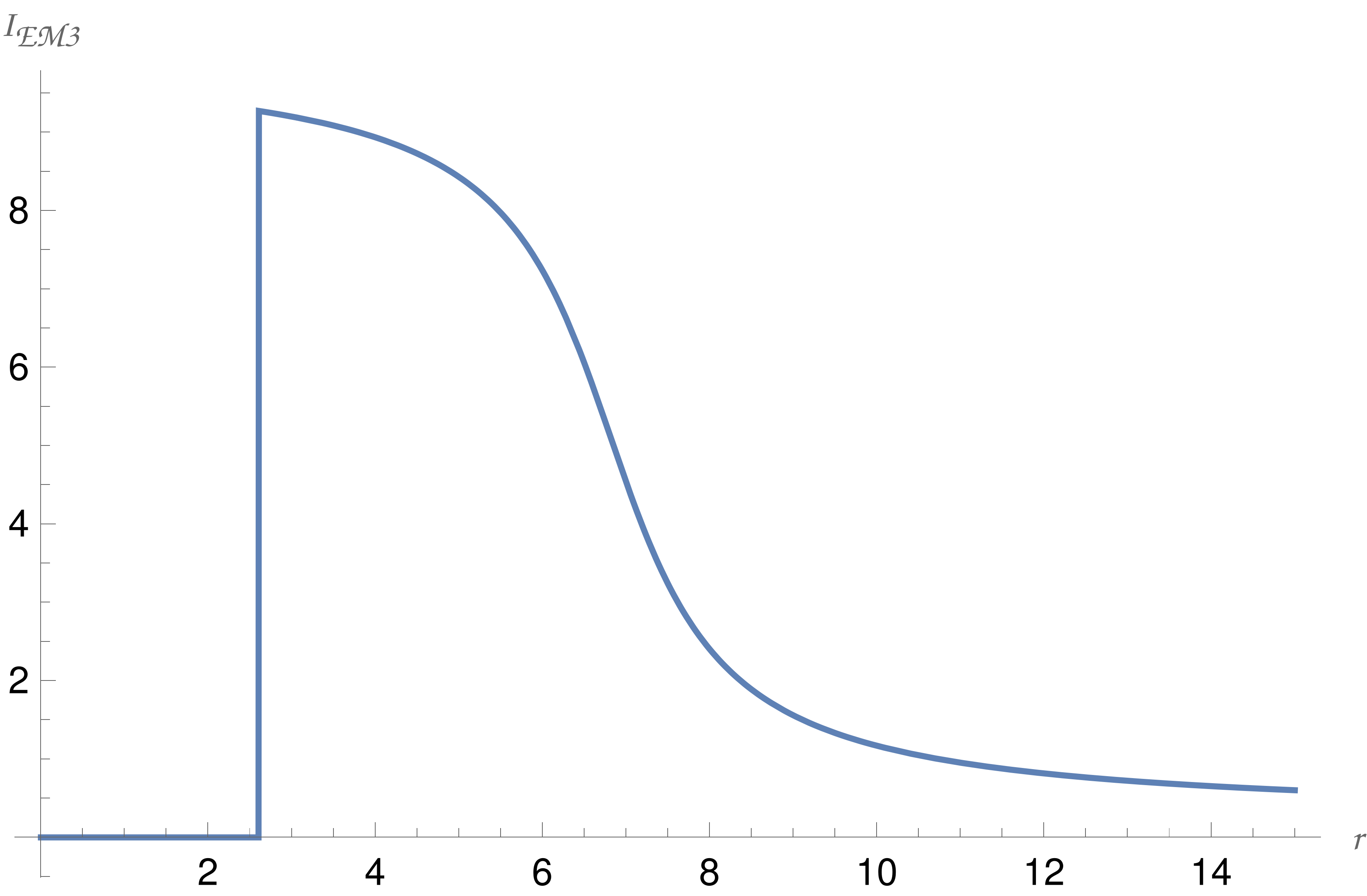}
    \includegraphics[width=.3\textwidth]{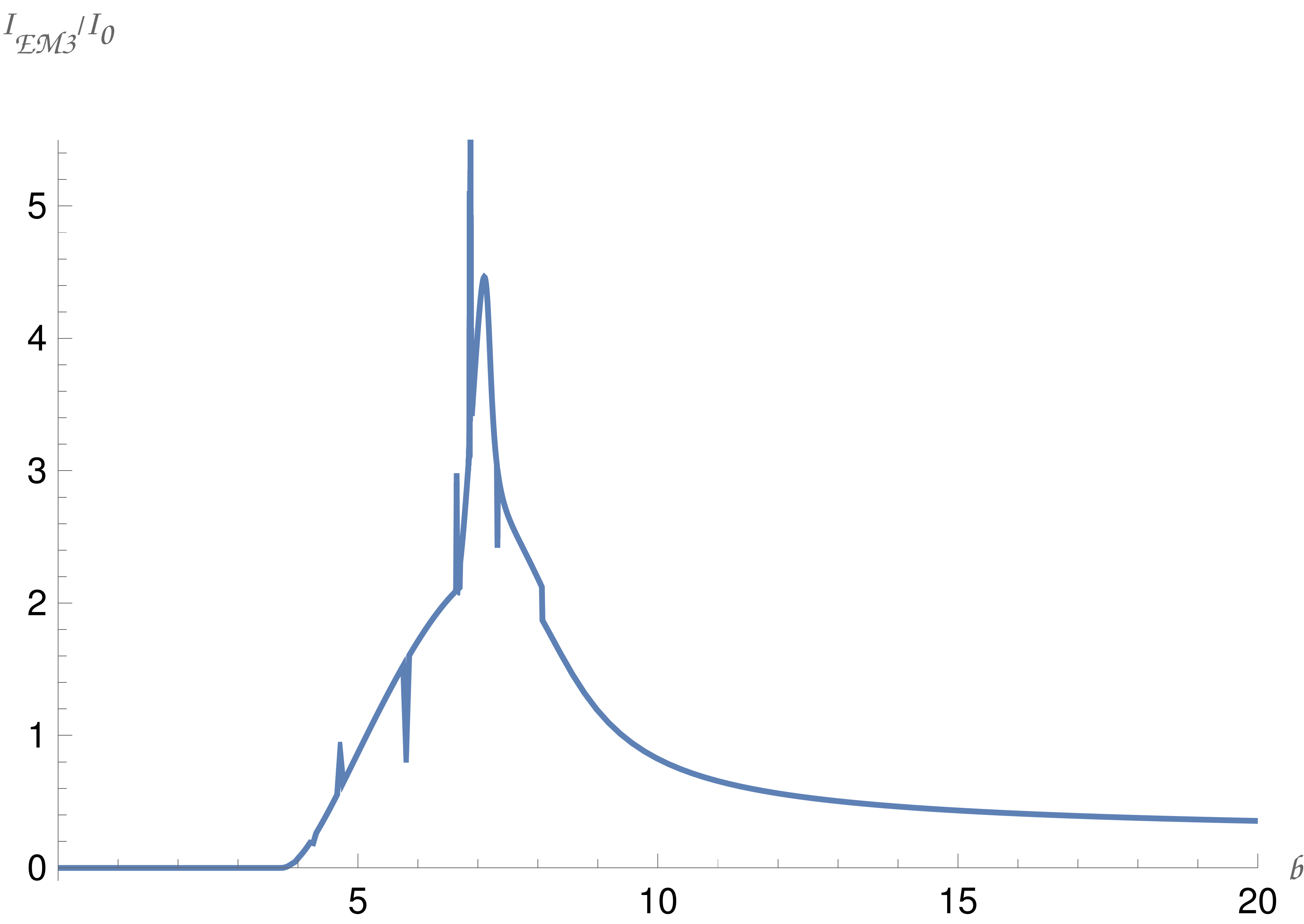}
    \includegraphics[width=.3\textwidth]{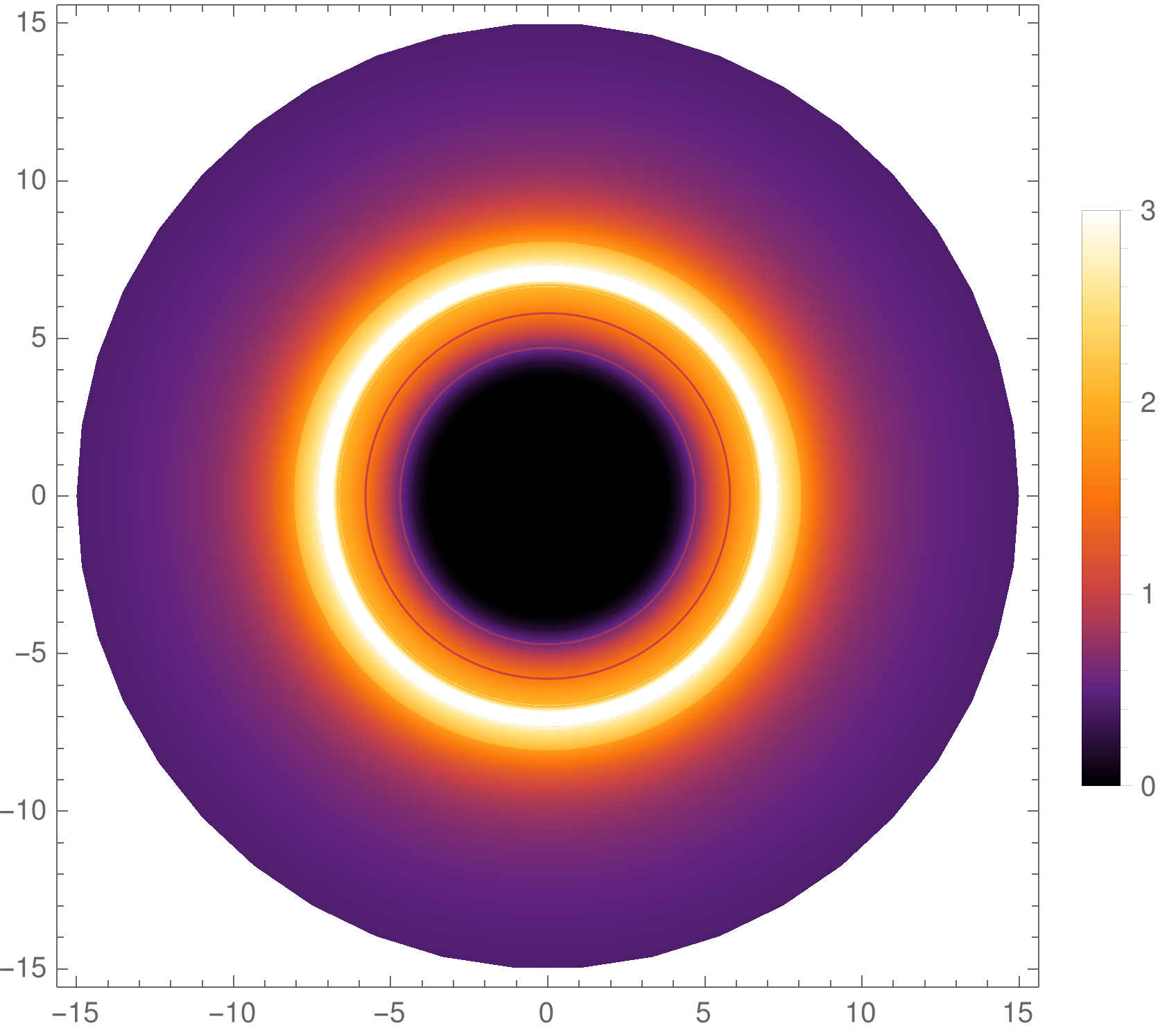}
    \caption{The appearance of a face-on observed thin disk with varying emission profiles for $q_m=0.6$ and $\beta=0.1$. Each row corresponds to a different model: top row - model $1$, second row - model $2$, and third row - model $3$. The emitted and observed intensities ($I_{\mathrm{EM}}$ and $I_{\mathrm{obs}}$) in the plots are normalized to the maximum emitted intensity value outside the horizon ($I_0$).}
    \label{fig:17}
\end{figure*}

\begin{figure*}[htbp]
    \centering
    \includegraphics[width=.3\textwidth]{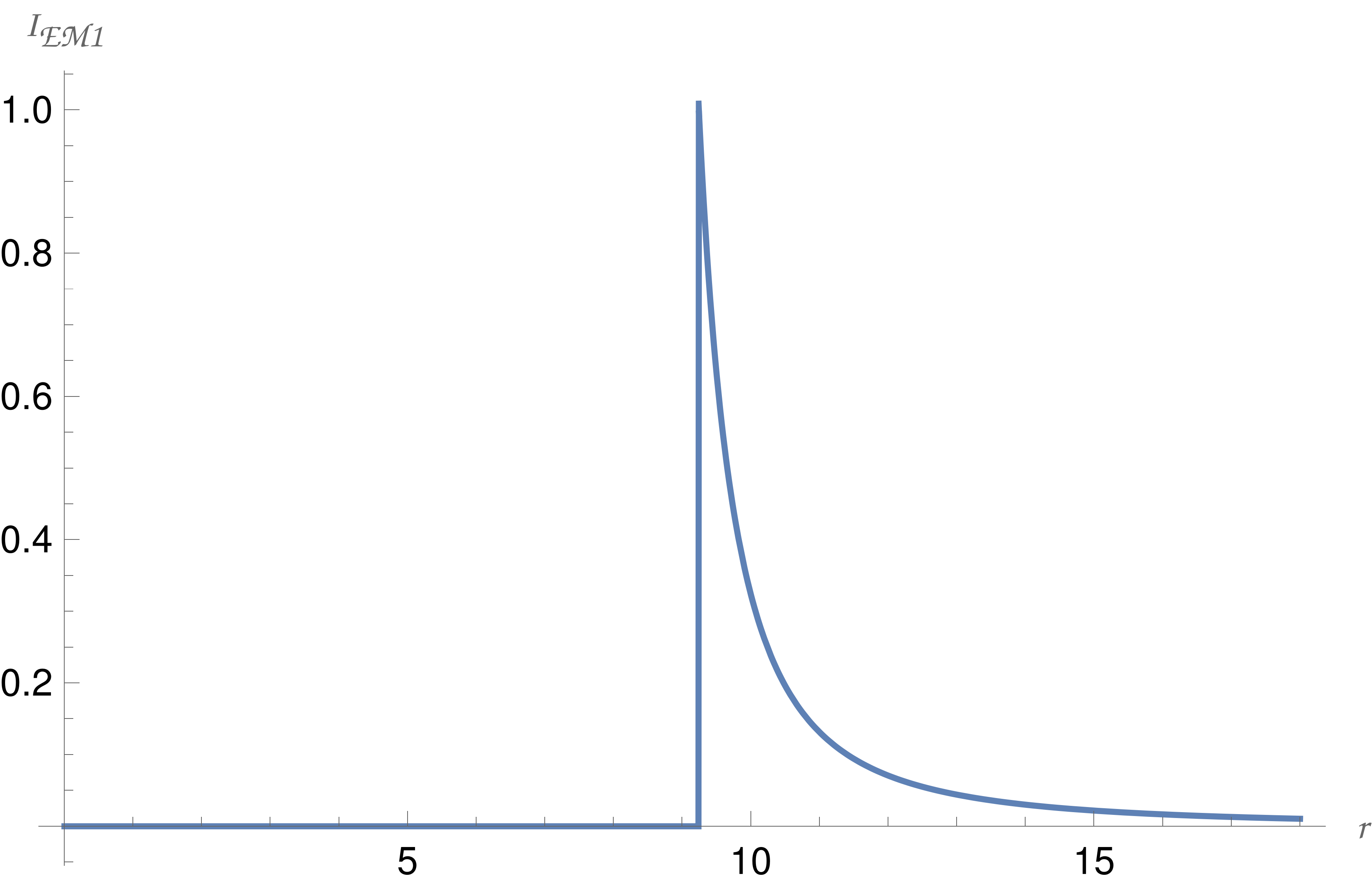}
    \includegraphics[width=.3\textwidth]{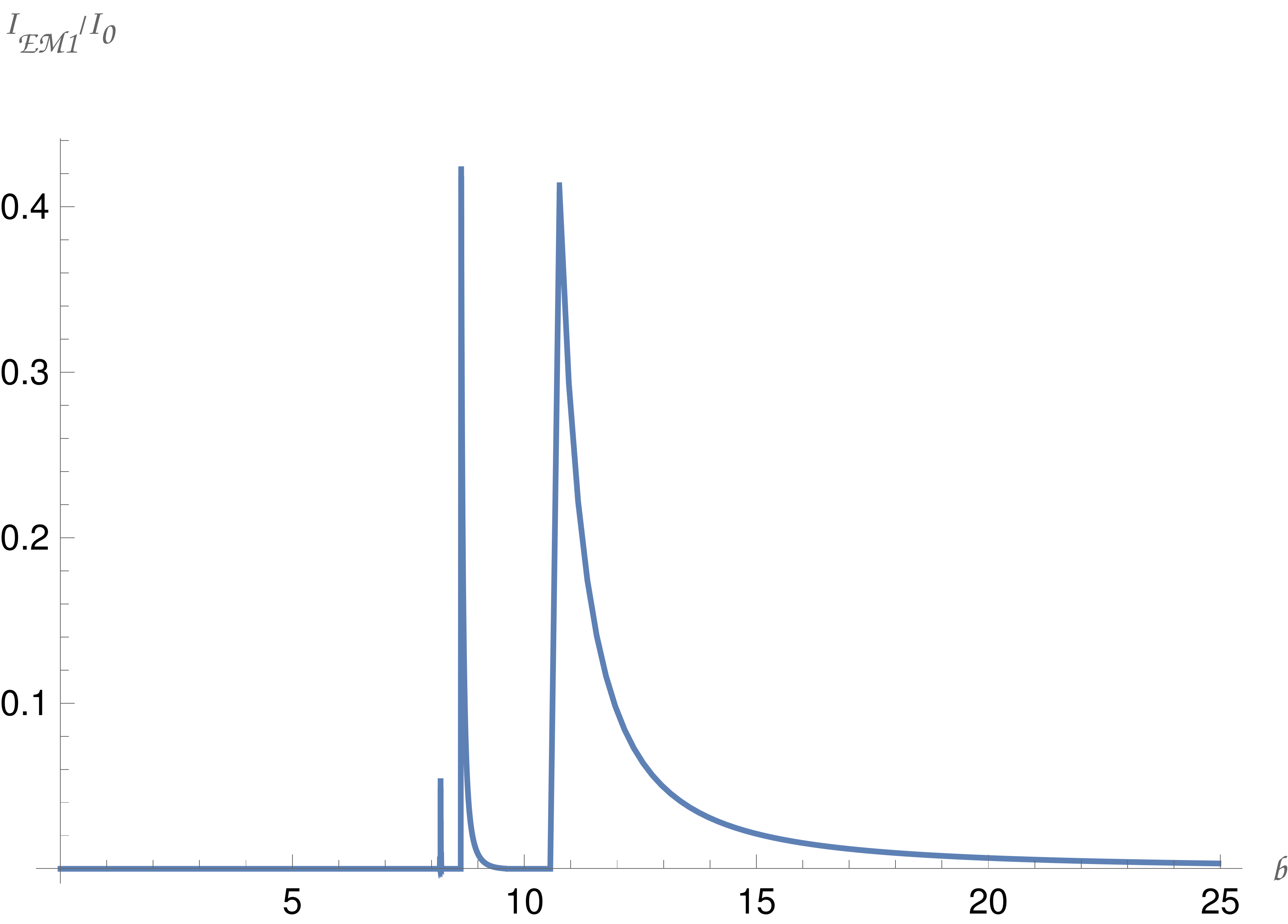}
    \includegraphics[width=.3\textwidth]{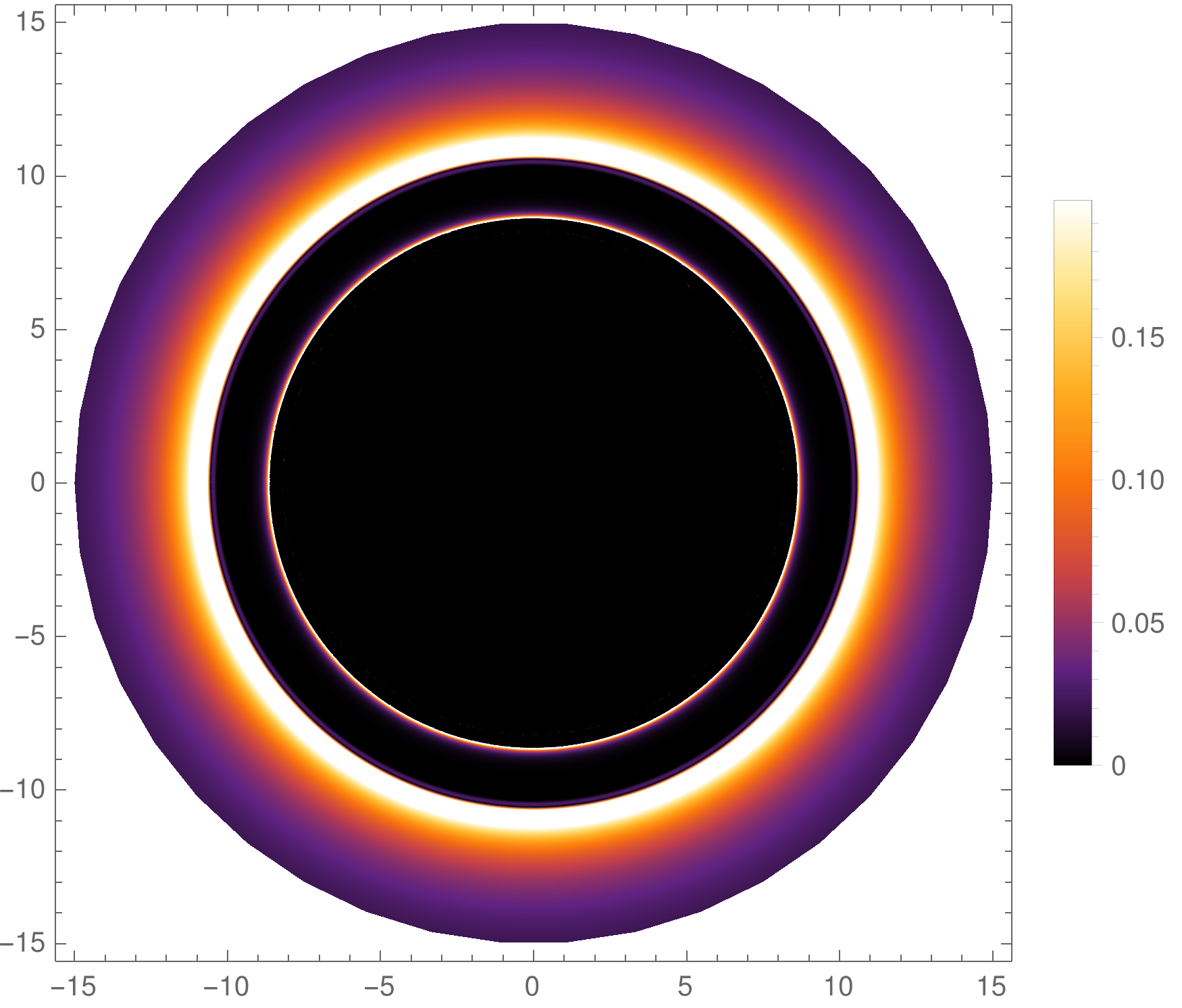}
    \vfill
    \centering
    \includegraphics[width=.3\textwidth]{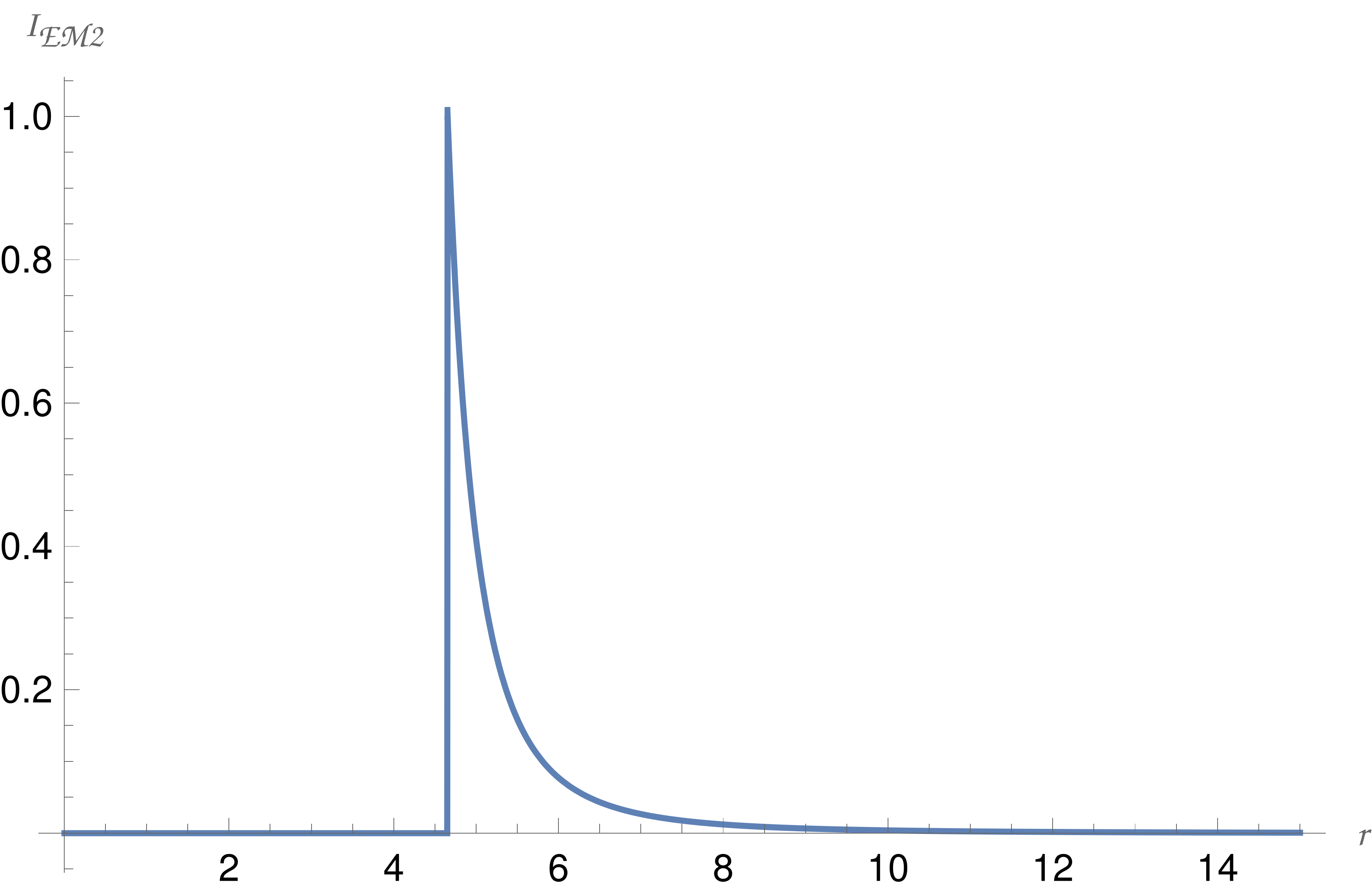}
    \includegraphics[width=.3\textwidth]{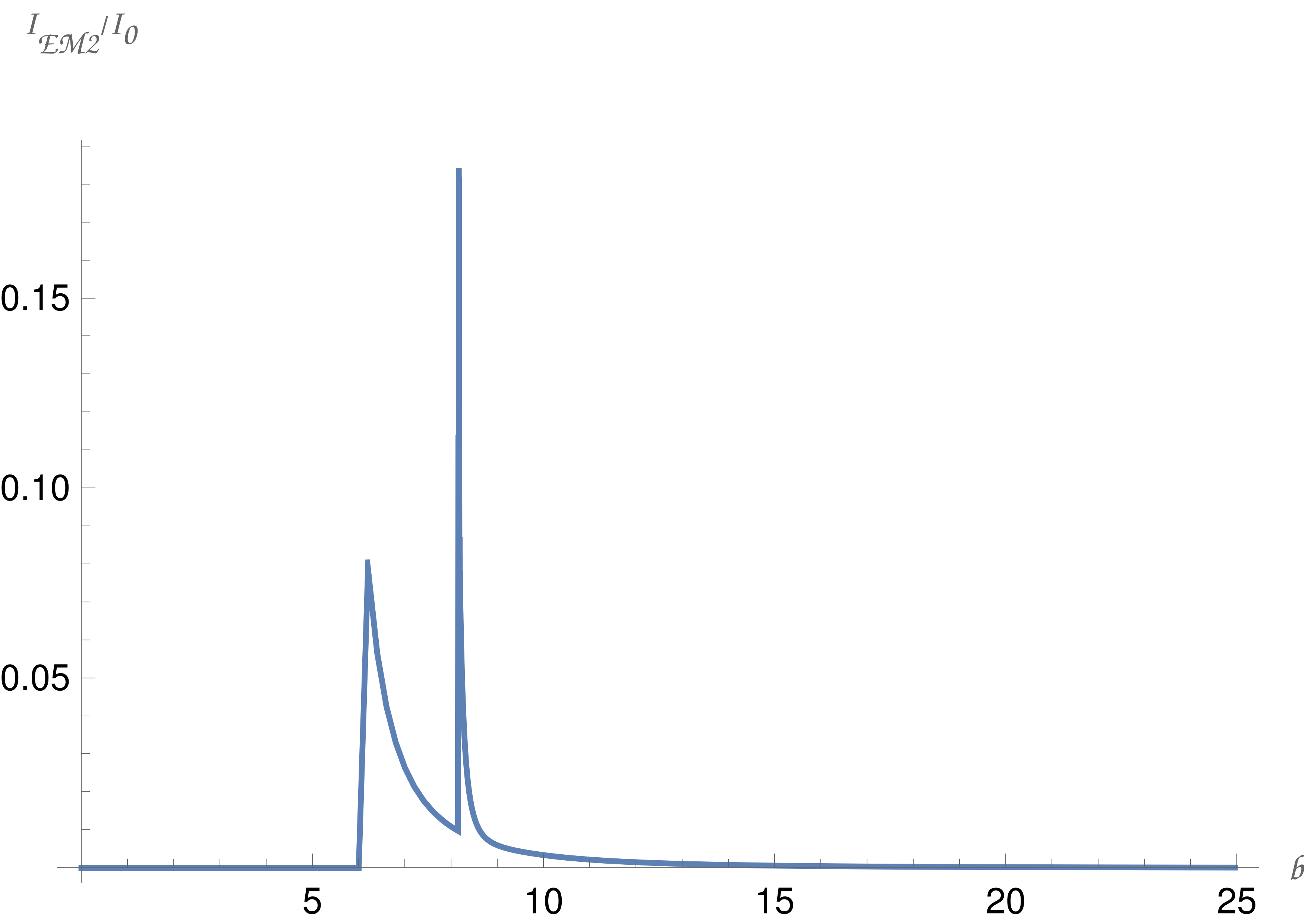}
    \includegraphics[width=.3\textwidth]{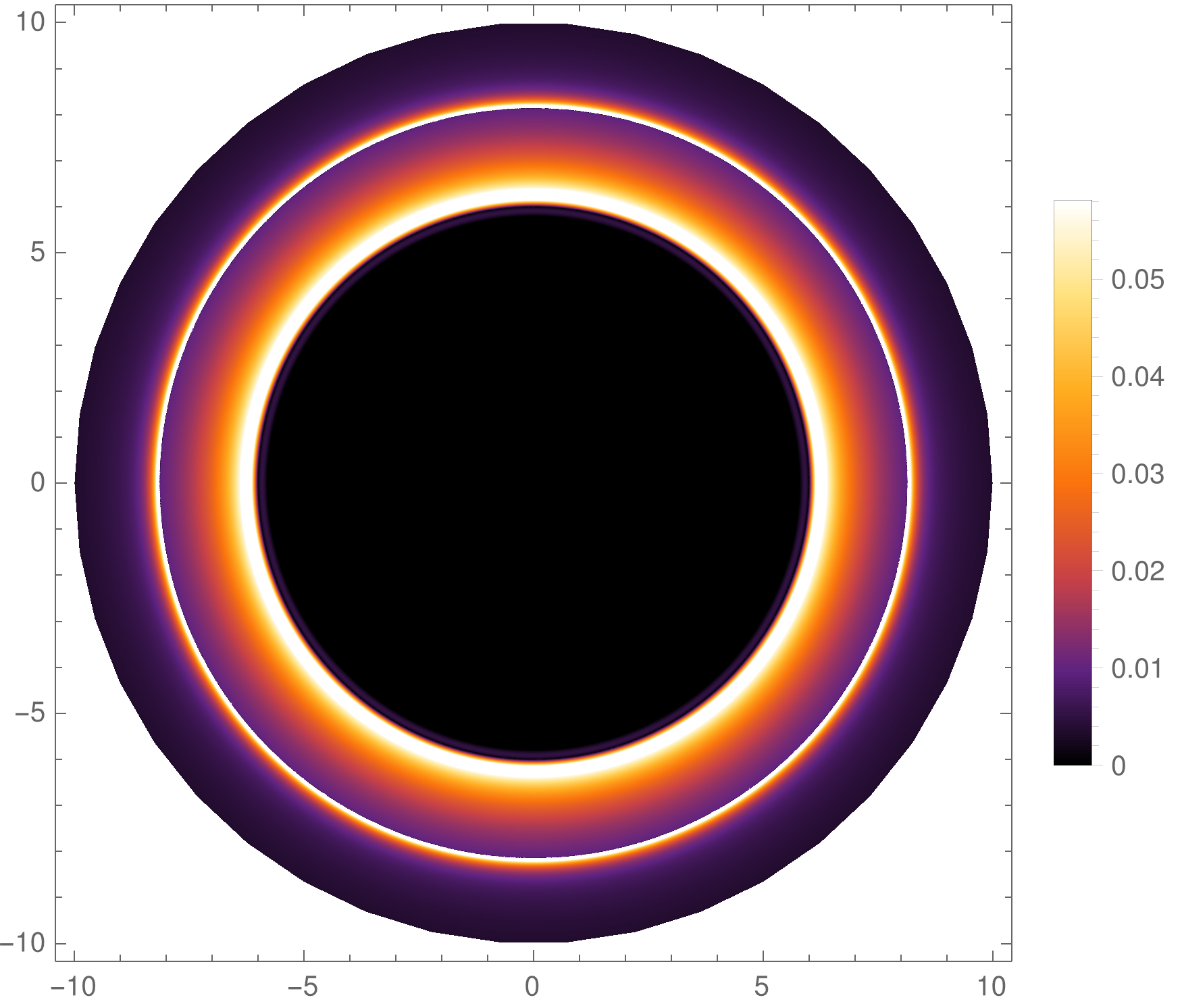}
    \vfill
    \centering
    \includegraphics[width=.3\textwidth]{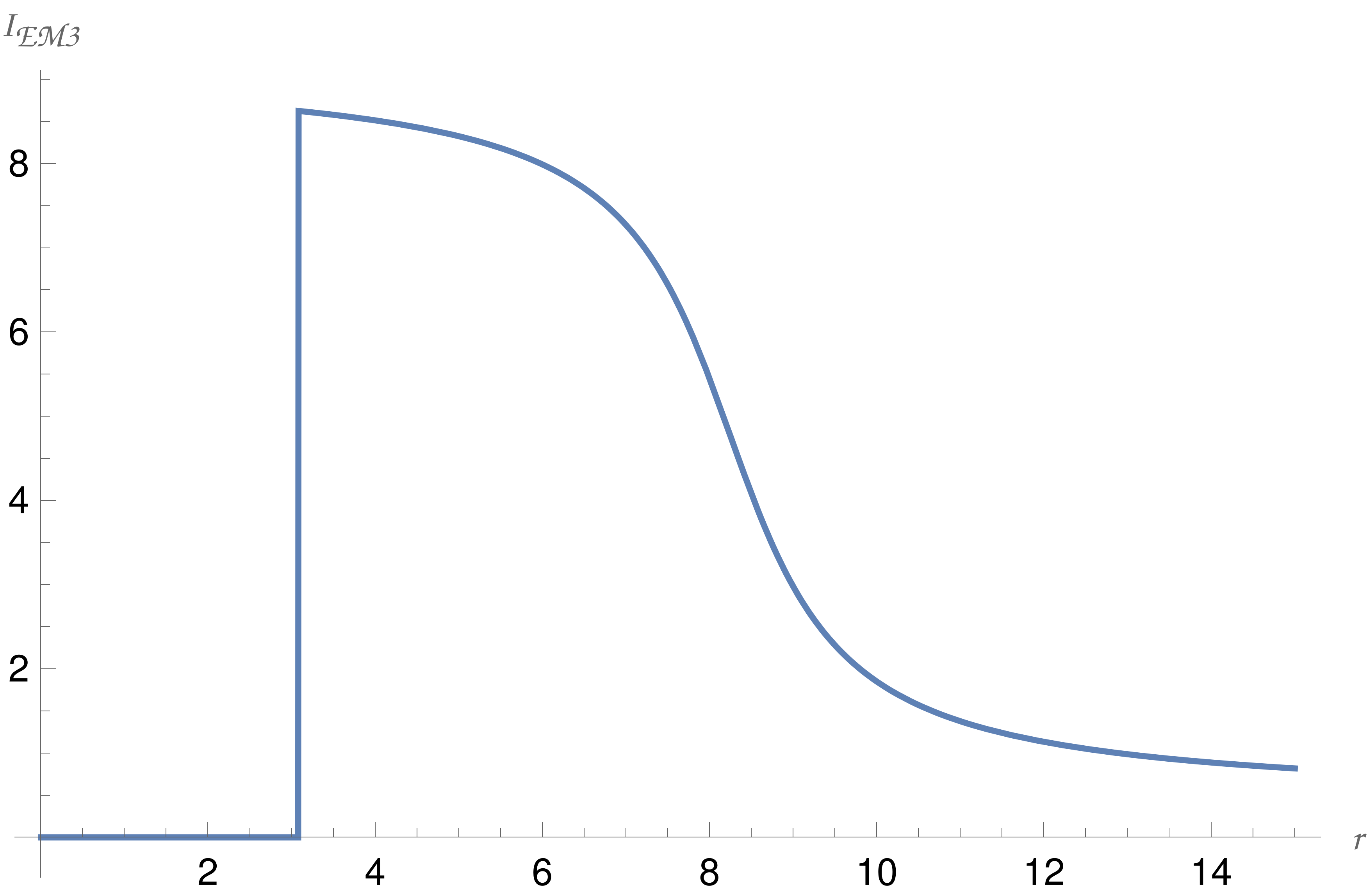}
    \includegraphics[width=.3\textwidth]{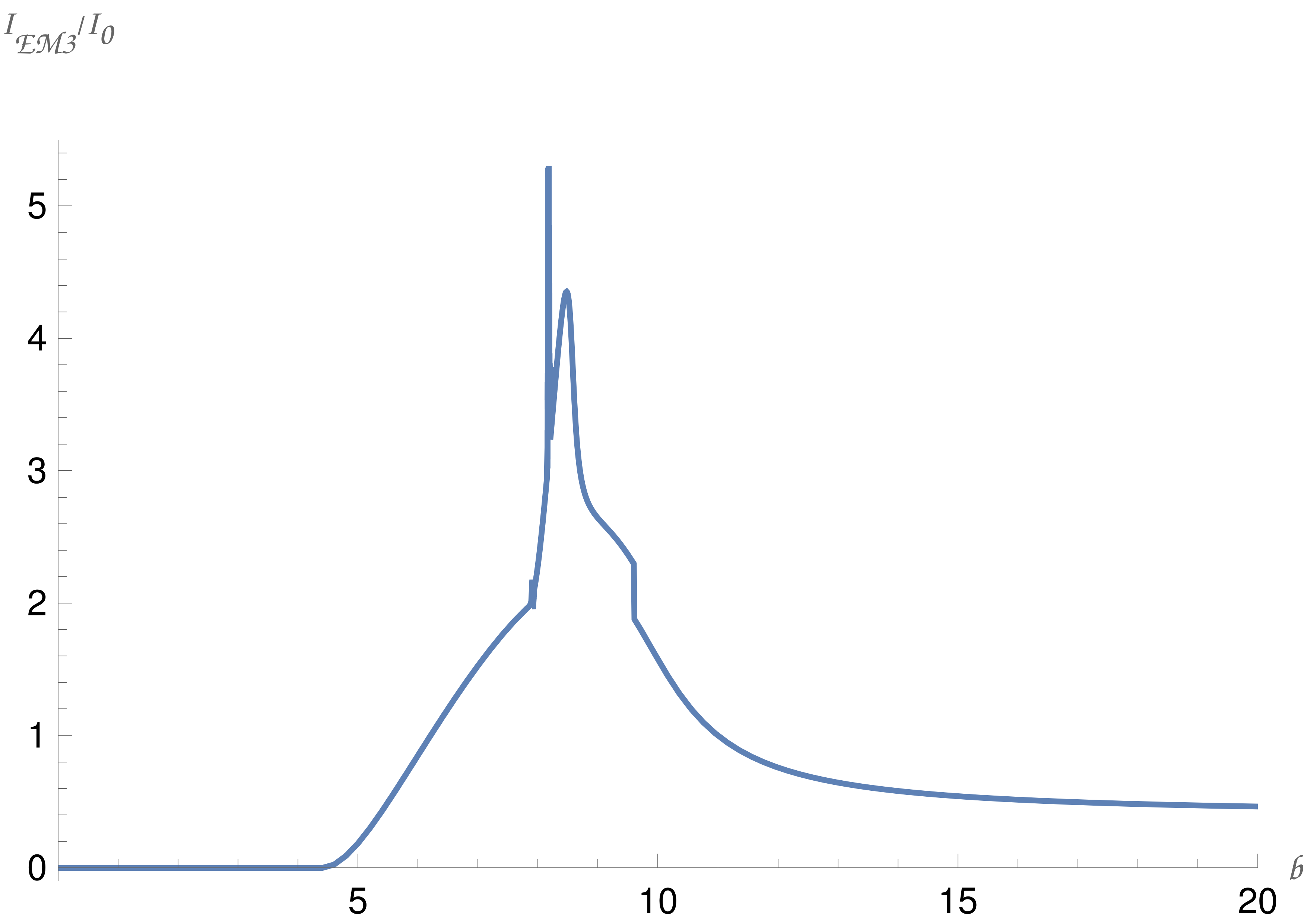}
    \includegraphics[width=.3\textwidth]{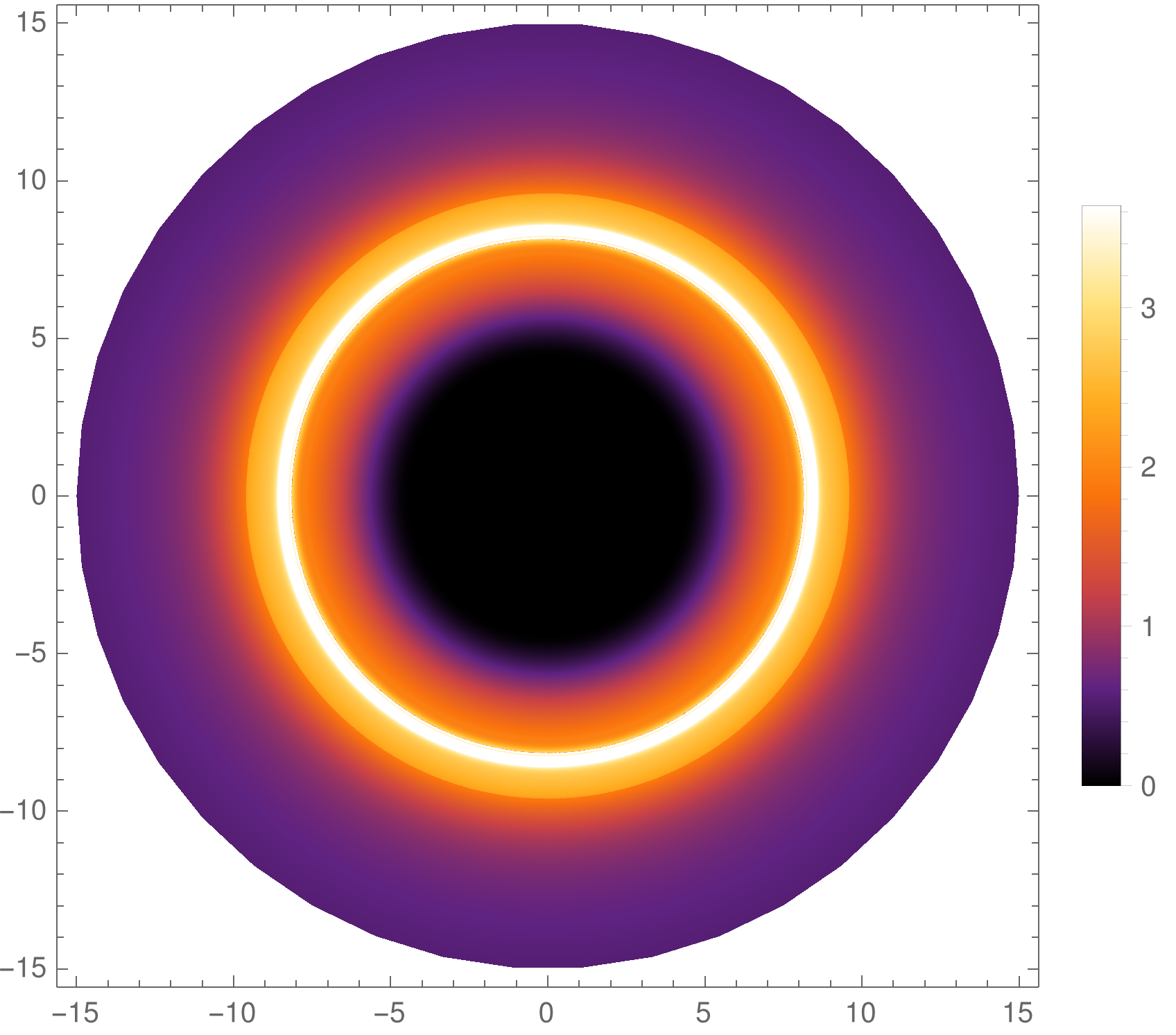}
    \caption{The face-on observed thin disk for $q_m=0.8$ and $\beta=0.1$.
    %The appearance of a face-on observed thin disk with varying emission profiles for $q_m=0.8$ and $\beta=0.1$. 
    %Each row corresponds to a different model: top row - model $1$, second row - model $2$, and third row - model $3$. The emitted and observed intensities ($I_{\mathrm{EM}}$ and $I_{\mathrm{obs}}$) in the plots are normalized to the maximum emitted intensity value outside the horizon ($I_0$).
    }
    \label{fig:18}
\end{figure*}
\begin{figure*}[htbp]
    \centering
    \includegraphics[width=.3\textwidth]{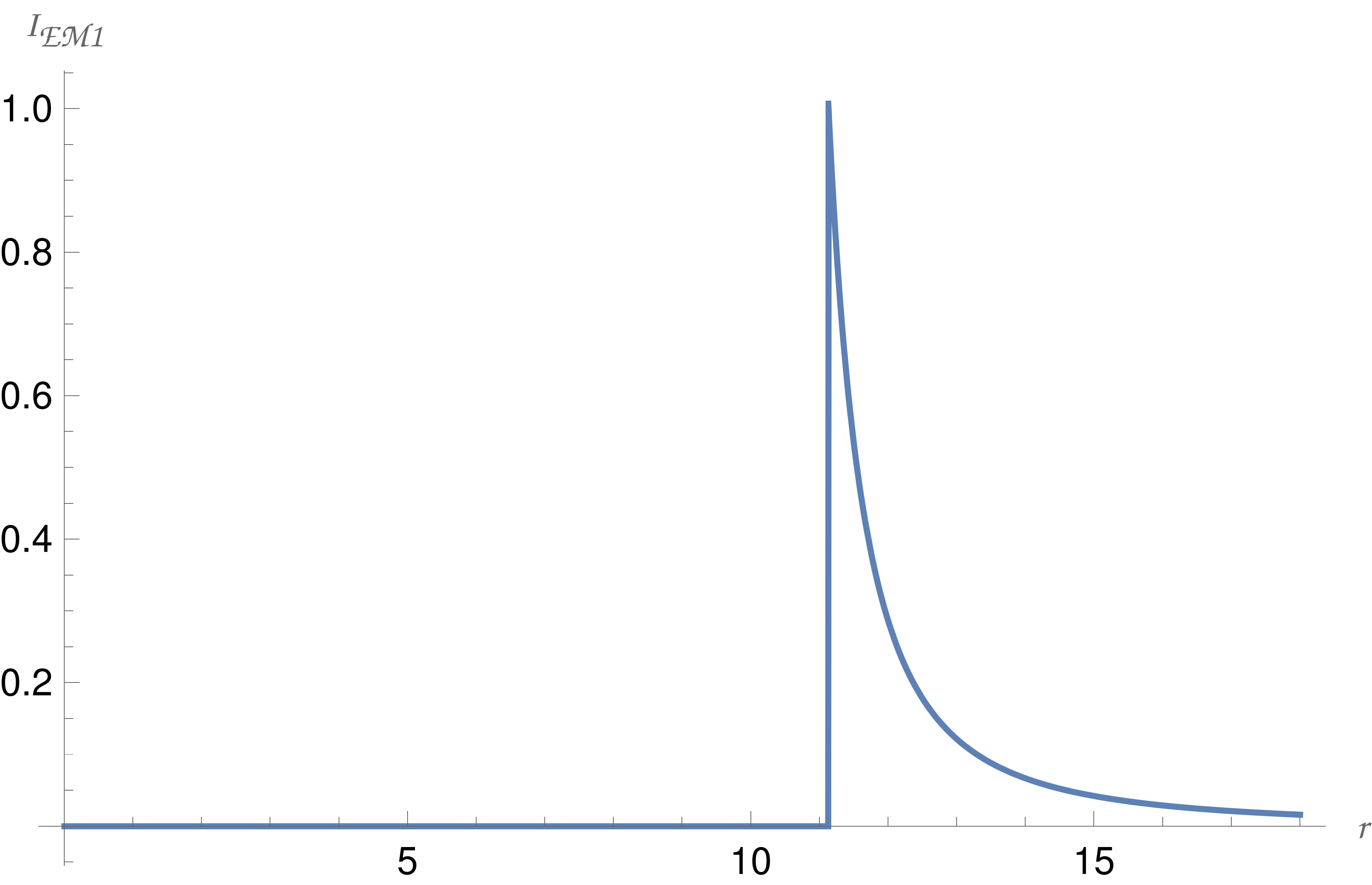}
    \includegraphics[width=.3\textwidth]{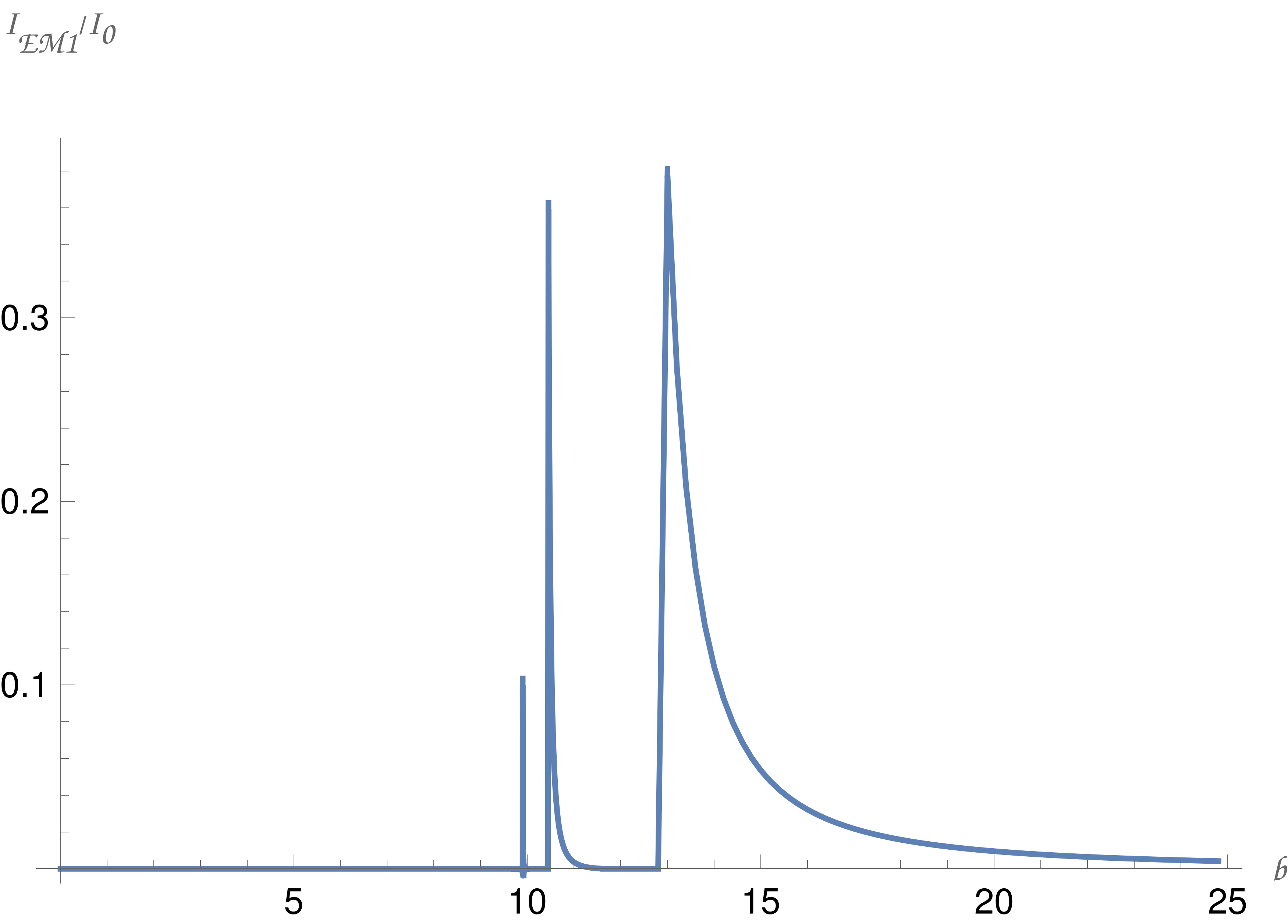}
    \includegraphics[width=.3\textwidth]{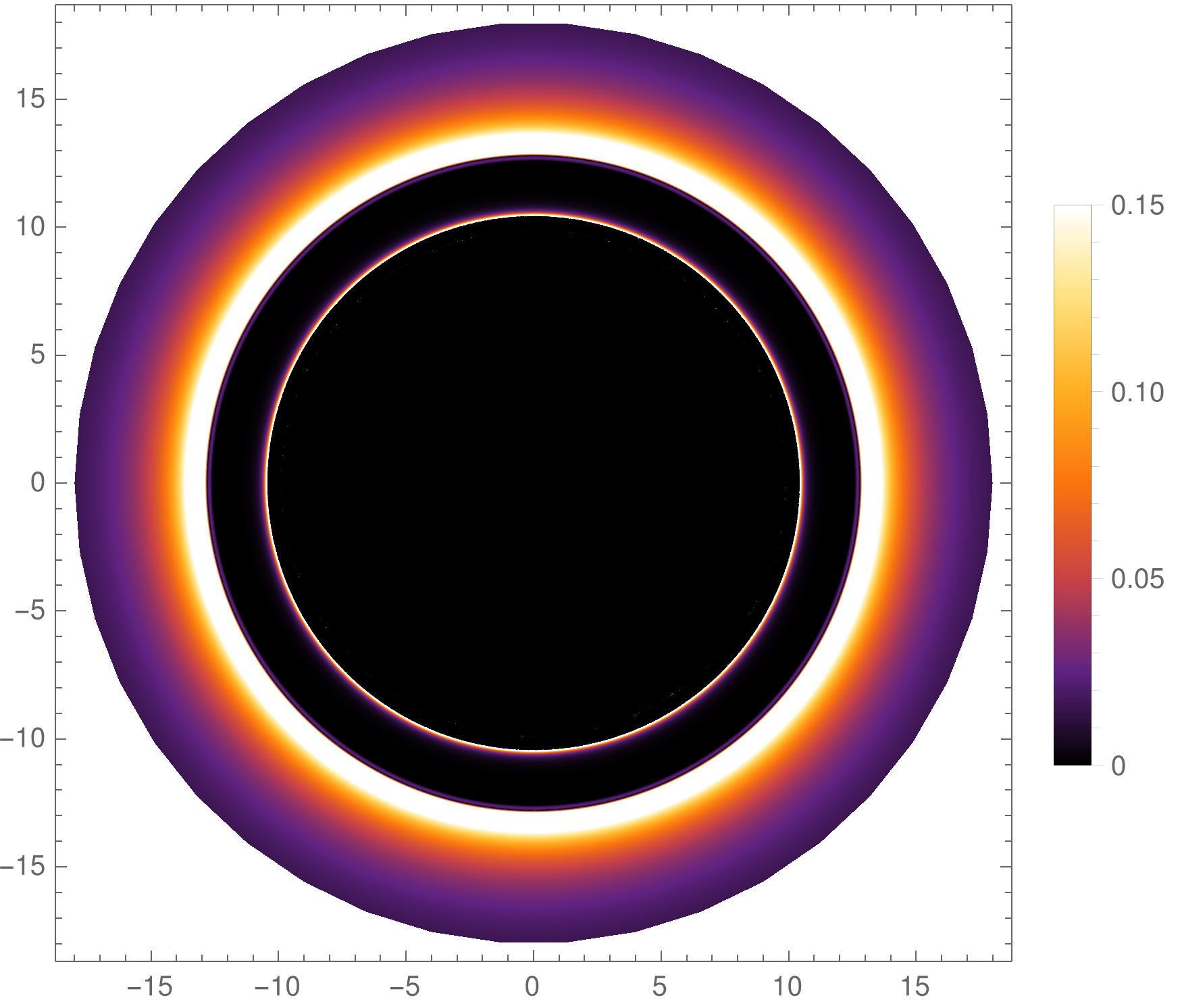}
    \vfill
    \centering
    \includegraphics[width=.3\textwidth]{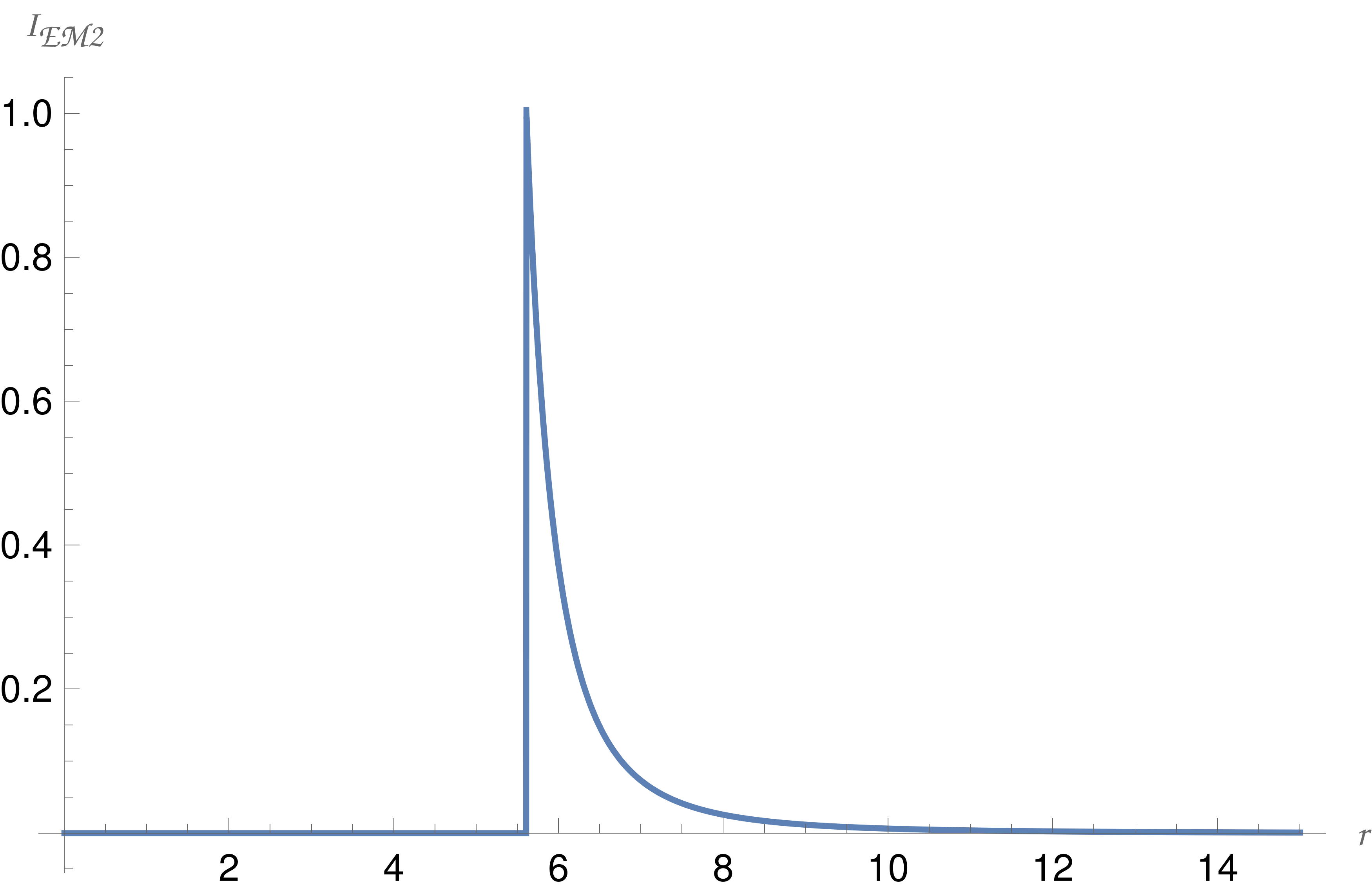}
    \includegraphics[width=.3\textwidth]{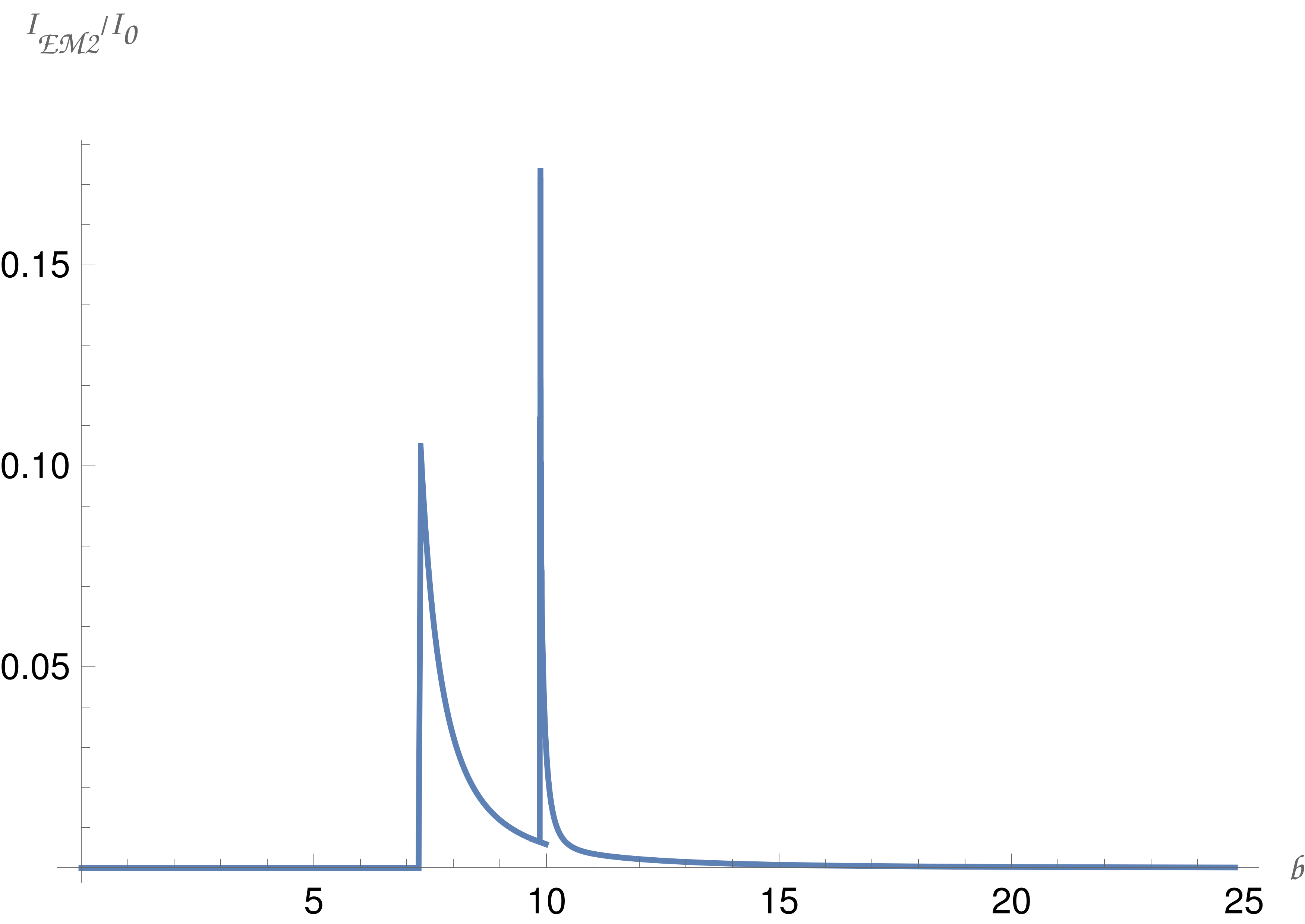}
    \includegraphics[width=.3\textwidth]{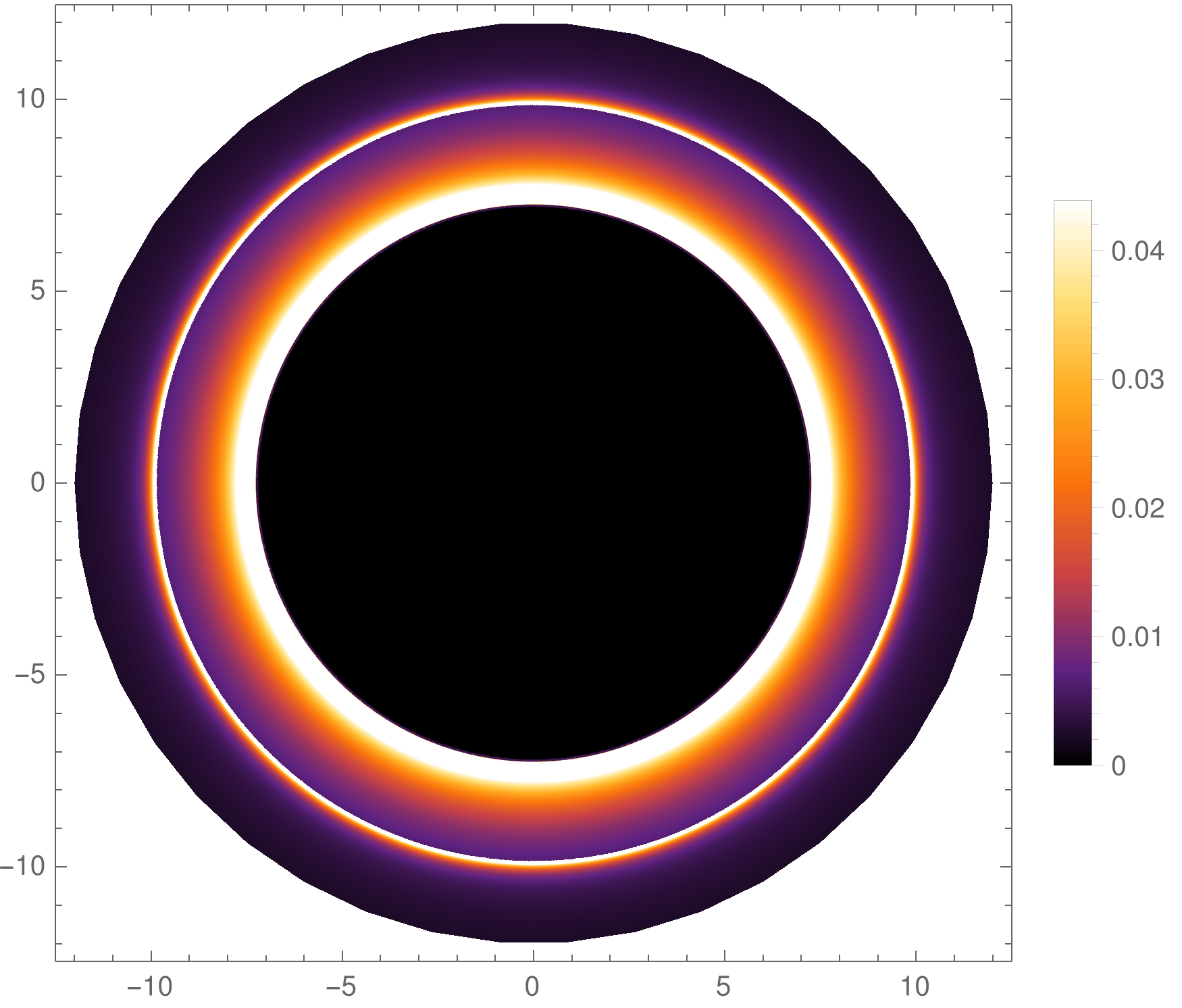}
    \vfill
    \centering
    \includegraphics[width=.3\textwidth]{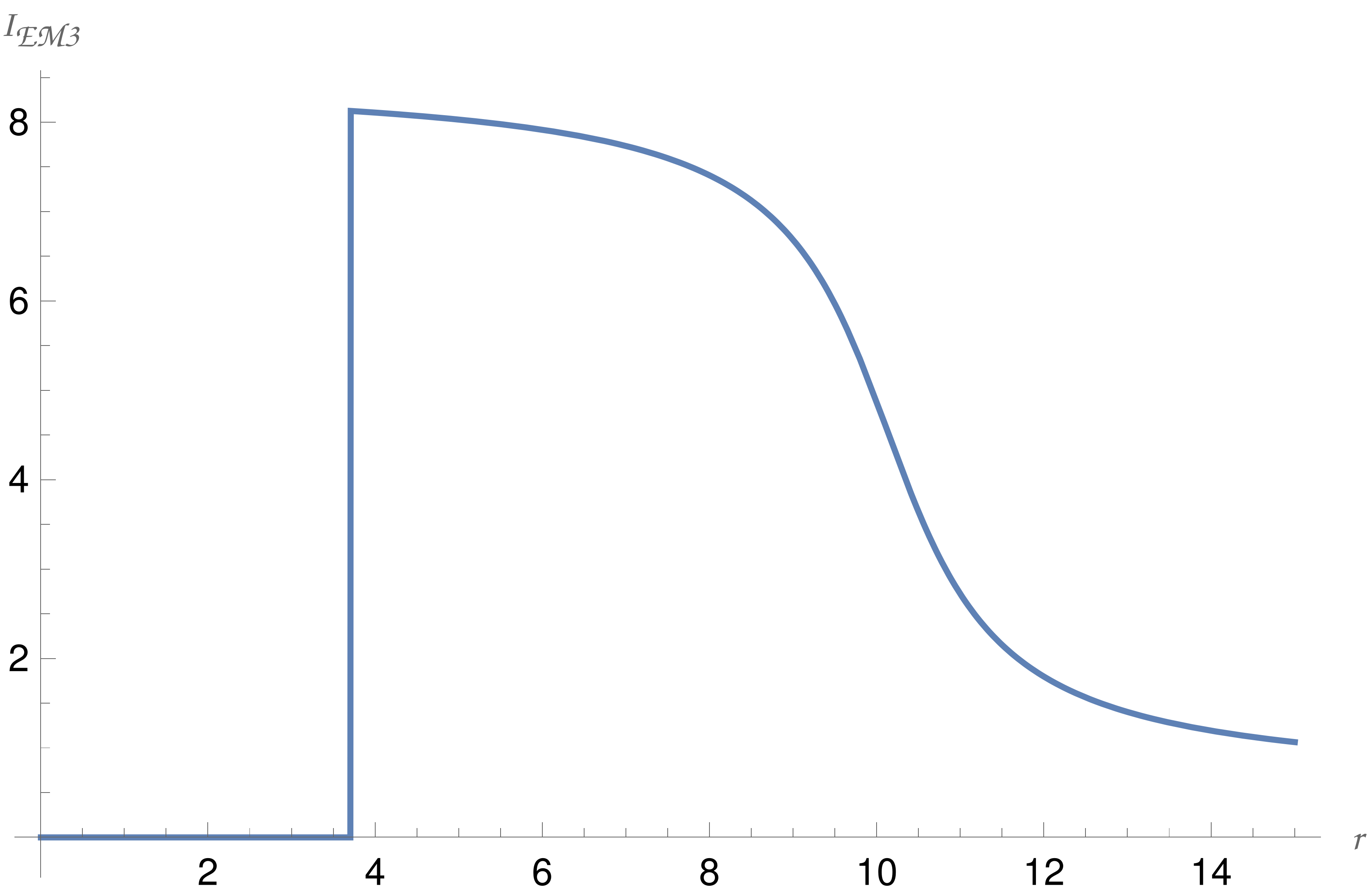}
    \includegraphics[width=.3\textwidth]{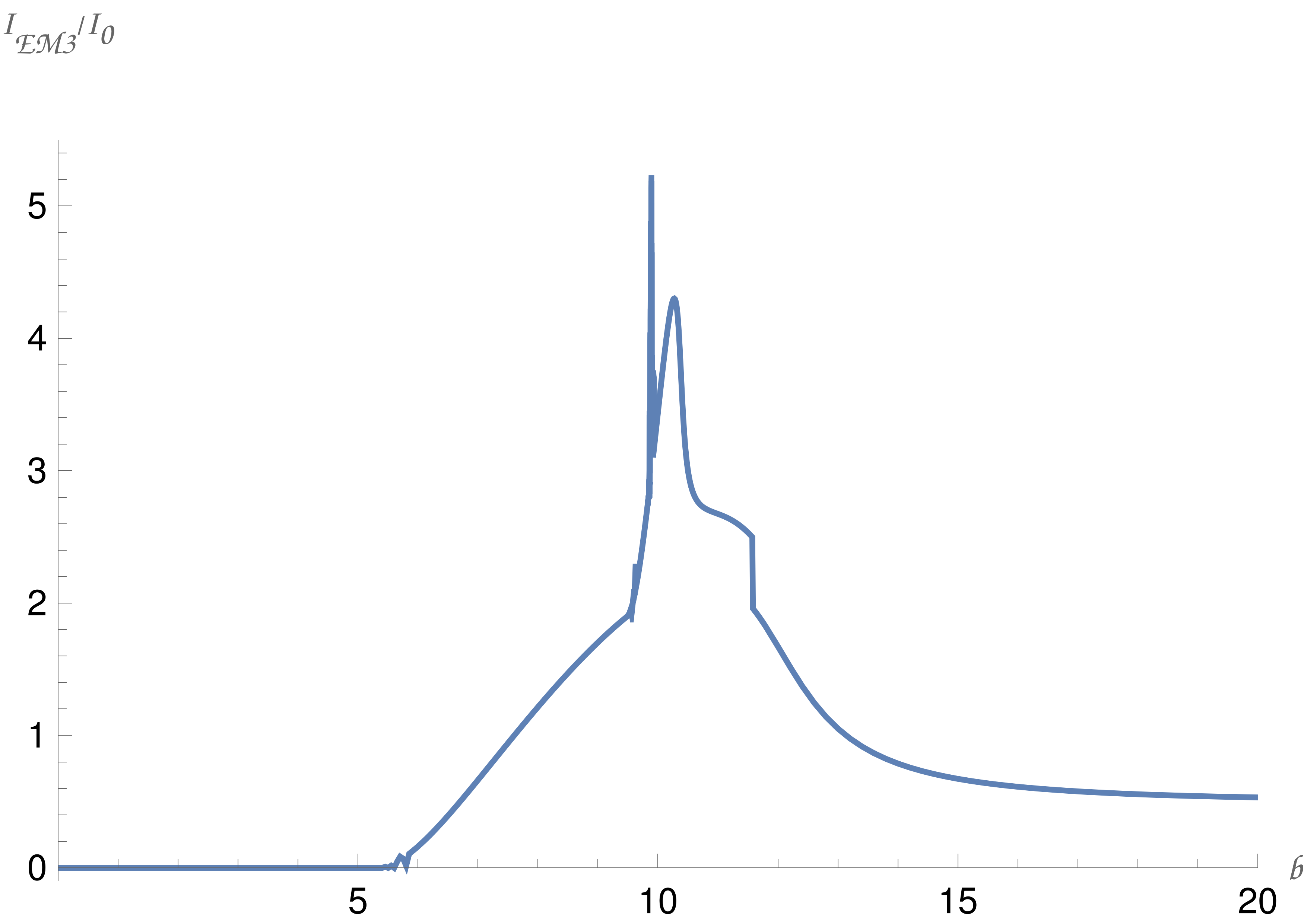}
    \includegraphics[width=.3\textwidth]{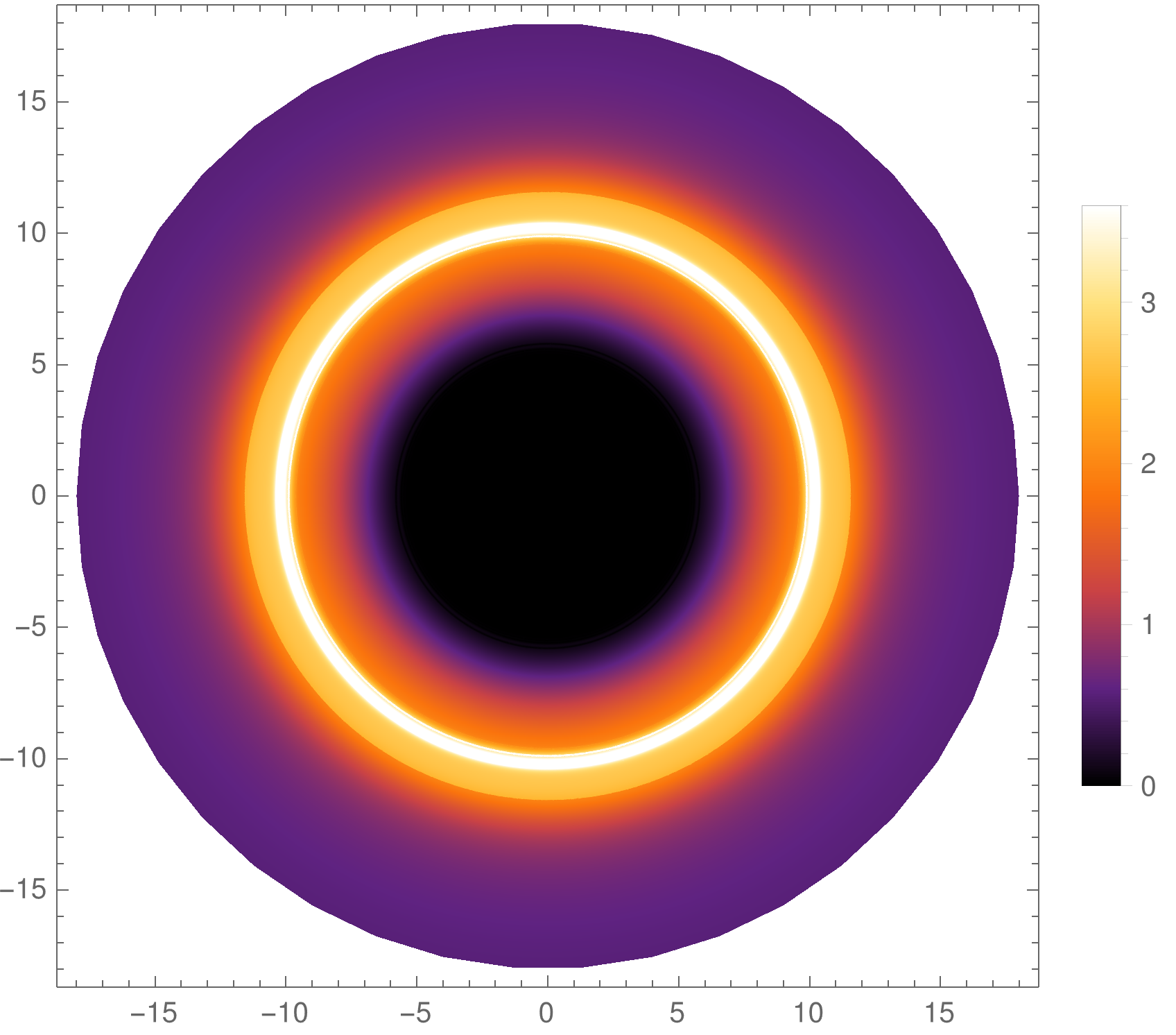}
    \caption{The face-on observed thin disk for $q_m=1.0$ and $\beta=0.1$.
    %The thin disk's visual appearance was observed in a face-on orientation with various emission profiles for $q_m=1.0$ and $\beta=0.1$. Model $1$ "model $2$ and "model $3$ from section VI were used for the upper, second, and third rows, respectively. The emitted and observed intensities ($I_{EM}$ and $I_{obs}$) in the plots are standardized to the highest emitted intensity outside the horizon ($I_0$).
    }
    \label{fig:19}
\end{figure*}

Figures \ref{fig:17}, \ref{fig:18}, and \ref{fig:19} were plotted to understand the behavior with fixed $\beta=0.01$ and varying $q_m=0.6,0.8$ and $1.0$, for the described emission profiles emitted from the thin accretion disk around the BH. Each row in the figures corresponds to a different model.

For model 1, the emission profile starts from the ISCO and decays as the radial distance increases. The observed intensity exhibits multiple peaks corresponding to direct emission, lensing, and photon rings. The photon and lensing rings' peaks are narrower and smaller than the direct emission peak. The two-dimensional shadow image in the first row and the third column shows a single bright ring inside the accretion disk's inner boundary, which is formed due to the existence of the photon sphere inside the ISCO. Direct emission contributes significantly to the brightness, and the shadow image displays additional rings resulting from lensing and photon ring effects.

In model 2, the emission intensity profile starts from the photon sphere position and decreases with increasing radial distance. The observed emission profile also exhibits multiple peaks, with a prominent narrow peak and a subsequent decrease with the impact parameter. The shadow image in the second row, third column, shows the superimposed contribution of lensing and photon rings on the direct emission, leading to an increased observed intensity area. Despite this, direct emission remains dominant.

Model 3 displays a peak at the horizon with a subsequent slow decrease to a certain value as the radial coordinate increases. The observed intensity has a large observational area due to the superimposed contribution of photon and lensing rings on the direct emission. The lensing ring's contribution increases, but direct emission still dominates, and the photon ring has minimal influence. The two-dimensional shadow image in the third row, third column, demonstrates this behavior.

As $q_m$ increases with a fixed $\beta=0.1$, the positions of the peaks shift significantly, distinguishing it from the Schwarzschild spacetime \cite{Gralla:2019xty}. The background metric plays a crucial role in determining the peak and deflection angle, influencing the observational features of the BH. In the next section, we will explore the observational features of the infalling accreting matter.

\section{Observational features of the shadow with Infalling Spherical Accretion}\label{sec:signaturesInfall}

In this section, we explore the observational characteristics of the BH accretion disk with infalling matter in the context of the effective metric background. We adopt the optically thin accretion disk model, where matter falls directly into the BH with negligible angular momentum \cite{Bambi:2013nla}. During this spherical infall, the gas heats up and emits radiation due to the strong gravitational field around the BH. Consequently, we consider the observed specific intensity from the perspective of an observer at infinity as 
\begin{equation}
 I_\text{obs}= \int_\Gamma g^3 j(\nu_e) \ed l_\text{prop},
 \label{is}
\end{equation}
where $g$, $\nu_e$, and $\nu_{\mathrm{obs}}$ are known as the redshift factor, the photon frequency at emission, and the observed photon frequency at infinity, respectively. To compute the aforementioned expression, it is necessary to define the emissivity per unit volume, $j(\nu_e)$, from the emitter's rest frame. For this work, we adopt a $1/r^2$ profile, represented as $j(\nu_e) \propto {\delta (\nu_e-\nu_f)}/{r^2}$. It is important to note that the expression contains the delta function $\delta$, where the frequency of the radiative light ($\nu_f$) is considered monochromatic in nature, and $\ed l_\text{prop}$ denotes an infinite proper length. The redshift factor can be expressed as 
\begin{equation}
g=\frac{\mathcal{K_{\rho}}u_{o}^{\rho}}{\mathcal{K_{\sigma}}u_{e}^{\sigma}},
\end{equation}
where $\mathcal{K^{\mu}}$ represents the four-velocity of the photon, and $u_o^{\mu}=(1,0,0,0)$ is the four-velocity of the static observer at infinity. In this scenario, the components of the four-velocity of the infalling matter in a static and spherically symmetric $(-A(r), B(r), C(r),sin^2\theta C(r))$ can be expressed as shown in \cite{KumarWalia:2022ddq, Shaikh:2018lcc} 
\begin{equation}
u^t_e=A(r)^{-1},\quad 
u^r_e=-\sqrt{\frac{1-A(r)}{A(r)B(r)}},\quad 
u^{\theta}_e=u^\phi_e=0.
\end{equation}
Therefore, the four-velocity components for photons originating from the spherical disk are
\begin{equation}
\frac{\mathcal{K}_r}{\mathcal{K}_t}=\pm \sqrt{\frac{B(r)}{A(r)}-\frac{b^2 B(r)}{C(r)}}.
\end{equation}
Here, the $+$ and $-$ signs correspond to the photons traveling towards or away from the BH. Hence, the redshift factor for the infalling accretion is given by
\begin{equation}
g=\left[ u_e^t + \left( \frac{\mathcal{K}_r}{\mathcal{K}_t} \right) u^r_e \right]^{-1}.
\end{equation}
In this case, the proper distance has the following expression:
\begin{equation}
\ed l_\text{prop}=\mathcal{K}_{\mu}u^{\mu}_e \ed\lambda = \frac{\mathcal{K}_t}{g |\mathcal{K}_r |}\ed r.
\end{equation}
Using all these expressions, we can now calculate the observed intensity for the static observer at infinity by integrating Eq. (\ref{is}) over all frequencies, which yields
\begin{equation}
I_\text{obs} \propto \int_{\Gamma} \frac{g^3 \mathcal{K}_t \ed r}{r^2 |\mathcal{K}_r |}.
\end{equation}
Using this expression, we can explore the effect of the effective metric background on photons reaching the observer at infinity. In Fig. \ref{fig:compp}, we compare the observed intensity ($I_\text{obs}$) profiles for the effective metric case with the Schwarzschild spacetime. In all cases, the intensity increases to a peak value at $b \sim b_{\mathrm{ph}}$ and then decreases as $b$ increases, eventually reaching zero. We also observe that the intensity decreases with the effective metric, but the observational area increases, indicating a decrease in brightness in the shadow image. In the figure, we fixed $\beta=0.1$ and varied $q_m$ along with Schwarzschild $\beta=q_m=0$ (in black), $q_m=0.6$ (in green), $q_m=0.8$ (in black), and $q_m=1.0$ (in red). To observe this effect in the two-dimensional shadow plot, we presented Fig. \ref{fig:21}. A similar trend is evident, where increasing $q_m$ results in decreased brightness but an increased observational area.
\begin{figure}[htbp]
    \centering
    \includegraphics[width=0.6\linewidth]{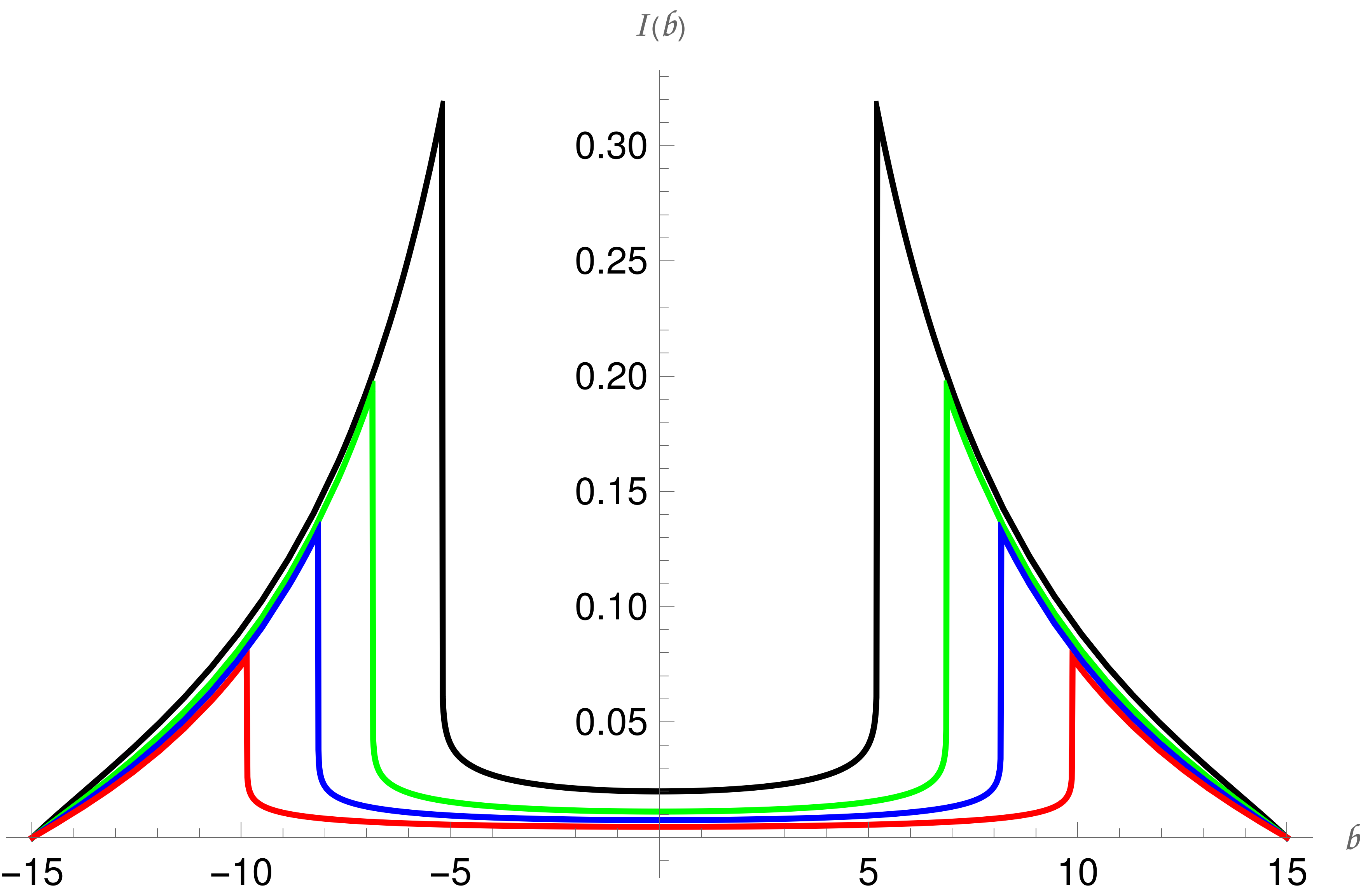}
    \caption{The intensity of infalling spherical accreting matter observed at various values of $q_m$, each represented by a different color. The color scheme includes Schwarzschild in black, $q_m=0.6$ in green, $q_m=0.8$ in black, and $q_m=1.0$ in red, all with $\beta=0.1$.
    %The intensity of infalling spherical accreting matter was observed at varying values of $q$, represented by different colors. The values include Schwarzschild in black, $q=0.6$ in green, $q=0.8$ in black, and $q=1.0$ in red with $\beta=0.1$.
    } 
    \label{fig:compp}
\end{figure}
\begin{figure}[htbp]
    \centering
    \includegraphics[width=.3\textwidth]{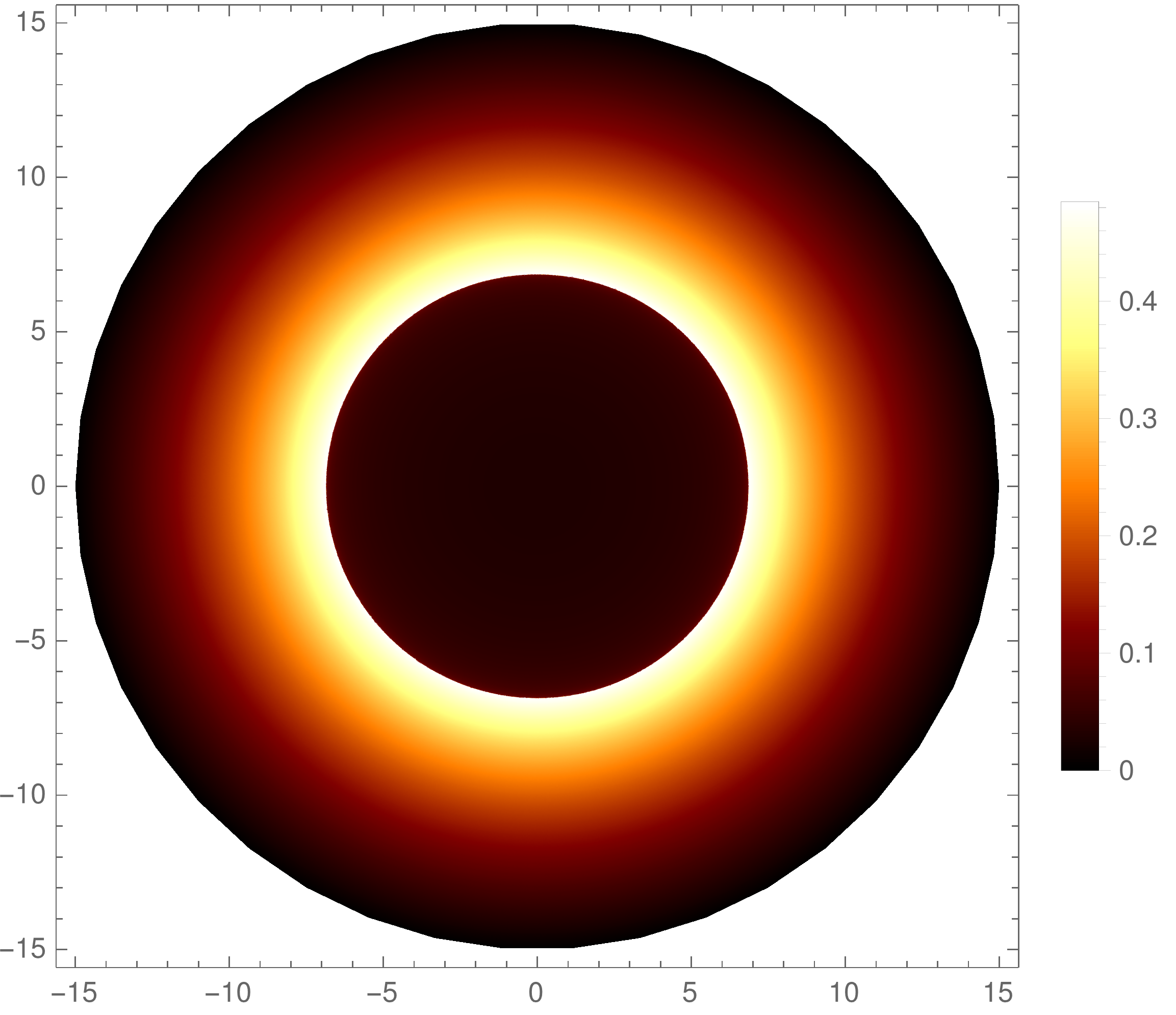}
    \includegraphics[width=.3\textwidth]{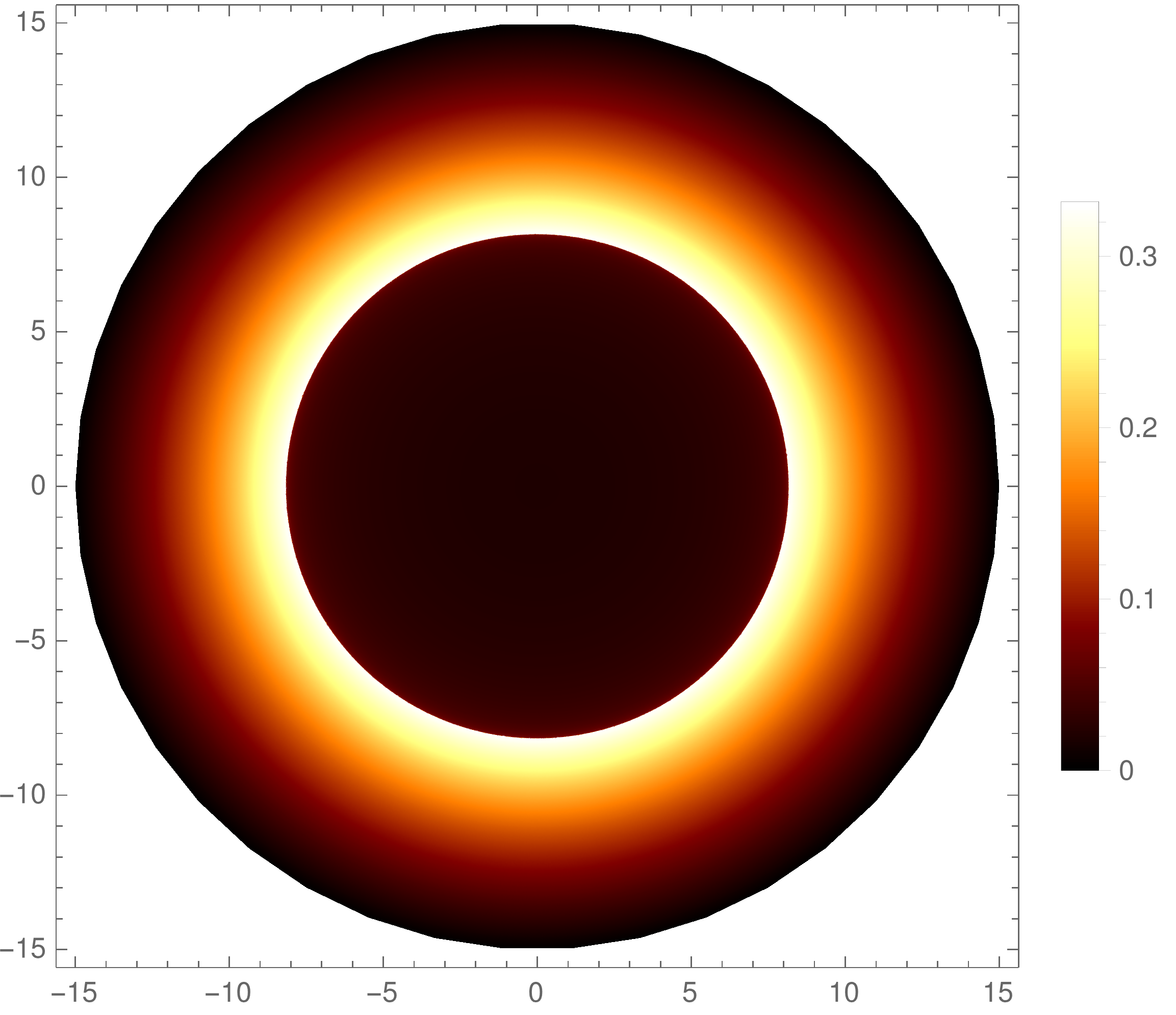}
    \includegraphics[width=.3\textwidth]{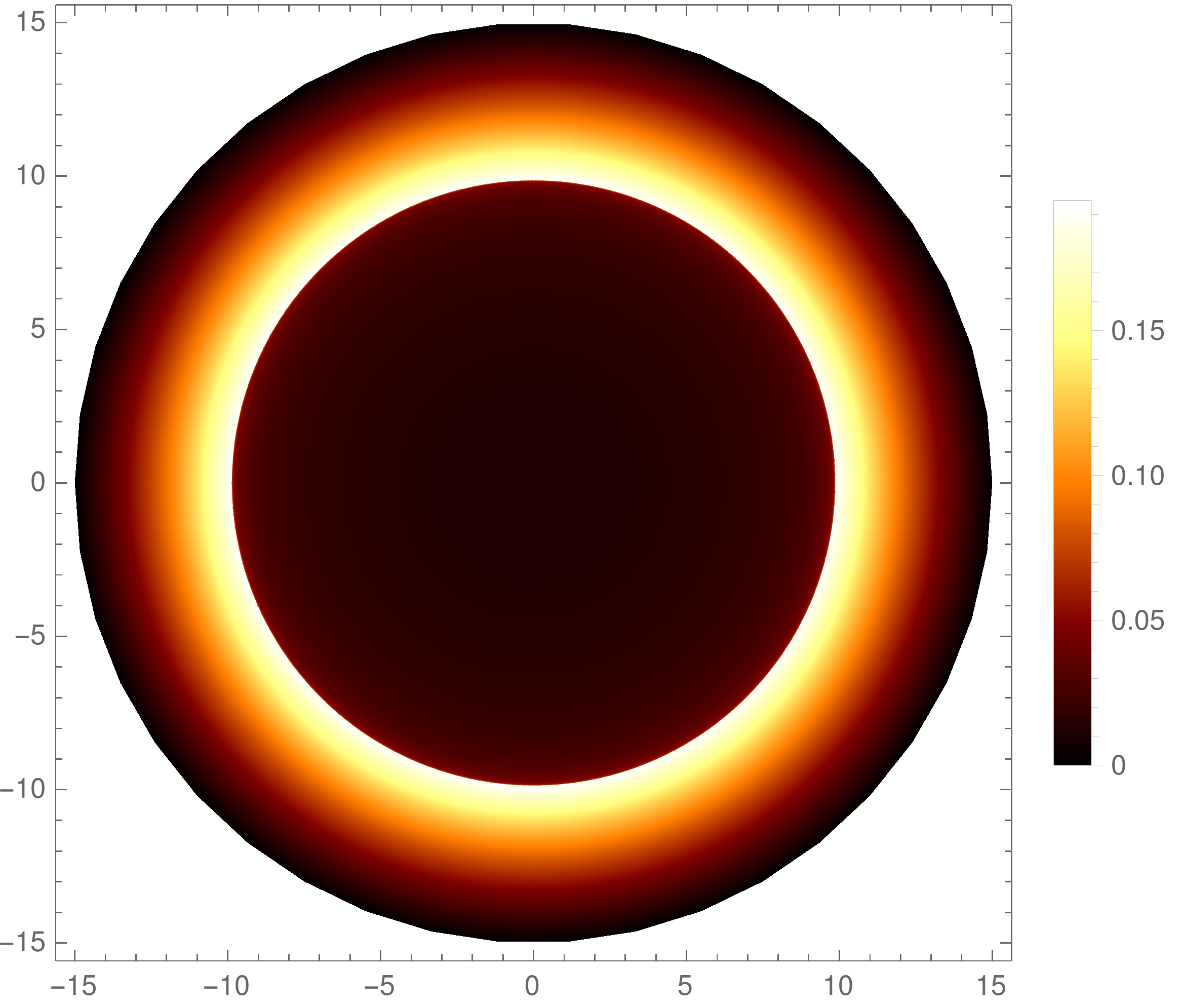}
    \caption{The BH shadow with infalling spherical accretion shown for $q_m=0.6, 0.8$ and $1.0$, all with $\beta=0.1$.
    %An image was captured of the BH shadow with infalling spherical accretion, with different values of $q$ represented by $0.6$, $0.8$, and $1.0$, respectively with $\beta=0.1$.
    } 
    \label{fig:21}
\end{figure}

%%%%%%%%%%%%%%%%%%%%%%
\section{Summary and Conclusions}\label{sec:conclusions}

%\textcolor{Orange}{Estoy aqui!$\rightarrow$}\\\\

In this paper, we examined a spherically symmetric NED BH solution. We considered the deviation of light rays from conventional null geodesics in NED, as they instead follow the null geodesics of the effective metric, obtained through modifications to the original spacetime. Consequently, we derived the effective metric, and in Fig. \ref{fig:1aa}, we demonstrated that the minimum allowed condition for the radial coordinate ($r_{\mathrm{eff}}$) in our specific solution lies inside the horizon ($r_\mathrm{h}$). Subsequently, we investigated the photon trajectories around the BH, whose spacetime is defined by the effective background geometry for the light rays. In Fig. \ref{fig:15}, we presented two-dimensional images of the photon trajectories for a specific value of the NED parameter $\beta=0.1$ and different magnetic charge values of $q_m=0.6,0.8$ and $1.0$. We observed that as the magnetic charge increases, the BH horizon expands, along with an increased range for the direct emission, lensing, and photon rings. Furthermore, we constrained the spacetime parameters using data from the EHT for M87* and Sgr A* in Fig. \ref{fig:shacons}. It was found that Sgr A* provides a more stringent constraint on the parameters. In our study, we selected parameter values within the $1\sigma$ and $2\sigma$ uncertainties and investigated the emission profiles around the BH. Subsequently, before delving into the observational signatures of the BH, we focused on the analytical derivation of the deflection angle for the passing light rays. This step is important as it forms the foundation of the gravitational lensing process. We used the constraints obtained earlier for the magnetic charge, but we varied the NED parameter to emphasize its effect on the deflection angle, especially for larger values. This allowed us to gain insights into how the NED parameter influences the bending of light around the BH and its impact on the overall gravitational lensing phenomenon. 

Regarding the observational signatures, we focused on the scenario where the BH is illuminated by an optically thin accretion disk. We studied three specific models, as described in Section \ref{sec:signatures}, for the emission profiles. For each model, we calculated the observed profile for the observer at infinity, while keeping the NED parameter fixed and varying the magnetic charge. The results were presented in Figures \ref{fig:17}, \ref{fig:18}, and \ref{fig:19}.
%with $\beta=0.1$ fixed, while varying the magnetic charge $q=0.6,0.8,1.0$ in Figs. \ref{fig:17}, \ref{fig:18}, and \ref{fig:19}. 
The observed emission profiles were then depicted in the two-dimensional images. In all the models, the direct emission consistently appeared prominent. However, in models 2 and 3, the contribution of the lensing ring increased in the observed emission, while the photon ring remained relatively small. These characteristics were evident in the form of brightness and rings in the respective figures. The results indicate that the direct emission plays a significant role in the observed brightness, while the lensing ring's contribution increases, and the photon ring remains relatively minor in all three models. This outcome is consistent across different values of the magnetic charge. Furthermore, we investigated the spherical infalling accretion around the NED BH and compared it with the Schwarzschild case. In Fig. \ref{fig:compp}, we demonstrated that the observed intensity peak is higher for the Schwarzschild BH compared to the NED BH; however, the observed angular region is larger for the NED BH. In all cases, the intensity peak occurs at $b \sim b_{\mathrm{ph}}$, and as the impact factor $b$ increases, the intensity gradually decreases until it eventually reaches zero. The impact of this effect can also be observed in the two-dimensional shadow images shown in Fig. \ref{fig:21}, where we fixed the $\beta$-parameter and varied the magnetic charge.

It is worth noting that the model retains its realism even when the NED parameter and magnetic charge are set to zero, as it corresponds to the well-known Schwarzschild case. However, the impact of NED on particle dynamics and light propagation in BH spacetimes presents an intriguing and potentially fruitful area for further exploration. Particularly fascinating is the investigation of strong gravitational lensing effects in NED BHs with accretion disks, especially as seen from the perspective of edge-on observers. Such research involves meticulous examination of higher-order photon rings for both static and stationary NED BHs. To ensure astrophysical reliability, all these studies must consider the constraints provided by the outcomes of the EHT. Subsequently, comparing the lensing features of NED BHs, including caustics, with observational expectations will be a crucial step. These investigations are left as topics for future works.

%%%%%%%%%%%%%%%%
%%%%%%%%%%%%%%%%
\section*{Acknowledgement}

The paper was funded by the National Natural Science Foundation of China 11975145. The work of S.C. is supported by the SERB-MATRICS grant MTR/2022/000318, Govt. of India. M.F. acknowledges financial support from Vicerrector\'{i}a de Investigaci\'{o}n, Desarrollo e Innovaci\'{o}n - Universidad de Santiago de Chile (USACH), Proyecto DICYT, C\'{o}digo 042331CM$\_$Postdoc. A. {\"O}. would like to acknowledge the contribution of the COST Action CA21106 - COSMIC WISPers in the Dark Universe: Theory, astrophysics and experiments (CosmicWISPers) and the COST Action CA22113 - Fundamental challenges in theoretical physics (THEORY-CHALLENGES). A.{\"O}. is funded by the Scientific and Technological Research Council of Turkey (TUBITAK).

\bibliography{references}
\end{document}